\documentclass[10pt,sigconf]{acmart}

\usepackage{CJKutf8}

\usepackage{booktabs} %
\usepackage[inline]{enumitem} %
\usepackage{caption}
\usepackage[multidot]{grffile}
\usepackage{graphics}
\usepackage{tikz-imagelabels}
\imagelabelset {
  border thickness = 0.4pt,
  arrow thickness = 0.3pt,
  arrow distance = 0pt,
  tip size = 0.4pt,
  outer dist = 2mm,
}
\usepackage{float}
\usepackage{subfig}
\usepackage{hyperref}
\usepackage{amsmath}
\newcommand{\nquad}[1]{\hspace*{#1em}\ignorespaces}
\newcommand{\qqquad}{\nquad{3}}
\newcommand{\qqqquad}{\nquad{4}}
\newcommand{\qqqqquad}{\nquad{5}}
\usepackage{xspace}
\usepackage{algpseudocode}

\usepackage{tikz}

  \setcopyright{rightsretained}
  \copyrightyear{2022}
  \acmYear{2022}
  \acmDOI{}
  \acmConference{Technical Report}{Feb.~2021}{Marina del Rey, California, USA}
  \acmISBN{}


\makeatletter
\newcommand\EatSpacesHack{\@bsphack\@esphack}
\makeatother

  \renewcommand\comment[1]{\EatSpacesHack}
  \newcommand\sx[1]{\EatSpacesHack}
  \newcommand\PostSubmission[1]{\EatSpacesHack}
  \newcommand\reviewfix[1]{\EatSpacesHack}

\def\Snospace~{\S{}}

\hypersetup{
	colorlinks,%
	citecolor=[rgb]{0.7,0,0},
	filecolor=[rgb]{0.2,0.2,0.2},
	linkcolor=[rgb]{0,0,0.7},
	urlcolor=[rgb]{0.7,0,0},
}

  \newcommand{\OurU}{USC\xspace}
  \newcommand{\AnonOrClear}[2]{#2}

  \newcommand{\SitesNum}{4\xspace}
  \newcommand{\SitesWord}{four\xspace}
  \newcommand{\SitesList}{ejnw\xspace}
  \newcommand{\CSBlockNumLow}{168k\xspace}
  \newcommand{\CSBlockNumHigh}{420k\xspace}

\newcommand{\RelaxFloats}{
	\renewcommand{\topfraction}{0.9}
	\renewcommand{\floatpagefraction}{0.9}
	\renewcommand{\textfraction}{0.1}
}
\RelaxFloats

\title[Measuring the Internet during Covid-19 to Evaluate Work-from-Home]{Measuring the Internet during Covid-19 \\ to Evaluate Work-from-Home}
  \subtitle{Technical Report: arXiv:2102.07433v5 [cs.NI] \\ released 2021-02-17, updated 2022-06-10}

\author{Xiao Song}
\affiliation{
	\institution{University of Southern California}
	\country{United States}
}

\email{songxiao@usc.edu}

\author{John Heidemann}
\affiliation{
	\institution{University of Southern California}
	\country{United States}
}
\email{johnh@isi.edu}

\begin{document}

\begin{abstract}
Covid-19 has radically changed our lives,
  with many governments and businesses mandating work-from-home (WFH)
  and remote education.
\reviewfix{}
However, whether or not work-from-home policy being taken is not always known publicly,
  and even when enacted, compliance can vary.
These uncertainties suggest a need to \emph{measure} WFH
  and confirm actual policy implementation.
We show new algorithms that detect WFH from
  changes in Internet address use during the day.
We show that \emph{change-sensitive} networks
  reflect mobile computer use, 
  detecting WFH from changes in network intensity, 
  the diurnal and weekly patterns of IP address response.
\reviewfix{}  
Our algorithms provide new analysis of
  existing, continuous, global
  scans of most of the responsive IPv4 Internet (about 5.1M /24 blocks).
Reuse of existing data 
  allows us to study the emergence of Covid-19, revealing global reactions.
We demonstrate our algorithms in networks with known ground truth,
  evaluate the data reconstruction and design choices with studies of real-world data,
  and validate our approach by testing random samples against news reports.
In addition to Covid-related WFH,
  we also find other government-mandated curfews.
Our results show the first use of
  network intensity to infer-real world behavior and policies.
\end{abstract}

\keywords{Active Internet Measurement, Covid-19, Work-from-Home}

\maketitle

\section{Introduction}
	\label{sec:intro}

In 2020, the Covid-19 pandemic has completely changed our lives. 
Reports show that millions of people suddenly stay at home,
  because they are studying, working there (\emph{WFH}),
  or are unemployed~\cite{baek2020unemployment}. 

\reviewfix{}
However, reactions to Covid-19 are not always certain,
  as they may vary by locations and are not publicly stated. 
Public reports of Covid-19 responses (such as WFH)
  are not always timely---policies are not always reported publicly,
  and their implementation may be uneven.
Even when a business or government establishes a policy (WFH, curfew, or lockdown),
  people may embrace or reject that policy,
  so actual participation may diverge~\cite{Epstein20a}.
\reviewfix{}
For these reasons, we would like to study WFH practice. 
We want to confirm WFH happens where it's intended
  and
  verify when it happens where it's not publicly stated.

The Internet has proven increasingly important in
  supporting society as we manage Covid-19 and WFH~\cite{Kang20a}.
Early in the pandemic,
  streaming media providers reduced video quality to manage increased use~\cite{Brodkin20a}.
Facebook reported a sharp 20\% increase in traffic 
  in late March after widespread WFH~\cite{Boettger20a}.
Researchers saw a similar 15--20\% increase in traffic at IXPs in mid-March~\cite{Feldmann20a}.
Changes depend on the perspective of the observer:
  mobile (cellular) networks show a 25\% drop in traffic
  and a decrease in user mobility as people stay at home~\cite{Lutu20a}.
While the Internet changes vary,
  these changes show potential to
  \emph{observe} compliance with WFH orders,
  given careful inference from network observations.

In this paper, we show that \emph{changes in Internet address use correlate with WFH}.
Many people use mobile devices (laptops, tablets, smartphones) 
  at home and office, and these devices acquire IP addresses
  when they are using the Internet.
Often these addresses are dynamically allocated with DHCP\@.
While many networks today allocate from private address space~\cite{Rekhter96a},
  many public IPv4 networks reflect changes in human activity~\cite{Xie07a,Cai10a,Quan14c,Moura15a,Padmanabhan16a}.

The first contribution of our work is to define a new algorithm
  that identifies changes
  in diurnal network usage of each public, /24 IPv4 block,
  a signal that correlates with WFH (\autoref{sec:methodology}).
We examine the active state of /24 blocks by counting the number of active IPv4 addresses---those
  that reply to an ICMP echo request.
We identify \emph{change-sensitive} network blocks, 
  where the number of active IPv4 addresses reflects the number of
  individuals actively using the network in that location,
  with noticeable diurnal changes.
We then 
  determine long-term trends in the usage of change-sensitive blocks
  and detect changes in these trends
  that correlate with WFH\@.

To our knowledge we are the first
  to analyze network addresses, as a whole, to
  reveal the behavior of human populations.
Prior work has used addresses as samples of space~\cite{Schulman11a},
  or to spread load~\cite{quan2013trinocular}, or reduce 
  traffic~\cite{katz2008studying,quan2013trinocular}.
Other work showed they reflect human trends~\cite{Shafiq11a,Quan14c,Dhamdhere18a}
  and ISP policies~\cite{Xie07a,Cai10a,Moura15a,Padmanabhan20a}.
\reviewfix{}
\reviewfix{}
Like all approaches using active probing of public IP addresses,
  our approach cannot see behind Network Address Translation (NAT)
  or firewalls,
  but we track WFH in 168k to 420k blocks (\autoref{sec:how_many_blocks})
  globally (\autoref{sec:block_where}),
  and use our method to discover events (\autoref{sec:results}),
  showing values in even incomplete coverage,
  particularly for regions where WFH activity is uncertain.

\reviewfix{}  
Our approach respects individual privacy.
We track changes aggregated by /24 address block;
  we do not know the identities of individuals,
Our internal and sharing policies limit
  data about specific addresses and prohibit de-identification.
Our work is classified as non-human-subjects research
  by our university's 
  IRB (USC \#UP-20-00909);
  see \autoref{sec:research_ethics} for details.

Our algorithms apply new analysis to ongoing,
  active measurements of the Internet~\cite{quan2013trinocular}.
Reusing existing data is necessary to study the unanticipated onset
  of Covid-19.
Reuse also reduces the cost of ongoing probes
  covering each of 5 million /24 IPv4 networks multiple times an hour.
Re-analysis of existing data requires care,
  since the measurement is optimized for other purposes,
  and its varying rate affects accuracy for some blocks.
We propose additional measurements to improve future reconstruction
  in \autoref{sec:alg_reconstruction}.

The second contribution of our work
  is to validate our algorithms against real-world data.
Known cases of Covid-related WFH
  with ground truth (\autoref{fig:isi_lab_example})
  motivate our design. %
We systematically evaluate
  design decisions in our algorithms (\autoref{sec:validation}).
Finally, we validate the detection of inferred WFH
  with random sample blocks (\autoref{sec:event_validation})
  and locations (\autoref{sec:validating_discoverability}).
We show that network changes near known confirmed Covid-WFH dates
  are usually captured by the algorithms (precision 93\%),
  and two randomly selected locations both show network changes near their WFH dates.

Our final contribution is to use our algorithms to study \CSBlockNumLow to \CSBlockNumHigh
  change-sensitive blocks worldwide 
  for the first six month of 2020 %
(\autoref{sec:results}).
We show global data and evaluate known, real-world events
  like Wuhan in January 2020.
\reviewfix{}  
We then demonstrate the potential of our approach to \emph{discover} changes,
  find WFH in
  the Philippines in March 2020,
  and non-Covid-related Internet shutdowns
  in India in Feb.~2020.

Data from our work is available at no cost~\cite{ANT22a},
  and we plan to release our analysis software with our paper publication.

This paper was first released as a technical report in Feb.~2021.
In May 2022 we updated all sections of the paper, including many additions:
  a summary of external probing, whit a clearer evaluation of data sources and alternatives in \autoref{sec:probing_addresses};
  description of geographic aggregation and limitations in \autoref{sec:detecting_changes};
  a new approach to adding additional observations in \autoref{sec:alg_reconstruction};
  a summary of how we share results in \autoref{sec:sharing_results};
  major revisions to validation in  \autoref{sec:validation},
    including more datasets and improved and corrected statistical analysis;
  explicit evaluation of reconstruction (\autoref{sec:reconstruction});
  evaluation of target sensitivity and change over time (\autoref{sec:how_many_blocks});  validation by both random blocks (\autoref{sec:event_validation})
    and random locations (\autoref{sec:validating_discoverability});
  new data on overall trends (\autoref{sec:overall_trends})
    and updated case studies (\autoref{sec:results});
  explicit related work (\autoref{sec:relatedwork});
  and supporting detail to much of the validation and results in appendicies.

\section{Methodology}
   \label{sec:methodology}

We detect changes in IP address use
  and infer WFH
  with the following steps:
\begin{enumerate*}
\item Probe IP addresses for activity,
\item Reconstruct active addresses,
\item Identify change-sensitive blocks,
\item De-trend address usage,
and
\item Detect changes in usage.
\end{enumerate*}
In addition, we combine data from multiple observes
  and plan for additional probing improve reconstruction.

We illustrate our approach with an example block shown in \autoref{fig:isi_lab_example}.
This block is at \OurU,
  so we know the start of WFH is on 2020-03-15,
  and the address changes we observe correspond to people at work.
\reviewfix{}
We provide other example in \autoref{sec:case_study_specific_blocks},
  and we verify their behavior with manual examination of more than 2200 blocks.  %

\subsection{Probing IP Addresses For Activity}
	\label{sec:probing_addresses}

Our approach observes active probing of the IPv4 address space.
Many active IP addresses are probed,
  reporting results as positive or non-replies. %
Addresses must be probed multiple times per day to
  show diurnal changes in responses.

\textbf{Data Sources:}
In principle, any frequent public scan of the IP space can serve as input.
We re-analyze data from Trinocular~\cite{quan2013trinocular} for three reasons.
\reviewfix{}
First, it provides continuous data since Nov.~2013,
  allowing us to study the months before and after Covid-19 swept the globe.
Second, it covers nearly all of the responsive Internet,
  with more than 5M IPv4 /24 blocks (achieving 96\% is possible with active methods~\cite{baltra2020improving}).
Finally, its 1 to 16 probes per 11 minutes
  are frequent enough that we can usually identify diurnal behavior.
However, re-purposing it to our ends requires reconstruction (\autoref{sec:accumulating_addreses}) and
  integration (\autoref{sec:multiple_observers}).
We show additional probing is sometimes helpful (\autoref{sec:alg_reconstruction}).

\reviewfix{}
We explored multiple alternative data sources.
USC Internet surveys~\cite{Heidemann08c} provide
  more frequent coverage (about $256\times$ more),
  with data for all IP addresses
  in many /24 blocks every 11 minutes,
  but they are spatially and temporally incomplete,
  covering only 4k blocks, with data only 4 weeks of every quarter.
We use survey data for validation in \autoref{sec:reconstruction}.
ZMap~\cite{zmap182948} has quick, complete scans of IPv4,
  but it preserves only positive replies.
Our algorithms require timing of both positive and negative replies,
  and reconstruction of ZMap is impossible because it
  does not preserve probe order (or initial seed).
Censys scans~\cite{censys15} emphasize daily services and certificates on hosts,
  not reachability many times per day,
  and bulk data is not currently available.
CAIDA's Archipelago~\cite{caidaArk} covers all routed /24 blocks,
  but it far too slow to track diurnal trends.
Its 3 teams of 17-18 probers cover all routed blocks every 2-3 days,
  about 1/256th the rate of Trinocular.
Thus while these systems are tuned for their goals,
  none can be adapted to support diurnal analysis of most of IPv4
  or of 2020,
  although they may be extended for future use.

Another option is to deploy new, dedicated probing.
We propose additional probing in \autoref{sec:alg_reconstruction}
  to improve coverage for some blocks.
However,  
  to see what happened when Covid-19 first spread in
  2020 requires the use of data taken at that time.
In addition, reuse avoids the small but non-zero cost
  that active probing places on the world's networks.
Reuse also avoids duplication of existing opt-out mechanisms and abuse handling,
  a noticeable operational cost of sustained probing.

\textbf{Specific datasets:}
We list specific datasets covering October 2019 
  through June 2020 in an appendix (\autoref{tab:datasets}).
This data is available at no cost to researchers.
\reviewfix{}
\reviewfix{}
We consider data from six locations
(coded c: Colorado; e: Washington, DC; g: Greece; j: Tokyo; and n: Netherlands; w: Los Angeles).
These sites provide very diverse perspectives,
  since each has different upstream ISPs
  and they are on three continents.

Prior work describes Trinocular collection in detail~\cite{baltra2020improving,quan2013trinocular},
  to summarize:
  each site probes about 5M /24 IPv4 address blocks with ICMP echo request messages.
Every 11 minutes each site probes from 1 to 16 targets in each block,
  drawing them from a list with a  
  pseudorandom order that is fixed for each quarter.
\reviewfix{}
Targets are limited to
  addresses that have ever responded to a complete scan sometime in the last
  three years, written as $E(b)$. 
Probe rates are low enough that rate limiting is unlikely~\cite{Guo18a}.
Although we consider data from all locations, we discard sites c and g in 2020
  because of hardware problems.
\reviewfix{}  
We merge data across sites as described next
  in \autoref{sec:accumulating_addreses}.

\textbf{IPv6:} 
Our detection depends on seeing diurnal changes
  in address usage.
Such changes exist in IPv6 and
  are prominent in Google's IPv6 reports~\cite{Google21a}.
However,
  the size and design of IPv6 addressing prevent exhaustive probing~\cite{Gont14a,Gont16a}
  and no IPv6 data is available to us today.
IPv6 address measurement is an active area of research~\cite{Gasser16a,Foremski16a,Murdock17a,Beverly18a}
  and we hope to explore it in future research.

\subsection{Reconstructing Active Addresses}
	\label{sec:accumulating_addreses}

\reviewfix{}

Our data source %
   scans the visible Internet incrementally in rounds,
   so our first step is to reconstruct the state of the Internet
   from these incremental results.

\reviewfix{}
Each Trinocular round occurs every 11 minutes and
  probes 1 to 16 addresses.
We ingest results from each round
  and count which addresses are active.
When all ever-active addresses ($E(b)$) have been observed,
  we have a complete \emph{reconstruction} of the block.
We report this value and then update incrementally each subsequent round.

\reviewfix{}
Accumulating state over multiple rounds
  leverages the fixed probing order.
We assume addresses do not change state until they are re-scanned.
\reviewfix{}
This assumption holds if we scan faster than addresses change.
If we scan 15 addresses per round and have 256 addresses to cover,
  then 17 rounds guarantee a complete picture.
Combining data from six observes (described below \autoref{sec:multiple_observers})
  will yield a full scan in only 3 rounds (33 minutes)
However, in the worst case, we observe only one address per round,
  so 6 observers complete a scan in 8 hours.
The size of a complete block ($E(b)$) varies by history (updated each quarter),
  and the probe rate per round varies over the day based on responses (updated dynamically),
  so upper and lower bounds are atypical.
In \autoref{sec:scan_time}, we examine scan times in practice,
  and in \autoref{sec:reconstruction} we consider how accumulation
  affects accuracy.

\reviewfix{}
We update our response estimate incrementally,
  as each additional round provides more data.
For example, an address reporting negative in the previous round
  that then receives a positive response 
  increments the number of active addresses of the block.
Thus, we generate new estimates of the number of active addresses 
  every 11 minutes, but each estimate reflects observations from multiple prior rounds.

\autoref{fig:isi_diurnal_block_80099000} shows counts of active addresses for
  our example block.
While 88 addresses in the last 3 years
  ($|E(b)|$, the top red line),
  but only 8 to 18 are active during these three months 
  (the lower blue line).

\subsection{Identifying Change-Sensitive Blocks}
  \label{sec:Identifying_Change_Sensitive_Blocks}
To detect WFH,
  we must have blocks where address changes
  reflect people's daily schedules.
We call such blocks \emph{change-sensitive},
  and identify them by two characteristics: 
  first, they show a regular, \emph{diurnal} pattern;
  second, the \emph{swing} (high and low count)
  over 24 hours must be large enough to detect its disappearance with confidence.

\begin{figure}
\subfloat[Active addresses over 3 months (input data).]{
    	\label{fig:isi_diurnal_block_80099000}
    	\hspace*{-5mm}\mbox{\includegraphics[width=0.85\columnwidth,trim=20 35 45 40]{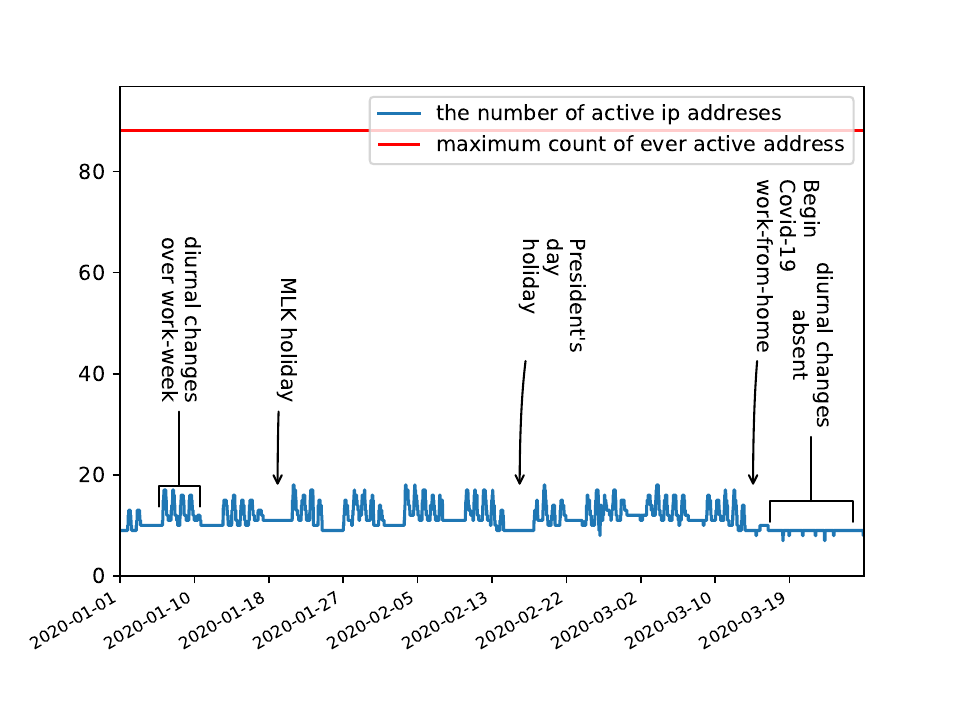}}}  \\
\subfloat[Active-address trend (top), its seasonal component (middle), and residual (bottom), from STL decomposition.]{
	\label{fig:decomp_80099000}
	\mbox{\includegraphics[width=0.84\columnwidth,trim=70 30 90 150,clip]{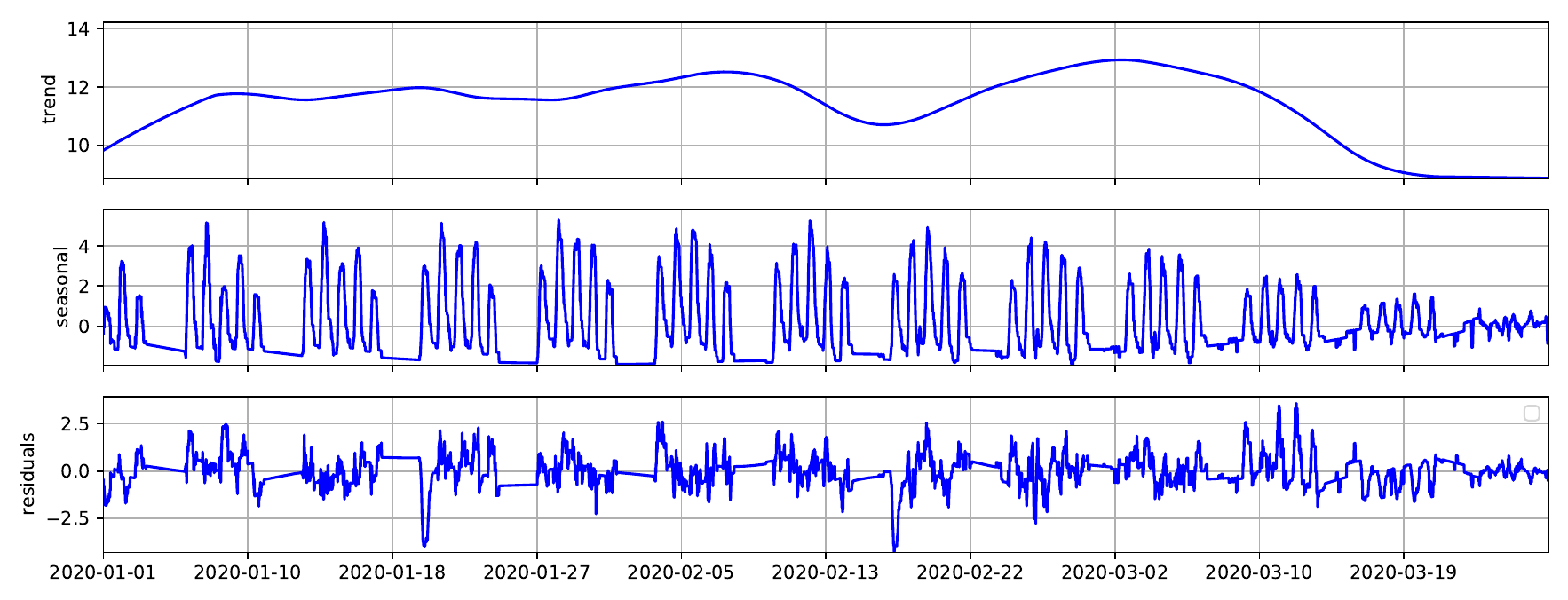}}} \\
\subfloat[CUSUM change detection on 2020-03-15 showing start, alarm, and end (arrows and dot on top), with threshold values (bottom) for rise (yellow +) and fall (purple -).]{
	\label{fig:CUSUM_detrend_block_80099000_a39w}
	\mbox{\includegraphics[width=0.8\columnwidth,trim=0 10 10 130,clip]{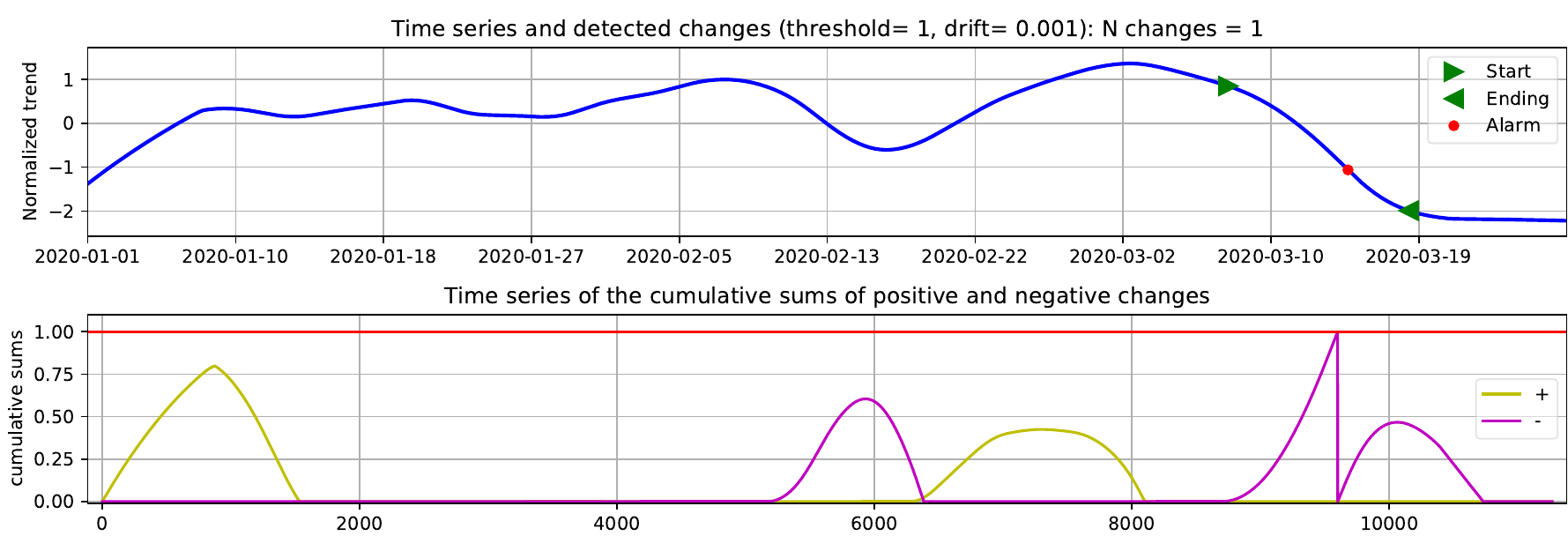}}
}
\caption{A block (\AnonOrClear{192.0.2.0/24}{128.9.144.0/24})
    illustrating address usage changes due to confirmed WFH.
}
	\label{fig:isi_lab_example}
\end{figure}

Our example block (\autoref{fig:isi_diurnal_block_80099000})
  is a known change-sensitive block.
The active addresses over time (the blue line)
  usually shows groups of five bumps, corresponding to the work-week,
  followed by two days of flat behavior over the weekend.
We also see known holidays on 2020-01-20 and -02-17.

\textbf{Diurnal blocks:} 
We identify diurnal blocks by taking the FFT of the
  active-address time series and looking for energy
  in frequencies corresponding to 24~hours, or harmonics of that frequency.
This approach follows prior work~\cite{Quan14c}, 
  which shows that IP addresses often reflect diurnal behavior,
  particularly in Asia, South America, and Eastern Europe.

The sudden absence of the diurnal pattern indicates a network change
  consistent with WFH\@.
The block of \autoref{fig:isi_diurnal_block_80099000}
  is diurnal from 2020-01-01 to 2020-03-15
  and becomes flat after 2020-03-15,  corresponding to the start of WFH\@.

\textbf{Persistent daily swing:}
We look for blocks that have a ``wide'' daily swing in addresses,
  as described below.
We define the daily swing as the range of addresses
  (maximum seen minus minimum)
  over midnight-to-midnight UTC\@.

\begin{figure*}
	\centering
	\begin{minipage}[b]{.32\linewidth}
		\includegraphics[width=1\columnwidth]{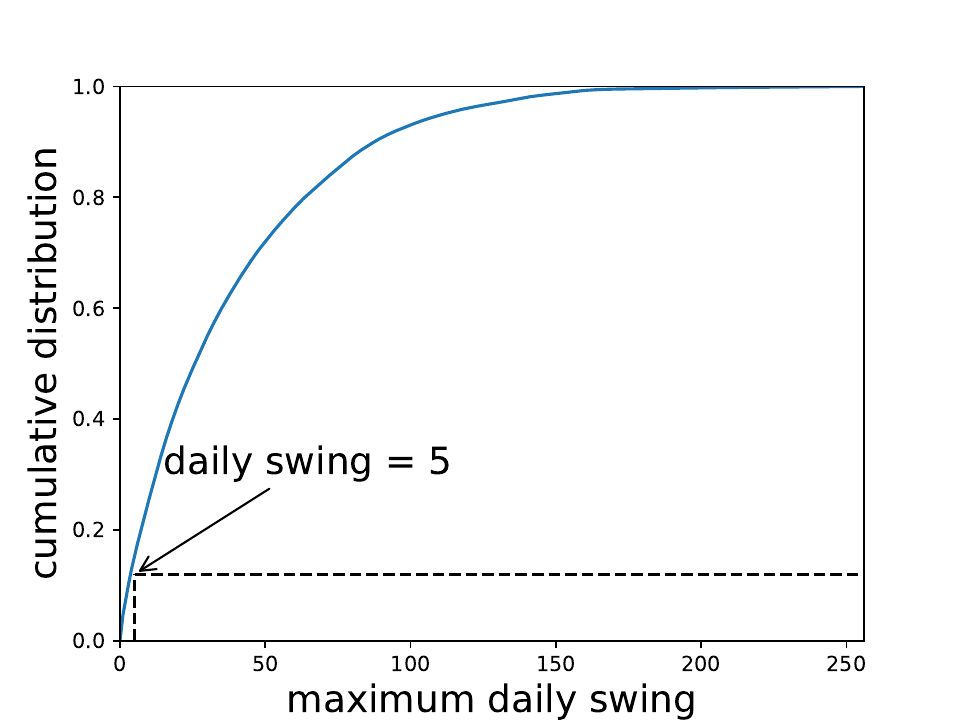}
		\caption{Cumulative distribution of maximum daily swing of all diurnal blocks in 2020q1.}
		\label{fig:cdf_daily_swing_diurnal}
	\end{minipage}
	\hfil
	\begin{minipage}[b]{.32\linewidth}
		\includegraphics[width=1\columnwidth]{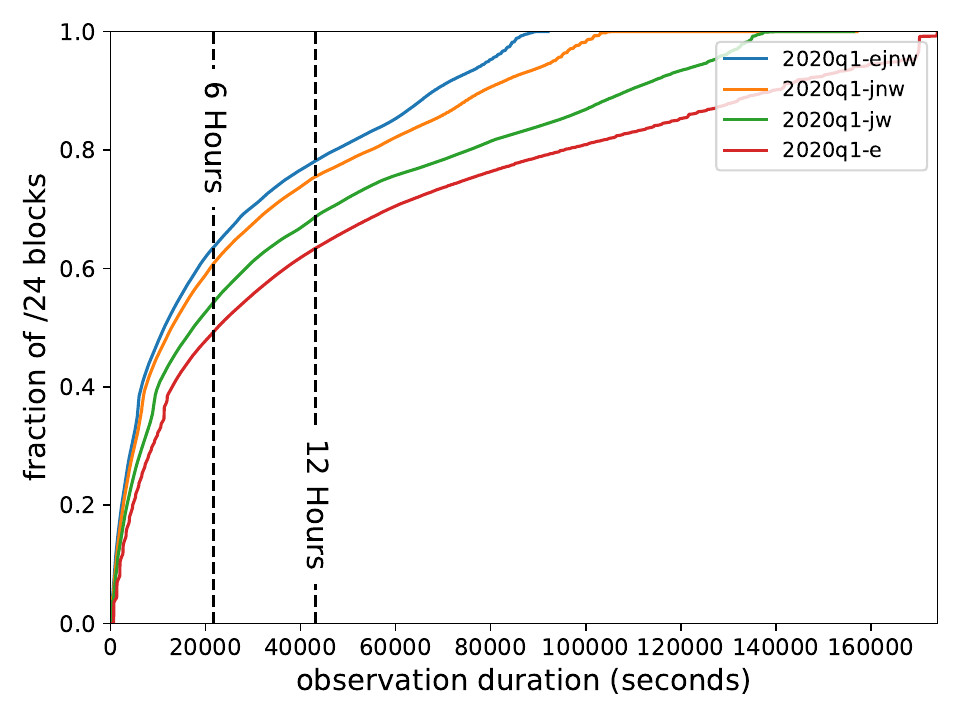}
		\caption{ {\small Cumulative distribution over all blocks of time to complete a scan of all known active addressees in 2020q1.}
		}
		\label{fig:cdf_fbs_time}
	\end{minipage}
	\hfil
	\begin{minipage}[b]{.32\linewidth}
		\mbox{\includegraphics[width=1\columnwidth,trim=10 30 40 40,clip]{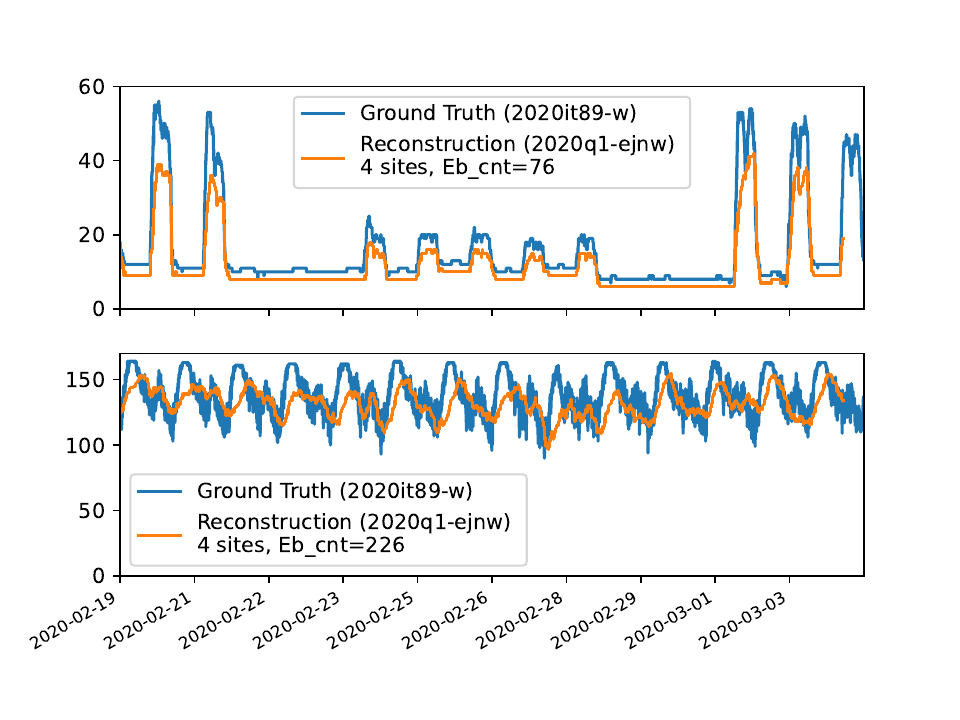}}
		\caption{Comparing two reconstructed /24 blocks with their ground truth. 			
		}
		\label{fig:trin_survey_block_cbc3c100_c136bb00}
	\end{minipage}%
\end{figure*}

A wide swing is a change of more than $s$ addresses per day.
We currently use a threshold $s=5$,
  based on the evaluation of the data.
\reviewfix{}  
Too large a threshold will reduce the number of accepted blocks,
  but too small makes the algorithm vulnerable to noise such as individual computer restarts.
\reviewfix{}
\reviewfix{}
We select 5 as the minimum value that
  tolerates uncorrelated outages caused by a few computers 
  (for example, due to maintenance).
\reviewfix{}
\autoref{fig:cdf_daily_swing_diurnal} shows that around 90\% of
  blocks that have daily swings exhibit a swing greater than 5. 

\reviewfix{}
Finally, the swing must be \emph{persistent}
  and reflects a \emph{work week}.
Changes need not occur every day,
  since many blocks (like \autoref{fig:isi_diurnal_block_80099000})
  show use primarily during the work-week and not on weekends and holidays.
We require blocks to have a wide 
  daily swing for at least 4 of 7 consecutive days
  for at least one week in the observation period.
\reviewfix{}
We use a 7-day window since work activity usually follows weeks,
  and a 4 day minimum to tolerate 3-day weekends
  (for example, the week of 2020-01-20 in \autoref{fig:isi_diurnal_block_80099000}).

\textbf{Coverage:}
We detect WFH in change-sensitive blocks as 
  the sudden disappearance of diurnal changes.
We must therefore identify   
  a pre-Covid baseline of change-sensitive blocks.
In \autoref{sec:how_many_blocks} we show that Jan.~2020
  provides a good baseline.
In \autoref{sec:how_many_blocks} we also show that
  between \CSBlockNumLow and \CSBlockNumHigh blocks meet these two requirements,
  depending on how much and how long we collect data.
We ignore non-change-sensitive blocks
  since their operation (perhaps firewalls or NAT)
  hides WFH changes, as we discuss in \autoref{sec:detecting_changes}.

\subsection{De-trending Address Usage}
	\label{sec:detrending}

Although the diurnal changes in our example (\autoref{fig:isi_lab_example})
  can be seen visually,
  in many blocks 
  daily fluctuations make it difficult to detect changes in use.
We expect WFH will either
remove the diurnal swing,
as occurs in \autoref{fig:isi_diurnal_block_80099000} after 2020-03-15,
\emph{or} decrease overall use (and possibly also the size of the swing),
as fewer people come into work.
These signals are properties of the general baseline of active addresses
  and are obscured by daily changes and day-to-day variation,
  so we need to extract this trend from noise.

We track the underlying baseline
  by applying a standard seasonality model to the data.
Seasonality models decompose the signal into a baseline
  convoluted with a daily and possibly weekly signal.
We considered two models:
  the ``naive'' seasonality model~\cite{seabold2010statsmodels}
  and Seasonal-Trend decomposition using LOESS (STL)~\cite{cleveland1990stl,seabold2010statsmodels}.
\reviewfix{}
Although both are similar, we adopted the STL 
  for our work after comparing the two and finding it
  more robust to outliers.

\autoref{fig:decomp_80099000} shows the decomposition from our sample block
  (\autoref{fig:isi_diurnal_block_80099000})
  into trend, seasonal, and residual components.
\reviewfix{}
The seasonal component (middle) models daily and weekly changes,
  while the trend (top) captures the long-term mean value,
  and residual (bottom) shows any remaining error.

\subsection{Detecting Changes in Usage}
	\label{sec:detecting_changes}

Finally, we detect changes in the trend.
We apply a standard change-point detection algorithm, CUSUM~\cite{Duarte2020,gustafsson2000adaptive}.
CUSUM looks for changes in the baseline,
  flags when the upward or downward trend begins
  and the time of largest change.

Before applying CUSUM,
  we normalize the STL trend to its $z$-score
  by subtracting the mean and dividing by standard deviation.
This normalization allows
  us to use the same CUSUM parameters for every block (threshold: 1, drift: 0.001).

\autoref{fig:CUSUM_detrend_block_80099000_a39w} continues our
  example block showing CUSUM detection.
The bottom graph shows the cumulative increase and decreases (in dark purple and light yellow).
The upper graph shows the normalized trend
  with start and end of the detected change
  as arrows on 2020-03-08 and -18.
The point of change is 2020-03-15,
  which we confirm as to when WFH began.
This change is detected from the fall in normalized trend,
  reflecting the drop in address activity
  due to the absence of diurnal address usage
(\autoref{fig:isi_diurnal_block_80099000}).

\textbf{Geographic Aggregation:}
We geolocate all blocks (using Maxmind GeoLite~\cite{Maxmind20a}).
We count blocks showing a decreasing trend
  (the purple line in \autoref{fig:CUSUM_detrend_block_80099000_a39w})
  in each $2 \times 2^{\circ}$ geographic region.
\reviewfix{}
\reviewfix{}
This downward trend reflects a decrease in aggregate use of
  those networks,
  and 
  our definition of change-sensitive reflects the work-week,
  so a large number of blocks showing a downward trend suggests increased WFH.
Possible future work is to use bumps per week to 
  distinguish work from home networks.

\textbf{Limitations and Other Sources of Change:}
CUSUM can automatically find behaviors often indicating WFH
  in change-sensitive blocks.
In \autoref{sec:event_validation} we validate this claim,
  and \autoref{sec:Some_Events} shows
  events we discovered with geographic aggregation.
But WFH can happen for reasons other than Covid,
  and network changes for reasons other than WFH (like \autoref{sec:CaseStudy:NewDelhi}).

Network outages are an additional possible source of downward changes in usage.
An outage will be a downward change, followed by an upward change
  when the network recovers.
Since outages are usually short (minutes or a few hours)
  relative to Covid-WFH (weeks or months),
  we treat closely timed down and upward changes as outages.

Our approach is limited to detecting WFH in change-sensitive blocks.
We cannot see changes that occur behind firewalls or NAT.
Our results will be less successful in countries
  where most individuals are behind always-on NAT devices,
  such as the U.S.~and western Europe.
In other countries, ISPs place wired customers behind CG-NAT,
  sometimes with multiple layers (we have seen this in Iran).
\reviewfix{}  
We study locations of change-sensitive blocks in \autoref{sec:block_where},
  but given the complexities of NAT in multiple prior studies~\cite{Kreibich10a,Grover13a,Richter16c},
  detailed quantification of its relationship with WFH detection is future work.

\reviewfix{}
Carrier-Grade NAT (CG-NAT) is widely used for mobile phones,
  and also by many wired ISPs~\cite{Richter16c}.
Like home-based NAT, CG-NAT may hide diurnal trends,
  but CG-NAT assignment strategies such as paired pooling
  may directly expose diurnal trends
  by assign individuals temporary
  public IP addresses when they are actively using the Internet.
The relationship between mobile phones and WFH is further complicated
  by opportunistic use of wifi instead of the cellular network when at home~\cite{Lutu20a}.
Our work helps to motivate
  future exploration of how CG-NAT and mobile networks interact with public IP address use.

The success of our approach depends on seeing some users in many locations,
  and it does not require seeing all users everywhere.
We show that 
  our approach provides broad coverage,
  we see \CSBlockNumLow to \CSBlockNumHigh change-sensitive blocks (\autoref{tab:block_count})
  in many countries (see \autoref{sec:block_where}).
\reviewfix{}
Although most U.S.~home users are behind always-on routers with NAT
  and we therefore cannot detect increase in home use,
  we see WFH in U.S.~universities (\autoref{sec:some_university_blocks}).
Our widespread but incomplete coverage
  suggests that our results are best used to estimate \emph{trends}
  in WFH and not absolute counts of individual choice.

\subsection{Using Multiple Observers}
	\label{sec:multiple_observers}

We estimate network usage in \autoref{sec:accumulating_addreses}
  with observations from a single observer.
However, we have data available from multiple observers
  in different geographic locations (\autoref{sec:probing_addresses}).
\reviewfix{}
Each observer probes the same targets in the same order,
  but they start independently and run unsynchronized,
  so they are almost always out of phase with each other.
  
We can strengthen our results by combining data from multiple observers,
  either to accumulate observations rapidly, 
  or to validate the results of one against the others.

To provide more complete data,
  we can combine multiple results from all observers
  to update block status more rapidly.
As described in \autoref{sec:accumulating_addreses},
  blocks update at different rates,
  in some cases failing to track diurnal changes. 
Combining multiple observers reduces time to scan the full block
  as we will show in \autoref{sec:scan_time},
  improving reconstruction accuracy (\autoref{sec:reconstruction}).

Alternatively,
  separate observers can be treated as equivalent but independent,
  allowing us to 
  test their results against each other
  and detect bias specific to any vantage point.
(Locations sometimes see different replies~\cite{Heidemann08c, Wan20a}.)

\reviewfix{}
We follow the first approach, combining results from multiple observers,
  since the additional information improves reconstruction of the true signal
  (see \autoref{sec:reconstruction}),
  increases our coverage (\autoref{sec:results}),
  and reduces worst-case scan time for a full block.
Combining data reduces per-site bias,
  but we found that trusting all sites equally results in false results
  when one site has observation problems.
We check data health,
  and site-specific problems prompt us to discard data from two observers
  (c and g).
  We confirmed that these sites had hardware or network problems.

Our example in \autoref{fig:isi_lab_example}
  has a two-hour scan time with data from one observer.
\reviewfix{}  
\autoref{fig:trin_survey_block_cbc3c100_c136bb00} compares
  ground truth (purple, with survey data probing all addresses every 11 minutes)
  with a 4-observer reconstruction
  for two blocks (yellow).

\subsection{Adding Additional Observations}
	\label{sec:alg_reconstruction}

Reusing existing data (\autoref{sec:accumulating_addreses})
  means some reconstructed blocks are under-observed.
Beyond combining all observers (\autoref{sec:multiple_observers}),
  we next describe deploying an additional observer
  designed specifically to cover previously under-observed blocks.
Additional observations require two decisions:
  which blocks need additional observations,
  and how make those observations.

Blocks that need additional observations are those
  with many addresses which always respond,
  as our existing data source (Trinocular)
  stops probing on the first positive response for the block,
  as mentioned in \autoref{sec:accumulating_addreses}.
We identify such blocks by
  the block refresh rate (\autoref{sec:scan_time}),
  as estimated from each block's
  historical response rate and
  the number of addresses that will be scanned
  ($E(b)$ from \autoref{sec:accumulating_addreses}).

Additional observations are taken by a designed observer,
  given a list of blocks that would be under-observed.
This prober runs the standard Trinocular algorithm,
  but extends each round with
  up to four extra probes per round, even after a positive response.
We adjust the number of additional probes
  based on the current observation rate to meet our goal.
We then combine these additional observations
  with other observers as in \autoref{sec:multiple_observers}.
Together, aggregated observations
  will scan the worst-case block (256 addresses, all always responding)
  in 352 minutes,
  and these four complete scans per day
  allow detection diurnal behavior for all blocks.

\reviewfix{}
In \autoref{sec:reconstructed_quality_improvement} we show
  that this strategy identifies blocks that otherwise would
  have insufficient reconstruction.
We are currently testing implementation of additional probing;
  regrettably, it cannot be applied retroactively to our 2020 data.

\subsection{Sharing the Results}
       \label{sec:sharing_results}

Our WFH detection results are available on our
  website~\cite{Stutz21b},
  with Google-maps-style pan and zoom,
  as well as custom visualizations of time series and ISPs that
  change~\cite{Stutz21a}.
Our WFH detection data is available to researchers at no cost~\cite{ANT22a}.
%
%


\section{Validating Design Choices}
	\label{sec:validation}

We next evaluate the algorithm design using our datasets.
We begin with design decisions:
  Do we track block states quickly enough to see diurnal changes?
  How accurate is the reconstruction?
  How many blocks do we see, and where are they?
We then evaluate end-to-end results:  
  do detections and discoveries match reports of real-world WFH\@?

\subsection{Block Refresh Rate}
        \label{sec:scan_time}

We examine how quickly we complete scans of each /24 block,
  a \emph{full block scan} (FBS).
Prior work has suggested sparsely occupied blocks are fully scanned
  in about two hours in the worst case~\cite{baltra2020improving},
  but that analysis covered only the subset of all blocks 
  that were intermittently responsive.
Our new analysis here examines change-sensitive blocks,
  a different subset.

\reviewfix{}
Prior work explored the bound of FBS time
  as part of developing a new algorithm
  to address false outages in sparse blocks, 
  those with low response rates~\cite{baltra2020improving}.
\reviewfix{}
This analysis showed 3.1~hours (17 rounds) as an upper bound for scanning sparse blocks,
  based on 15 probes per round (a chosen design limit), over 11 minutes,
  and 256 addresses to cover in the block.
A more general worst case will consider blocks
  where all 256 addresses always respond,
  so only one address is probed per round and a full scan requires 256 rounds (1.8 days).
Such blocks are not change-sensitive,
  but this rare worst-case suggests we need to look at block scanning duration
  empirically.

We evaluate time to scan all ever-responsive addresses ($E(b)$) in each block
  across all change-sensitive blocks in 2020q1.
\autoref{fig:cdf_fbs_time} shows cumulative distributions
  for four cases: a single observer,
  then combined data from two, three, and \SitesWord observers,
  from bottom to the top respectively.
We see that about 65\% of change-sensitive blocks
  can be fully scanned in 6 hours or less (the left vertical dashed line),
  when data is combined from \SitesWord observers.
By contrast, 
  6 hours provides about 48\% of blocks with one.
Given 12 hours, \SitesNum observers cover 78\% of blocks,
  compared to 61\% with one.
This result shows that multiple observers are important to 
  see most diurnal changes (\autoref{sec:multiple_observers}),
  and those additional observations are required
  to handle the tail of challenging blocks (\autoref{sec:alg_reconstruction}).

\subsection{Reconstruction Quality}
 \label{sec:reconstruction}

\reviewfix{}
Change sensitivity requires block reconstruction that is sufficient
  to detect diurnal changes and swing.
To evaluate when and why reconstruction is sufficient,
  we next compare
  our reconstruction to ground truth from complete data.
Our estimates of block refresh rates (\autoref{sec:scan_time})
  suggest bounds on what we can measure
  (a refresh rate of 24 hours is below the Nyquist rate
  and so cannot track diurnal changes),
  but observation and the underlying changes are both non-linear
  and so it provides a pessimistic bound.

We compare reconstruction against ground truth:
  Internet address surveys (2020it89-w in \autoref{tab:datasets})
  scan all addresses in a block every 11 minutes
  for two weeks.
Surveys cover about 2\% of the responsive blocks in the IPv4 Internet;
  we find 32,437 blocks overlap between surveys and our data.
\autoref{fig:trin_survey_block_cbc3c100_c136bb00}
  shows two representative blocks (from the 5,440 overlap)
  to illustrate how block refresh rate and sampling rate
  affect reconstruction quality. 

\begin{table}
	\begin{footnotesize}
		\hspace*{-2ex}\begin{tabular}{l |  r r  r  r r}
			\textbf{Dataset:} & \begin{tabular}{@{}c@{}}2020it89 \\-w \end{tabular} & \begin{tabular}{@{}c@{}}2020q1 \\ -w \end{tabular} & \begin{tabular}{@{}c@{}}2020q1 \\ -\SitesList \end{tabular}  & \begin{tabular}{@{}c@{}}2020m1 \\ -\SitesList \end{tabular} & \begin{tabular}{@{}c@{}c@{}} 2020it89\\-match-\\ \SitesList \end{tabular} \\ 
			\hline
			duration (weeks) &  2  &  12 & 12 & 4 & 2 \\
			completeness (sites) &  full & 1 site & \multicolumn{3}{c}{\ldots \SitesNum sites \ldots} \\
			& & \multicolumn{4}{c}{(*intersected with it89-w) } \\
			\hline
			responsive  &  \multicolumn{5}{c}{32,437 }   \\
			\hline
			\quad      not diurnal & 25,170 & 30,137 & 29,493 & 29,049 & 27,674\\
			\quad      diurnal &  7,257 & 2,300 & 2,944 & 3,388 & 4,763\\
			\hline
			\quad      narrow swing & 15,104 & 12,597 & 11,112 & 11,112 & 11,112 \\
			\quad      wide swing &  17,333 & 19,840 & 21,325 & 21,325 & 21,325 \\
			\hline
			\quad      not change-sensit.  & 26,997 & 30,630 & 30,434 & 29,890 & 28,643\\
			\quad      \textbf{change-sensitive} &  5,440 & 1,807 & 2,003 & 2,547 & 3,794\\
		\end{tabular}
	\end{footnotesize}
	\caption{The number of each type of /24 blocks detected by Trinocular reconstruction and Internet address surveys.}
	\label{tab:reconstructed_vs_survey}	
\end{table}

\subsubsection{Quantifying Reconstruction Success}
	\label{sec:algorithm_success}

We first measure the success rate of block reconstruction 
  across all blocks with ground truth.
\autoref{tab:reconstructed_vs_survey}
  compares block counts for our algorithms (change detection)
  and its components (diurnal and wide swing).
First, two weeks of full survey data (2020it89-w)
  define ground truth (probing all addresses every 11 minutes).
We intersect the data with four reconstruction options:  
  one observer for a quarter (2020q1-w),
  \SitesWord observers for a quarter (2020q1-\SitesList),
  \SitesWord observers for a month (2020m1-\SitesList),
  and \SitesWord observers for two weeks (2020it89-match-\SitesList).
The last dataset shares the same starting time and duration as the survey data.
We expect more observers to provide better quality reconstruction.

Overall, of the 5,440 change-sensitive blocks in ground truth,
  the \SitesNum-site, 2-week reconstruction discovers 3,794, 70\% of truth.
Below, in \autoref{sec:reconstruction_causes}, 
  we show that the main reason it misses blocks is that they 
  do not appear to be diurnal in the reconstruction.
We then show that using a shorter duration and more sites helps to get better reconstruction.

First, we see that a shorter observation detects more change-sensitive blocks:
  comparing 2020m1-\SitesList to 2020q1-\SitesList (the third and fourth columns, 
  one month against three months),
  shows one month detects 2,547 change-sensitive blocks 
while three months reduces that to 2,003.
While reducing duration to 2 weeks, 
  2020it89-match-\SitesList finds the most change-sensitive blocks,
  confirming the observation.
\reviewfix{}
In the first quarter of 2020, we speculate that these changes may be due to Covid-WFH\@.

Second, we see that combining data from \SitesWord sites improves detection
  of change-sensitive blocks, confirming that decision (\autoref{sec:multiple_observers}).
Comparing 2020m1-\SitesList or 2020q1-\SitesList (the third and fourth columns)
  to 2020q1-w (the second column):
  \SitesWord sites provide roughly \SitesWord times more observations,
  allowing better reconstruction of address usage over the day
  and therefore more frequent detection of diurnal blocks 
  (2,547 or 2,003 instead of only 1,807).

\subsubsection{Causes of Imperfect Reconstruction}
	\label{sec:reconstruction_causes}

To understand why shorter duration and combining multiple sites help,
  we next look at the two components of change sensitivity:
  checks for diurnal behavior and consistent swing.
The middle rows of \autoref{tab:reconstructed_vs_survey}
  show how many blocks pass each of these checks in blocks
  that are present in the ground truth (2020it89-w) and our four reconstruction options.

Duration of observation strongly affects diurnal detection:
  2020m1-\SitesList finds 3,388 diurnal blocks (47\% of ground truth), 
  but reconstructions using all three months detect 2,944 
  or 2,300 (\SitesNum and 1 site, finding 41\% and 32\% of ground truth).
With the same duration that the survey data has, 
  2020it89-match-\SitesList finds 4,763 diurnal blocks, 
This decrease supports our claim that diurnal behavior changed for some blocks
  throughout 2020q1.
Data with a longer duration reduces detection of diurnalness
  for blocks that are diurnal over short period but not over longer period.

However, reconstruction seems to increase the amount of swing:
  it finds 19.8k to 21.3k blocks with a wide swing
  compared to 17.3k in the ground truth (a 14\% to 23\% overestimate).

These differences between reconstruction and ground truth are because reconstruction
  is based on less data than ground truth.
Trinocular optimizations that minimize probing for outage detection  
  effectively apply a non-linear, low-pass filter
  over active addresses.

\subsubsection{Reconstruction case studies}
\label{sec:reconstructed_blocks}

We next examine two sample blocks in detail---reconstruction
  in the first is excellent, while the second diverges from the ground truth.
These blocks help understand the factors that affect the reconstruction quality.
  \autoref{fig:trin_survey_block_cbc3c100_c136bb00} shows two blocks,
  comparing the survey's ground truth (the dark blue line)
  against our reconstruction (the light orange line).

The block in the top graph
  has an accurate reconstruction---the Pearson's correlation coefficient is 0.89
  and the block is fully scanned in 3300\,s (just under an hour).
The quality of the shape matches well,
  with the orange reconstruction showing
  the same daily peaks matching working hours,
  the same sharp changes at the start of each day,
  and the same flat behavior between workdays.
The main difference is the maximum number of detected addresses
  in reconstruction is about 9\% lower than truth.
This shortfall is because Trinocular stops probing after a success,
  so the discovery of new addresses is slow.  
However, the reconstruction preserves change sensitivity.

The block in the lower figure shows a greater challenge for reconstruction.
This block is heavily used (120 to 160 active addresses),
  and it sees a large daily shift, with consistent changes every day of the week.
With so many active addresses, a full scan requires around 8 hours
  and the reconstruction (orange) lags the truth (blue).
Here the limited input to the reconstruction
  ``spreads out'' some of the true behavior of the block,
  flattening the peaks and raising the valleys.
The correlation coefficient is 0.40.

\reviewfix{}
Both blocks show imperfect reconstruction,
  but reconstruction is sufficient for change-sensitivity
  and these blocks are used in WFH detection.
These examples show the opportunity of additional probing
  and the reasons blocks can be misclassified
  in \autoref{tab:reconstructed_vs_survey}.

\subsubsection{Additional probing to improve reconstruction quality}
\label{sec:reconstructed_quality_improvement}

\begin{figure}
	\subfloat[The count of change-sensitivity failures in reconstruction compared to ground truth.]{
		\label{fig:reconstruction.failed}
		\mbox{\includegraphics[width=0.9\columnwidth,clip]{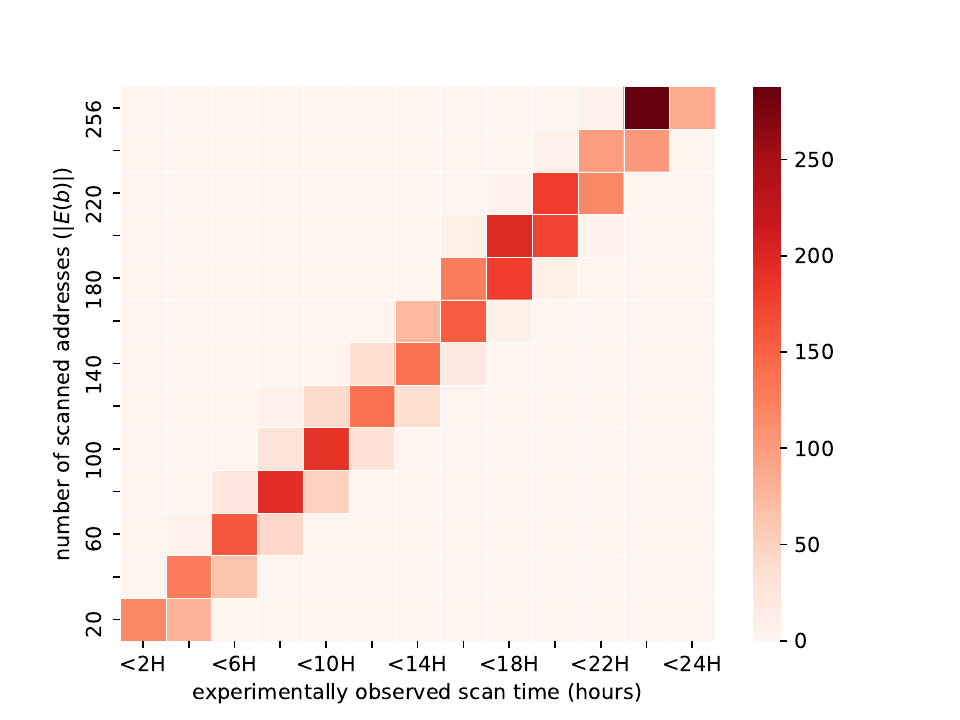}}
		} \\
	\subfloat[The count of responsive blocks in \autoref{tab:reconstructed_vs_survey}.]{
		\label{fig:reconstruction.responsive}
		\mbox{\includegraphics[width=0.9\columnwidth,clip]{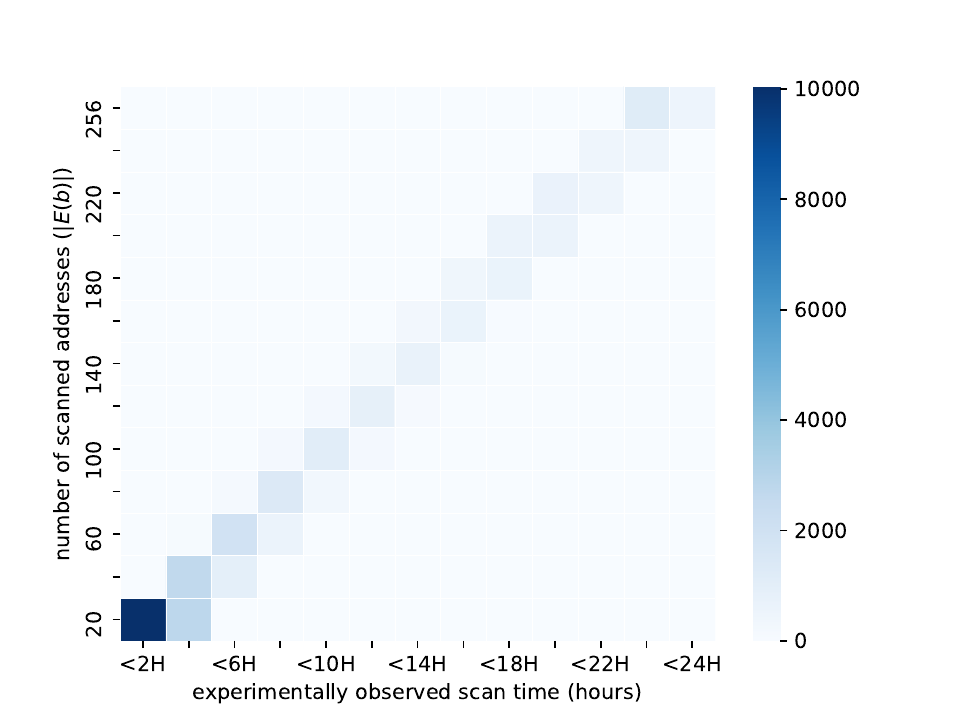}}
		}
	\caption{A heatmap of block counts, as a function of experimentally 
		observed full block scan time ($x$-axis) and scanning size ($|E(b)|$, $y$-axis). Dataset: 2020q1-ejnw and it89w. }
	\label{fig:reconstruction.compare}
\end{figure}

Reconstruction makes results defective from insufficient data,
  so additional probing (\autoref{sec:alg_reconstruction})
  is designed to fill this gap.
We next show that we can identify under-probed blocks
  (the tail of the distribution in \autoref{fig:cdf_fbs_time})
  with targeted additional probing.

\autoref{fig:reconstruction.failed} illustrates the absolute count of 
  known change-sensitive blocks that are not
  classified as change-sensitive in reconstruction.
It illustrates blocks in a heatmap binned
  by full-block scanning time ($x$-axis)
  and the number of scanned addresses ($y$-axis).
This data shows that problems occur
  in full blocks with long-scan-time (above and right of the origin).
\autoref{fig:reconstruction.responsive} shows the vast majority of blocks are near the origin,
  confirming failures occur in the tail, consistent within \autoref{fig:cdf_fbs_time}.

From these two figures, we derive our sliding-scale of additional probes:
  4 probes if scan time exceeds 24 hours,
  3 if over 18, 2 if over 12, and 1 if over 6.
\reviewfix{}
For each round, 
  additional probes do not stop even if encountering positive reply. 

With these additions, all blocks complete scanning in 6 hours or less.
\reviewfix{}
This data supports our claim
  that additional probing (\autoref{sec:alg_reconstruction})
  can augment basic Trinocular to provide good
  reconstruction for all responsive blocks.

\begin{table*}[t]
\begin{small}
	\begin{tabular}{l | r r r | r | r | r | r }
		\textbf{Dataset:} & \textbf{2019q4-w} & \textbf{2020q1-w} & \textbf{2020q2-w} & \textbf{2020h1-w} & \textbf{2020m1-w} & \textbf{2020h1-\SitesList} & \textbf{2020m1-\SitesList} \\ 
		\hline
		duration (weeks) &  \multicolumn{3}{c|}{12} & 24 & 4 & 24 & 4 \\
		\hline
		completeness (sites) &  \multicolumn{5}{c|}{1} & \multicolumn{2}{c}{\SitesNum} \\
		\hline
		allocated blocks &  \multicolumn{7}{c}{14,483,456} \\
		\quad  not routed & 3,388,236 & 3,361,864 & 3,333,669 &  3,333,669 & 3,361,864 & 3,333,669 & 3,361,864 \\
		\qquad  routed blocks & 11,095,220  & 11,121,592 & 11,149,787  & 11,149,787 & 11,121,592 & 11,149,787 & 11,121,592 \\
		\qqquad    not responsive &  7,049,754 & 5,922,911 & 5,925,238 & 6,024,576 & 5,922,911 & 6,024,576 & 5,922,911 \\
		\qqquad    responsive &  4,045,466 & 5,198,681 & 5,224,549 & 5,125,211 & 5,198,681 & 5,125,211 & 5,198,681 \\
		\hline
		\qqqquad      not diurnal & 3,631,272 & 4,799,382 &  4,849,579 & 4,877,967 & 4,796,117 & 4,889,070 & 4,652,108  \\
		\qqqquad      \textcolor{blue}{diurnal} & \textcolor{blue}{414,194} & \textcolor{blue}{399,299} &\textcolor{blue}{374,970} & \textcolor{blue}{ 247,244} & \textcolor{blue}{402,564} & \textcolor{blue}{236,141} & \textcolor{blue}{546,573} \\
		\hline
		\qqqquad      narrow swing & 1,375,566 & 2,170,901 & 1,700,045 & 2,269,473 & 2,977,067 & 1,825,465 &  1,656,565  \\
		\qqqquad      \textcolor{blue}{wide swing} &  \textcolor{blue}{2,669,900}& \textcolor{blue}{3,027,780} & \textcolor{blue}{3,524,504} & \textcolor{blue}{2,855,738} & \textcolor{blue}{2,221,614} & \textcolor{blue}{3,299,746} & \textcolor{blue}{3,542,116} \\
		\hline
		\qqqquad      not change-sensitive & 3,675,118 & 4,880,856 & 4,948,888 & 4,956,244 & 4,888,659 & 4,937,024 & 4,778,003 \\
		\qqqquad      \textcolor{blue}{\textbf{change-sensitive}} &  \textcolor{blue}{370,348} & \textcolor{blue}{317,825}& \textcolor{blue}{275,661} & \textcolor{blue}{168,967} & \textcolor{blue}{310,022}& \textcolor{blue}{188,187} & \textcolor{blue}{\textbf{420,678}} \\
	\end{tabular}
\end{small}
	\caption{
		Blocks before and after filtering (in /24s). 
		Change-sensitive is interpreted as /24 blocks 
		that are diurnal and with wide swing. Allocated addresses from IPv4 Address Space Registry~\cite{ianaipv4}; Routing data from Routeviews~\cite{apnic2019q4,apnic2020q1,apnic2020q2}.}
	\label{tab:block_count}
\end{table*}

\subsection{How Many Change-Sensitive Blocks?}
\label{sec:how_many_blocks}

Since our algorithms detect WFH only in change-sensitive blocks,
  we next ask:
How many change-sensitive blocks are there?
Does that number change over time?
We also revisit the effect of observation duration on the number
  of change-sensitive blocks.

\textbf{Decrease over time:}
The left three columns of 
\autoref{tab:block_count} show three-quarters of data from 2019q4 to 2020q2.
Each quarter uses a 12-week observation.
We see the number of change-sensitive blocks
  decreases somewhat over this period: from 370k to 327k to 275k.
We expect some of the decrease from 2020q1 to 2020q2 may
  reflect Covid-19 WFH
  as people move from universities and workplaces with more public IP addresses
  to homes where always-on modems shield their workday status.

\textbf{Churn:}
There is a fairly large rate of churn (turnover) in change-sensitive blocks.
We see that in comparing the first six months of 2020 (2020h1-w)
  with each three-month quarter (2020q1-w and 2020q2-w).
The number of change-sensitive blocks for 2020h1-w is
  the intersection of the two quarters, and with 169k blocks,
  it is only 53\% or 61\% of each quarter.
This drop is likely because when blocks are not consistently diurnal
  for the entire observation period we classify them as non-diurnal.
We see a similar drop to 199k when merging 2019q4 and 2020q1,
  consistent with duration as the primary factor.

Another possible reason for churn in the set of change-sensitive blocks
  in 2020 is the Covid-induced changes in network usage.
We show some examples in \autoref{sec:algorithm_success}.

\textbf{Input targets:}
A complicating factor in this analysis is that the underlying target list changes over time:
  in each quarter, the target list is updated to reflect currently responsive blocks,
  and in 2020q1 the target list was expanded from 4.0M blocks to 5.2M
  to take advantage of algorithm changes that correctly handle sparse blocks~\cite{baltra2020improving} 
  (compare the number of responsive blocks in 2019q4-w and 2020q1-w).
This expansion only modestly increases the number of change-sensitive blocks
   (it adds 11,547 or 3.6\% to 2020q1-w, and 16,852 or 3.8\% to 2020m1-\SitesList)
  because only a few of the newly added blocks are diurnal.
  
\textbf{Measurement Duration:}
We previously observed that longer periods decrease the number of
  diurnal blocks.
We see that effect again here,
  comparing 2020m1-w to 2020q1-w to 2020h1-w.
As discussed in \autoref{sec:algorithm_success},
  to factor out this change, 
  we detect change-sensitive blocks based on 2020m1-\SitesList
  and apply that to all of 2020h1-\SitesList for our data in \autoref{sec:results}.
(The longer duration in 2020h1-\SitesList would reduce
  the number of change-sensitive blocks by 42\%,
  in part because of the very changes in address use we are working to detect.)

\textbf{Implications:}
In spite of dynamics, we see
  \CSBlockNumLow to \CSBlockNumHigh change-sensitive blocks, as shown in \autoref{tab:block_count}.

\reviewfix{}
Any real-world system like the Internet will evolve over time,
  and the above factors of change of use, churn, and new allocations
  all contribute to such non-stationarity.
Non-stationarity is common in measurement
  and can be addressed by regular retraining,
  as is already done for input targets.
Our data shows coverage is sufficiently stable for the six-month period we analyze
  here;  we are exploring integration of retraining in ongoing work.

\subsection{Where are Change-Sensitive Blocks?}
	\label{sec:block_where}
Prior analysis of diurnal blocks has shown 
  their frequency varies by country~\cite{Quan14c}.
Countries have different amounts of IPv4 address space,
  and they also adopt different telecommunications and cultural policies
  about keeping devices ``always-on''.
Change-sensitive blocks occur when devices are directly attached
  to the public Internet,
  so we do not see change-sensitive blocks in
  ISPs where devices use private address space
  behind an always-on router on the public IPv4 Internet.
\reviewfix{}
IP address assignment and use have been
  a topic of considerable study~\cite{Xie07a,Cai10a,Quan14c,Moura15a,Padmanabhan16a,Richter16b,Richter16c,Padmanabhan20a},
Fully exploring the relationship between policy and change-sensitive blocks
  is future work,
  but we next summarize what we see.

\begin{figure*}
	\mbox{\includegraphics[trim=75 150 90 110,clip,width=1.5\columnwidth]{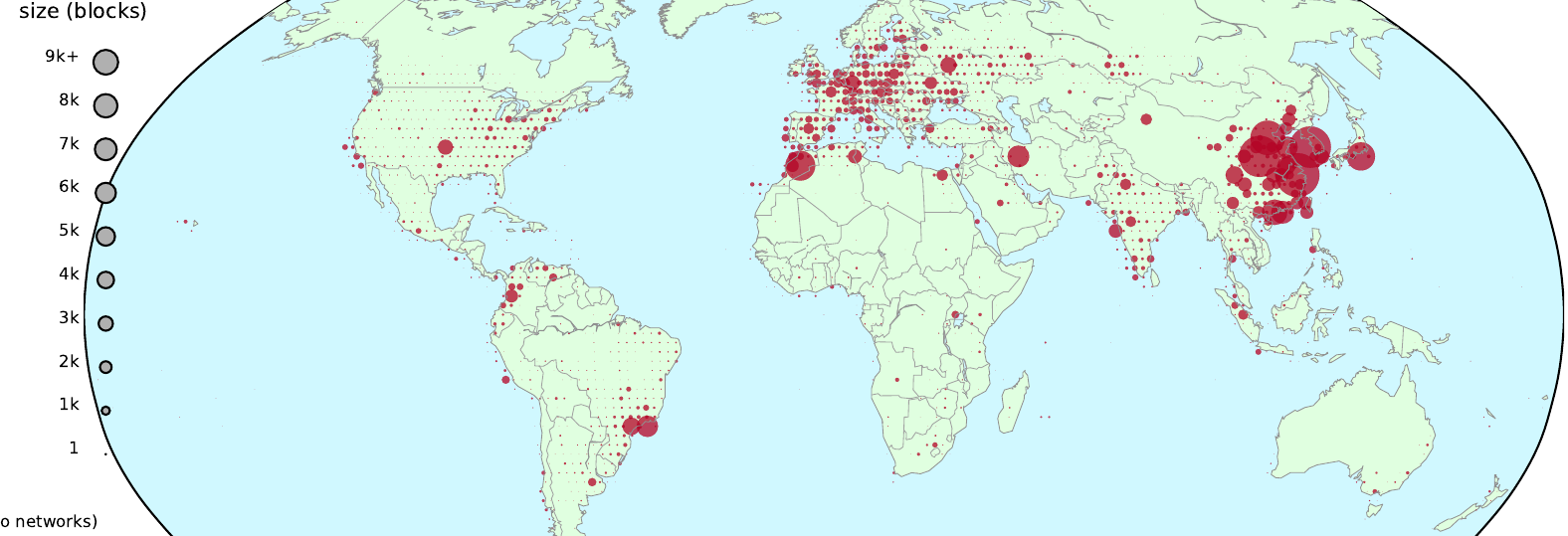}}
	\caption{
		Number of change-sensitive blocks (circle area)
                by geolocation (in a $2\times 2^{\circ}$ grid).  Dataset: 2020m1.}
	\label{fig:block_coverage}
\end{figure*}

\autoref{fig:block_coverage} shows the locations of all change-sensitive blocks
  we see in the 2020m1-\SitesList dataset.
This graph counts blocks in each $2\times 2^{\circ}$ latitude/longitude grid,
  showing the number of blocks as the area of a red circle in each grid cell.

We see that the best coverage is in East Asia (China, Korea, and Japan), 
  with moderate coverage in Brazil and North Africa,
  and sparse coverage in the United States, Europe, and India.
Coverage reflects the intersection of where IPv4 addresses are allocated
  and where users of those addresses turn off devices at night.
Sparse results from the U.S.~and Europe reflect
  widespread use of always-on home routers.
(Although such routers use many public IPv4 addresses,
  their 24x7 operation means they are not diurnal and so do not show when
  they are actually in use.)
If these users turn off their devices at night,
  that change is hidden behind those routers and NAT\@.
The diurnal blocks in the U.S.~and Europe that we see
  often correspond to universities
  where users occupy public IP addresses during the work-week (like \autoref{fig:isi_lab_example}).
Heavier presence in Asia is likely to reflect local
  ISP policies, with most users using dynamically assigned, public IPs. %
\reviewfix{}
\reviewfix{}
Future work could further explore correlations of change-sensitivity with network types.

\subsection{Validation by Sampled Blocks}
	\label{sec:event_validation}

We next validate the correctness of our algorithms
  by examining a random sample of blocks,
  looking for events in that block,
  then looking for ground truth about that event
  in public news sources.

\textbf{Defining correctness:}
We claim that changes in change-sensitive blocks can indicate WFH,
  but defining ``correct'' can be challenging.
WFH may occur for reasons other than Covid
  (see \autoref{sec:CaseStudy:NewDelhi} for example).
WFH is not the only cause of network changes;
  networks also suffer outages and shift users due to maintenance.
Networks might not show changes even
  though a region begins WFH---networks with NATs and firewalls will not show diurnal behavior
    and so cannot change during WFH\@.
Finally, our temporal precision is limited.
We detect changes daily, and must account for weekends,
  so detection may lag up to 4 days.

Our goal is that our algorithms are \emph{useful},
  as case studies in \autoref{sec:results} show.
To support that our algorithms should be trusted,
  we quantify correctness in two ways.
In this section we evaluate random blocks,
  confirming they show CUSUM changes on dates that match confirmed WFH reports.
We define block-level correctness
  as a WFH detection within four days of a public WFH report.
This correctness is not as strong as 1:1 mapping of events with WFH,
  but it suggests correlation.
  
In \autoref{sec:validating_discoverability}
  we examine random locations to see when groups of block-level changes
  correspond to WFH reports.
We define location-level discoverability as
  a noticeable number of block-level changes that correspond with a public WFH report.

Together these metrics suggest utility.

\begin{table}
\begin{tabular}{lr}
\textbf{dataset:} & \makebox[0pt][r]{\textbf{2020q1-\SitesList}} \\
change-sensitive blocks & 420,678 \\
\quad random selection &  50 \\
\qquad geolocatable & 50 \\
\qqquad no WFH in quarter & 6 \\
\qqquad WFH in quarter & 44 \\
\qqqquad CUSUM near ($\pm 4$d) WFH date & 14 \\
\qqqqquad manual confirmation \textbf{(TP}) & 13 \\
\qqqqquad apparent outage \textbf{(FP)} & 1 \\
\qqqquad no CUSUM near WFH date & 30 \\
\qqqqquad visual-detection near WFH \textbf{(FN)}& 5 \\
\qqqqquad CUSUM 5d from WFH & 1 \\
\qqqqquad CUSUM not related to WFH & 9 \\
\qqqqquad no CUSUM detections & 15 \\
\end{tabular}
\caption{Validation of sampled blocks.}
	\label{tab:validation_by_block}
\end{table}

\textbf{Methodology:}
Validation begins by selecting
  50 random blocks from
  all blocks that are change-sensitive in 2020q1 (the first three months of 2020) (see \autoref{tab:validation_by_block}).
This selection is unbiased and 
  large enough to evaluate statistically.
(Our case studies that examine locations (\autoref{sec:results})
  are also helpful, but will under-represent blocks in urban areas.)

We then geolocate each block to match it to ground truth news reports.
All blocks in our sample are geolocatable.
\reviewfix{}
We see the blocks are global, in 18 different countries or regions,
  following the distribution seen in \autoref{fig:block_coverage},
  with 22 in China, 5 in Russia, 4 in Malaysia,
  3 in India, 2 in Brazil and Hong Kong SAR (China),
  and remaining 12 are single blocks in single countries.

We then look at change events per block
  and search for public news reports about Covid-19 lockdown dates.
We use global Covid-19 lockdown dates from multiple media sources~\cite{Wuhan1,russia1,malaysia1,India1,Brazil1,hongkong1,uk1,ukr1,france1,iraq1,germany1,iran1,italy1,spain1,venezuela1,belgium1}.
Russian and Singapore lockdowns are not in this quarter
  (they are March 30 and April 7, but we cannot consider March 30
  because of overlap with transients at the change of quarter)
  so we discard those 6 blocks,
  leaving 44 with news reports.

\textbf{Correctness:}
For correctness,
  our algorithms detect changes in~14
  of these 44~blocks.
In 13 of those 14
  we confirm a CUSUM-detected change in the raw data within 4 days of the
  reported Covid-19 lockdown date (true positives),
  showing \emph{precision is 93\%}---the detections that we see are usually covid related.  %
We manually examined raw data for each of those blocks.
The 14th block %
  shows a change on 2020-03-21,
  during WFH---but the raw data suggests a network outage, not WFH\@.
One of the remaining 13 %
  shows a tiny visual change
  that are better detected by our algorithms detect changes,
  showing the importance of automatic, quantitative evaluation for accuracy.
The other 12 all show WFH changes analogous to our example (\autoref{fig:isi_lab_example}).

\textbf{Completeness:}
While we do not claim a 1:1 association of detected changes and WFH events,
  we suggest \emph{weak} completeness:
  do we detect all blocks that have changes?
To determine all blocks with changes (positives),
  we manually examine the 30 blocks without change in the WFH period
  for visual changes that are
  missed in CUSUM detection.
We find 5 blocks (of 30) are missed by our algorithms
  but could have been found; these represent false negatives
  that we could find by tuning detection parameters.
(The sixth block in China just misses our 4-day window,
  making it a true negative.)
With 13 true positive detections in 18 positive events,
  that implies \emph{recall is 72\%}
  based on weak completeness.

Finally, 9 blocks show CUSUM detections
  at dates distant from WFH reports.
Manual confirmation in raw data confirms these are real changes.
They may be WFH that we could not document;
  some other event in the region;
  or network maintenance actions,
  like moving users to new IP addresses,
  that appear to be WFH\@.
We expect regional events to affect many people,
  so we can examine look for  downward trends in other blocks in the same location,
  as we study in \autoref{sec:validating_discoverability}.  
Lack of other blocks with the same trend suggests network maintenance.
\reviewfix{}  
We examined the locations of these 9 blocks
  and only two had many other blocks triggering on the same day,
  suggesting that two downtrend events which could not document as Covid-related,
  and seven events which are consistent with small-scale network changes.
This analysis suggests that correlated changes in one location
  are better predictors than results of individual blocks.
%
%


\subsection{Validation by Location}
	\label{sec:validating_discoverability}

Examination of blocks suggests block-level precision is good,
  and our experiences (\autoref{sec:results})
  suggest it can discover events.
We next \emph{validate} the ability of our algorithms to assist discovery,
  examining the data behind two random locations selected
  from all grid cells with change-sensitive blocks.

\textbf{The United Arab Emirates:}
We randomly select grid cell (24N, 54E) in The United Arab Emirates,
  and 25 of its 230 change-sensitive blocks. %
This country started a Covid-cleaning campaign on 2020-03-22 and 
  then began a night curfew on 2020-03-26~\cite{uae1}.
  
As before, we validate all blocks by comparing detection dates to news reports
  and examining raw data.
Of the 25 blocks, 
  11 blocks have CUSUM-detected changes near the lockdown date. 
We confirm that all 11 blocks suggest Covid-related changes (true positives),
  showing \emph{precision is 100\%}. 
CUSUM changes peak on 2020-03-24 with 21.3\% of blocks changing,
  ten times more than any other day in 2020h1.
Four blocks show changes at other dates, but this huge peak focuses on the true WFH period.  
The other four blocks show changes in raw data but are not detected by CUSUM,
  suggesting 73\% recall at this location. %

\textbf{Slovenia:}
The second grid cell we randomly select is (46N, 14E) in Slovenia.
We examine 25 of the 936 change-sensitive blocks in the region.
Slovenia closed all educational institutions on 2020-03-16, %
  with additional suspensions later~\cite{svn1}.

Of the 25 randomly selected blocks,
  we find 7 blocks show changes near 2020-03-16 and confirm them to be Covid-related (true positives),
  showing \emph{precision is 100\%}. 
As with UAE, the peak of changes (here on 2020-03-16)
  is larger than other peaks,
  supporting filtering the 6 blocks that show changes
  on other dates as non-correlated.
Two blocks show changes in the raw data but are not detected by CUSUM,
  suggesting 77\% recall here. %

\textbf{Discussion:}
These two examples suggest
  that our approach finds outages at locations due to Covid,
  with detections at enough blocks to filter out
  non-correlated sources of network change.

\section{Results: Real World Events}
    \label{sec:results}
	\label{sec:Some_Events}

Finally, we use our approach to confirm and discover
  real-world WFH reactions to Covid-19.
We show overall statistics and then
  examine China as known example of curfew.

More interesting are events we discovered through our system.
We then turn to events in India and the Philippines that we discovered
  from our data.
India shows a confirmed non-Covid curfew.
For space, additional examples are in \autoref{sec:more_wfh_detections}.


\subsection{Overall Trends}
        \label{sec:overall_trends}

\reviewfix{}
\reviewfix{}
\reviewfix{}

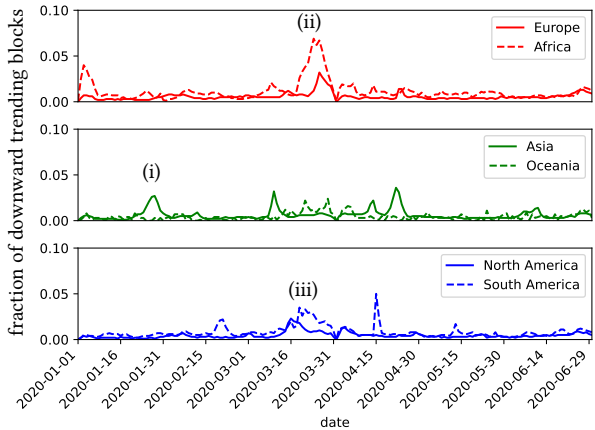
\begin{figure}
	\mbox{\begin{annotationimage}{width=0.95\columnwidth}{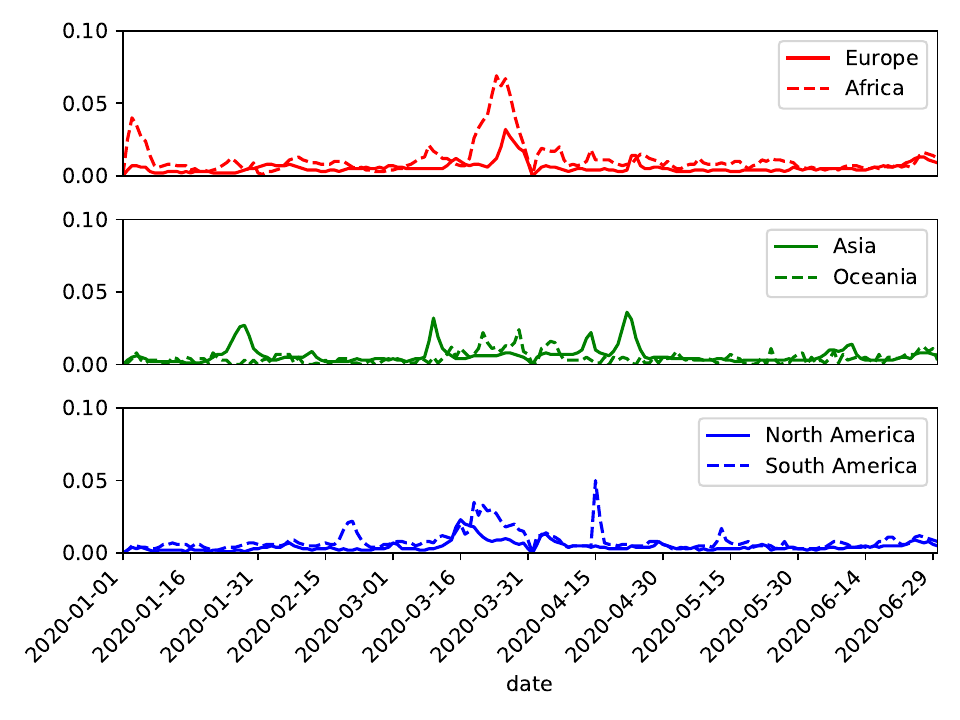}
          \draw[coordinate label text = black, coordinate label back = none, coordinate label font = \rmfamily\footnotesize, coordinate label = {\rotatebox{90}{fraction of downward trending blocks} at (0.03,0.6)}];
          \draw[coordinate label text = black, coordinate label back = none, coordinate label font = \rmfamily\footnotesize, coordinate label = {(i) at (0.25,0.6)}];
          \draw[coordinate label text = black, coordinate label back = none, coordinate label font = \rmfamily\footnotesize, coordinate label = {(ii) at (0.51,0.93)}];
          \draw[coordinate label text = black, coordinate label back = none, coordinate label font = \rmfamily\footnotesize, coordinate label = {(iii) at (0.5,0.34)}];
	\end{annotationimage}}
	\caption{WFH changes for 2020h1 by continent.}
    \label{fig:covid_2020h1}
\end{figure}

In this paper we examine all data for 2020h1.
\autoref{fig:covid_2020h1} shows the global count of downward trends
  in WFH changes for each continent over six months.
We geolocate blocks using Maxmind and assign each grid to a continent.
By continent, data is heavily aggregated;
  we find data exploration is easier in
  a $2\times2\deg$ grid in our interactive website~\cite{Stutz21b}.

Although aggregated, we see several trends.
First, the large percentage of changes in Asia around 2020-01-20 (at (i))
  corresponds to initial control measures taken in China.
Most of the rest of the world show large changes around 2020-03-20 (at (ii) and (iii)).
Low percentages in Oceania (dashed line, middle graph)
  show the success of their limits on international
  travel to control spread in this period.
The large percentage in Africa (at (ii))
  reflects the over-representation of Morocco in our data (see \autoref{fig:block_coverage})
  and their lockdown beginning 2020-03-20~\cite{Hekking20a}.
These trends show the opportunity for global analysis;
  we next examine specific localities.

\begin{figure*}
	\centering
	\begin{minipage}[b]{.25\linewidth}
		\subfloat[2x2$^{\circ}$ grid dotmap for 2020-01-27.]{
			\label{fig:global_2020-01-27}
			\mbox{\includegraphics[height=3cm,trim=40 125 825 130,clip]
			{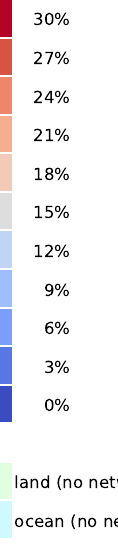}
			}
			\mbox{\begin{annotationimage}{width=1\columnwidth,trim=590 250 190 150,clip}
			{FIG/SWITCH/2020-01-27.fsdb.4sites.pdf}
				\path (0.25,0.55)node(x){Beijing}(0.69,0.785)node(y){}; %
				\draw[->,black,font=1pt] (x) -- (y);
				\path (0.35,0.2)node(x){Wuhan}(0.72,0.65)node(y){};
				\draw[->,black,font=1pt] (x) -- (y);
				\path (0.65,0.3)node(x){Shanghai}(0.78,0.65)node(y){}; %
				\draw[->,black,font=1pt] (x) -- (y);
			\end{annotationimage}}
		} \\
		\subfloat[Changes for (30N, 114E: Wuhan, top) and (38N, 116E: Beijing, bottom), 2020h1.]{
			\label{fig:summary_line_chart_Wuhan}
			\includegraphics[width=1.1\columnwidth]{
			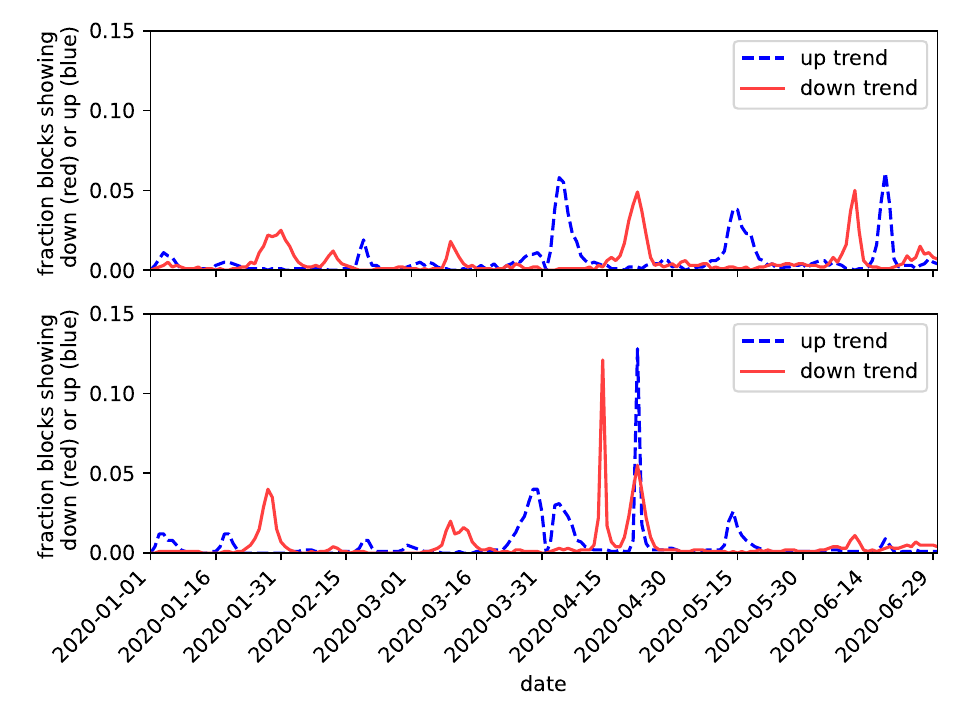}
		}
		\caption{China}
		\label{fig:china_changes}
	\end{minipage}
	\hfil
	\begin{minipage}[b]{.25\linewidth}
		\subfloat[A $2\times 2^{\circ}$ grid dotmap on 2020-02-28.]{
			\label{fig:global_2020-02-28_ND}
			\mbox{\includegraphics[height=3cm,trim=40 125 825 130,clip]
			{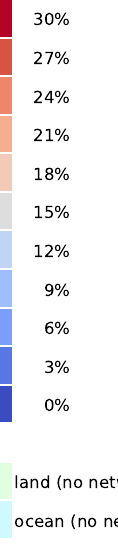}}
			\mbox{\begin{annotationimage}{width=1\columnwidth,trim=580 270 270 190,clip}{
			FIG/WORLDMAP/2020-02-28.fsdb.pdf}
			\path (0.2,0.45)node(x){New Delhi}(0.55,0.85)node(y){};
			\draw[->,black,font=1pt] (x) -- (y);
			\end{annotationimage}}
		} \\
		\subfloat[Changes for (28N, 76E: New Delhi) 2020h1.]{
			\label{fig:summary_line_chart_ND}
			\includegraphics[width=1.1\columnwidth]{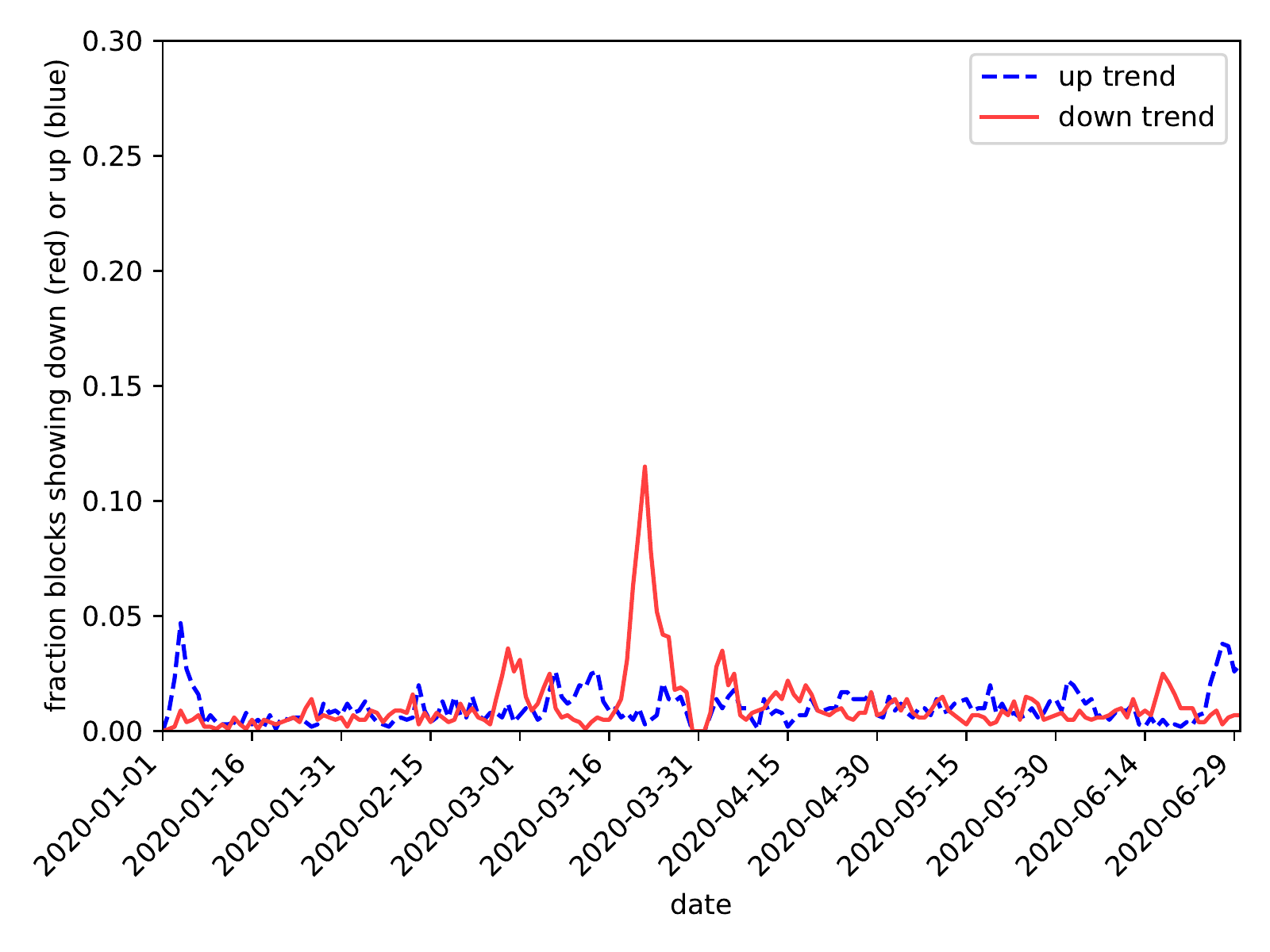}
		}
		\caption{India}
		\label{fig:india_changes}
	\end{minipage}
	\hfil
	\begin{minipage}[b]{.25\linewidth}
		\subfloat[2x2$^{\circ}$ grid dotmap for 2020-03-17.]{
			\label{fig:global_2020-03-17}
			\mbox{\begin{annotationimage}{width=1\columnwidth,trim=680 261.2 185 210,clip}
			{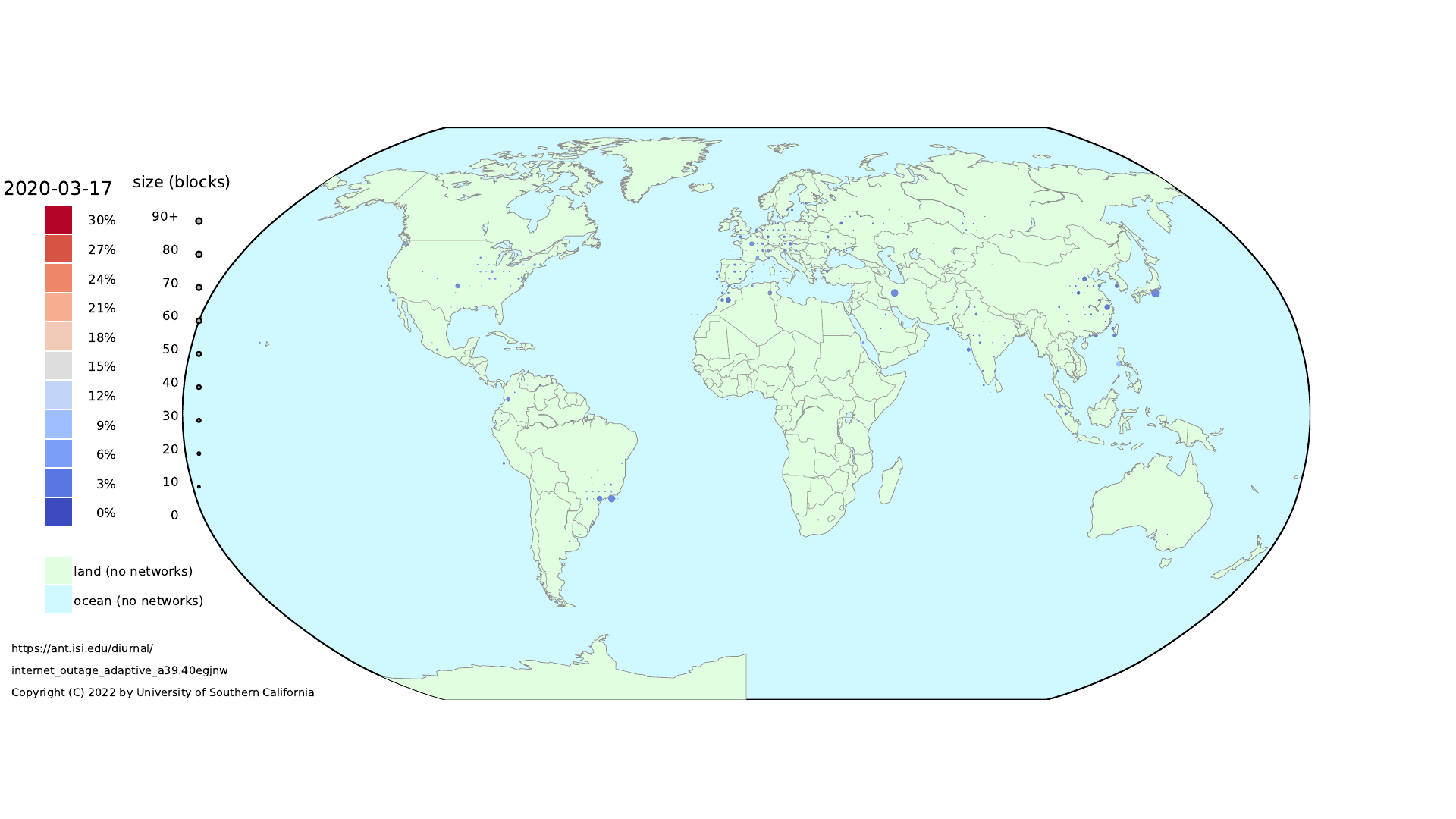}
			\path (0.25,0.2)node(x){Manila}(0.5,0.6)node(y){};
			\draw[->,black,font=1pt] (x) -- (y);
			\end{annotationimage}}
		} \\
		\subfloat[Changes for (14N, 120E: Manila) 2020h1.]{
			\label{fig:summary_line_chart_14N_120E}
			\includegraphics[width=1.1\columnwidth]
			{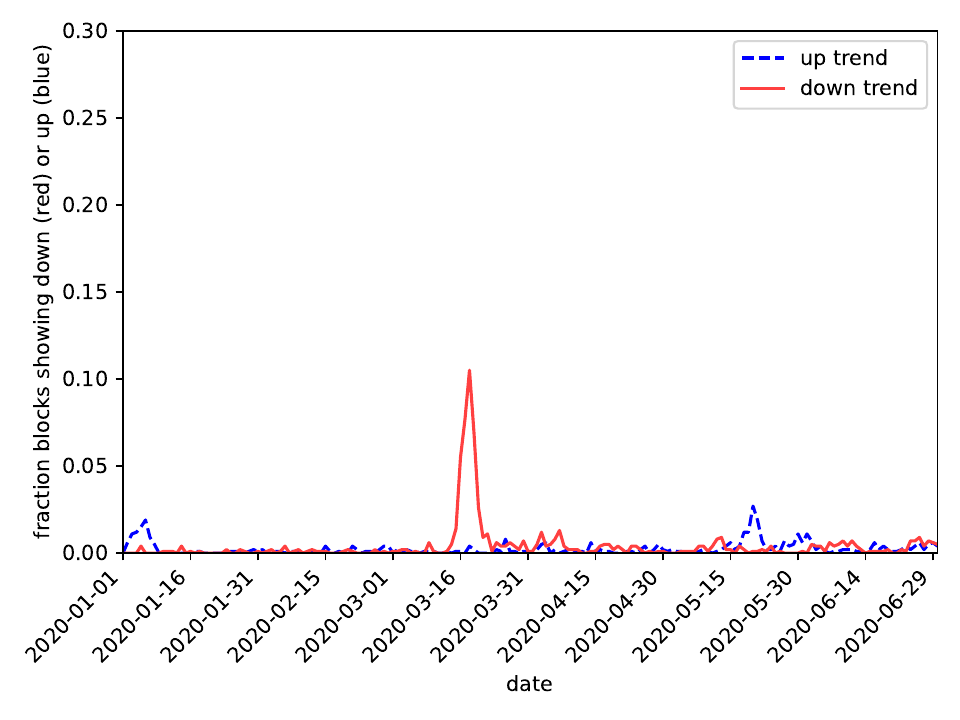}
		}
		\caption{%
			 The Philippines}
		\label{fig:Manila_changes}
	\end{minipage}%
\end{figure*}

\subsection{China on 2020-01-27}
	\label{sec:CaseStudy:Wuhan_China}

We first show the detection of network changes corresponding to Covid-related events.
According to media reports, Wuhan went into lockdown on 2020-01-23~\cite{Wuhan1}.

\autoref{fig:global_2020-01-27} shows downward network changes for all of China
  (in a $2\times 2^{\circ}$ grid) on 2020-01-27.
We see large downward trends in several cities,
  including Wuhan.

For Wuhan's (30N, 114E) grid cell,
  \autoref{fig:summary_line_chart_Wuhan} illustrates the number of 
  downward detections we observe for all blocks over six months. 
We see a peak in the first quarter around 2020-01-27,
  showing about 1.5\% or 53 of 3,589 change-sensitive blocks reduce usage that day.

We also see changes elsewhere in China at the same time.
We see small peaks in Shanghai
  (the light blue circle (30N, 120E) on the east coast of China,
  4.3\% or 1,280
  of 30,127 change-sensitive blocks)
  and Beijing
  (another light-blue circle at (38N, 116E), with
  3\% or 549 of 18,689 blocks).
Although a small percentage, both are large absolute changes.

Our data confirms we detect changes when Wuhan began WFH,
  consistent with widespread media reports~\cite{Wuhan1}.
The apparent WFH in Shanghai and Beijing surprised us,
  with no international reports,
  but we found Chinese-language news about a level-1 health alert
  and school postponement in these cities~\cite{Chinanews20a}.
This example shows the potential of measurements to discover
  under-reported events.

In addition to the January peak matching a known WFH event,
  \autoref{fig:summary_line_chart_Wuhan} shows large peaks
  in April and June.
As with Beijing and Shanghai in January,
  we cannot find media reports of Covid-related WFH in these months.
These events may be unpublicized or voluntary WFH,
  or possibly non-Covid events (described next).

\subsection{India on 2020-02-28}
\label{sec:CaseStudy:NewDelhi}

In our second case study, we examine India in February and March 2020.
In browsing our data, we noticed hot-spots of network changes
  starting on 2020-02-26 for several days.
We noticed that the east of New Delhi, where fewer networks
  made for a larger relative change.
As we refined our data processing, these changes were smaller than Covid-related
  lockdowns on 2020-03-23, but the 2020-02-28 events provide evidence for
  network changes that are not Covid-related.

\autoref{fig:global_2020-02-28_ND} shows our 2x2\,degree grid of 
changes on 2020-02-28,
  with a noticeable drop (50 blocks, 2.4\% of 1983) in usage.
The largest drop (154 blocks, 7.7\%)
  in this location occurs on 2020-03-22.

The larger drop in March corresponds to the first Covid-related curfew in India,
  the Janata curfew on 2020-03-22~\cite{India1}
  and a nationwide lockdown on 2020-03-24.
The February drop is correlated with riots over several days (-02-23 to -29)
  protesting changes in immigration law~\cite{Tanwar20a}.
We have no evidence of curfews, but there were calls for curfews
  and both police and army intervention, suggesting people chose to stay home.
These examples of Covid-related WFH in March,
  and non-Covid-WFH in February
  suggest that WFH can have multiple causes,
  but their outcome on the Internet is similar.
\autoref{sec:CaseStudy:Thailand} describes a second example of non-Covid-WFH in Thailand.

\subsection{The Philippines on 2020-03-17}
Our third case study examines the Philippines in March.
This event is a true-positive,
  but unlike Wuhan, we \emph{discovered} this event from our data,
  then confirmed it with news reports.
  
\autoref{fig:global_2020-03-17} shows a geographic map of
  downward changes on 2020-03-17.
We see a large, medium-blue circle in Manila (at 14N, 120E) 
  showing a large change on this date.
The timeline \autoref{fig:summary_line_chart_14N_120E}
  confirms that 10.5\% (90 of 854) change-sensitive blocks show changes on 2020-03-17,
  the largest of 2020h1.
This date is shortly after Manila's lockdown beginning on 2020-03-15
  and extended to the island of Luzon on the 17th~\cite{Metro1}.

\section{Related Work}
	\label{sec:relatedwork}

Given the impact of Covid-19 on our lives,
  and Internet's role in WFH,
  several studies consider their interaction.

\textbf{Covid and Network Traffic:}
Several groups have reported about how the Internet responded to changes
  during Covid-19.
Ukani et al. studied the network usage of university students 
  at the application level~\cite{UkaniIMC21Covid}.
Facebook reported large traffic increases following lockdown~\cite{bottger2020internet}.
Another study evaluated traffic changes from multiple networks, including an
  ISP, IXP, and educational network~\cite{feldmann2021year}.
Researchers examined the Italian internet during Covid-19,
  finding increased variability in latency~\cite{candela2020impact}.
ICANN examined the impact of a nationwide lockdown in France 
  on DNS, showing increases in overall DNS traffic~\cite{icann2020}.
Telefónica analyzed how the cellular network usage and performance shifted in UK~\cite{Lutu20a}.
Toorn et al. investigated how rDNS entries change 
  due to work-from-home measures~\cite{Toorn22Covid}.
Unlike above works, we use the Internet to understand the real world
  and Covid-triggered WFH.
  
Closest to our work is the use of Google Trends
  to relate public interest in Covid with observed cases~\cite{effenberger2020association}.
We too study Covid-related changes using the Internet,
  but we consider WFH as inferred from IP address usage.

\textbf{Active Internet Measurement:}
Several prior groups have studied the Internet with active probing
  of some or all of the Internet,
  often to study address use or outages.
USC began whole-Internet censuses in 2006
  to evaluate address usage~\cite{Heidemann08c}.
ZMap~\cite{Durumeric13a} and Massscan~\cite{Graham14a} emphasize scanning speed.
Other systems have leveraged active probing to detect outages:
Thunderping probes addresses in areas undergoing weather events
  to look for outages~\cite{Schulman11a}.
Trinocular pings millions of networks,
  inferring outages from  Bayesian inference~\cite{quan2013trinocular}.
Chocolatine employs SARIMA models to forecast 
  Internet Background Radiation (IBR) time series to detect outages~\cite{guillot2019chocolatine}.
  It uses data from the UCSD Network Telescope~\cite{caidaNetworkTelescope}.
Richter et al.~infer network disruptions
  from drops in traffic as seen by a major CDN~\cite{richter2018advancing}.
Disco monitors the bursts of TCP disconnects to 
  detect outages~\cite{shah2017disco}.
Dainotti et al. use BGP updates and IBR to 
  study the Internet outages caused 
  by censorship at the country level~\cite{dainotti2011analysis}. 
Hubble combines active probing with passive BGP monitoring 
  to detect Internet failures. 
  It also identifies network
  entities that might be the cause~\cite{katz2008studying}. 
Our work builds on these prior systems for active scanning
  and reuses data from Trinocular,
  but with the new algorithms and the new application of detecting WFH.
  \reviewfix{}

\textbf{Diurnal networks and trends:}
Finally, diurnal behaviors are common in many time series, including
  networking.
Because seasonal patterns exist in many different types of time series,
  there are several well-established mathematical methods to
  estimate and extract seasonal trends from the underlying data~\cite{hyndman2018forecasting,de2011forecasting}.
It is well known that network traffic is diurnal,
  but recent work showed that IP address usage
  often shows diurnal patterns~\cite{Quan14c}.
We use diurnal addresses usage to detect human activity,
  and established tools to extract underlying trends (\autoref{sec:detrending}).

\section{Conclusions}
    \label{sec:conclusion}

This paper has shown that we can use observations of the Internet address
  responsiveness to detect WFH during Covid-19.
Our algorithms allow us to reconstruct diurnal trends
  from existing data collected to detect Internet outages,
  possibly augmented by additional probing,
  then detect changes in daily IP address use.
We validate the algorithms through the study of their components,
  evaluation of randomly selected blocks,
  and end-to-end verification of observed changes
  and news events in multiple locations.
\reviewfix{}
These algorithms represent a first demonstration of the
  potential to infer a class of large-scale human events (work-from-home)
  from analysis of aggregate address responsiveness,
  an important example of using the Internet to understand our world.

\section*{Acknowledgments}
This work is partially supported by the project
``Measuring the Internet during Novel Coronavirus to Evaluate Quarantine (RAPID-MINCEQ)''
(NSF 2028279),
and John Heidemann's work is partially supported by the project 
``CNS Core: Small: Event Identification and Evaluation of Internet
Outages (EIEIO)''
(CNS-2007106).
We thank Guillermo Baltra for prototyping analysis of Trinocular 
addresses across a block, his involvement in early versions of this work,
and his comments on the work.
We thank Yuri Pradkin for collecting the Trinocular data that we use 
in our analysis.

\label{page:last_body}

\bibliographystyle{ACM-Reference-Format}
\begin{CJK*}{UTF8}{gbsn}
\bibliography{paper}


\begin{thebibliography}{95}


\ifx \showCODEN    \undefined \def \showCODEN     #1{\unskip}     \fi
\ifx \showDOI      \undefined \def \showDOI       #1{#1}\fi
\ifx \showISBNx    \undefined \def \showISBNx     #1{\unskip}     \fi
\ifx \showISBNxiii \undefined \def \showISBNxiii  #1{\unskip}     \fi
\ifx \showISSN     \undefined \def \showISSN      #1{\unskip}     \fi
\ifx \showLCCN     \undefined \def \showLCCN      #1{\unskip}     \fi
\ifx \shownote     \undefined \def \shownote      #1{#1}          \fi
\ifx \showarticletitle \undefined \def \showarticletitle #1{#1}   \fi
\ifx \showURL      \undefined \def \showURL       {\relax}        \fi
\providecommand\bibfield[2]{#2}
\providecommand\bibinfo[2]{#2}
\providecommand\natexlab[1]{#1}
\providecommand\showeprint[2][]{arXiv:#2}

\bibitem[\protect\citeauthoryear{{ANT Project}}{{ANT Project}}{2022}]%
        {ANT22a}
\bibfield{author}{\bibinfo{person}{{ANT Project}}.}
  \bibinfo{year}{2022}\natexlab{}.
\newblock \bibinfo{title}{ANT Covid-Work-from-Home Datasets}.
\newblock \bibinfo{howpublished}{\url{https://ant.isi.edu/datasets/covid/}}.
\newblock
\urldef\tempurl%
\url{https://ant.isi.edu/datasets/covid/}
\showURL{%
\tempurl}


\bibitem[\protect\citeauthoryear{APNIC}{APNIC}{2020a}]%
        {apnic2019q4}
\bibfield{author}{\bibinfo{person}{APNIC}.} \bibinfo{year}{2020}\natexlab{a}.
\newblock \bibinfo{title}{Routing Table Report - Japan view}.
\newblock
\newblock
\urldef\tempurl%
\url{https://mailman.apnic.net/mailing-lists/bgp-stats/archive/2019/10/msg00001.html}
\showURL{%
\tempurl}


\bibitem[\protect\citeauthoryear{APNIC}{APNIC}{2020b}]%
        {apnic2020q1}
\bibfield{author}{\bibinfo{person}{APNIC}.} \bibinfo{year}{2020}\natexlab{b}.
\newblock \bibinfo{title}{Routing Table Report - Japan view}.
\newblock
\newblock
\urldef\tempurl%
\url{https://mailman.apnic.net/mailing-lists/bgp-stats/archive/2020/01/msg00001.html}
\showURL{%
\tempurl}


\bibitem[\protect\citeauthoryear{APNIC}{APNIC}{2020c}]%
        {apnic2020q2}
\bibfield{author}{\bibinfo{person}{APNIC}.} \bibinfo{year}{2020}\natexlab{c}.
\newblock \bibinfo{title}{Routing Table Report - Japan view}.
\newblock
\newblock
\urldef\tempurl%
\url{https://mailman.apnic.net/mailing-lists/bgp-stats/archive/2020/04/msg00001.html}
\showURL{%
\tempurl}


\bibitem[\protect\citeauthoryear{Arends}{Arends}{2020}]%
        {icann2020}
\bibfield{author}{\bibinfo{person}{Roy Arends}.}
  \bibinfo{year}{2020}\natexlab{}.
\newblock \showarticletitle{Analysis of the Effects of COVID-19-Related
  Lockdowns on IMRS Traffic}.
\newblock \bibinfo{journal}{\emph{Roy Arends}} (\bibinfo{date}{Apr.~15}
  \bibinfo{year}{2020}).
\newblock
\urldef\tempurl%
\url{https://www.icann.org/en/system/files/files/octo-008-15apr20-en.pdf}
\showURL{%
\tempurl}


\bibitem[\protect\citeauthoryear{Baek, McCrory, Messer, and Mui}{Baek
  et~al\mbox{.}}{2020}]%
        {baek2020unemployment}
\bibfield{author}{\bibinfo{person}{ChaeWon Baek}, \bibinfo{person}{Peter~B
  McCrory}, \bibinfo{person}{Todd Messer}, {and} \bibinfo{person}{Preston
  Mui}.} \bibinfo{year}{2020}\natexlab{}.
\newblock \showarticletitle{Unemployment effects of stay-at-home orders:
  Evidence from high frequency claims data}.
\newblock \bibinfo{journal}{\emph{Review of Economics and Statistics}}
  (\bibinfo{year}{2020}), \bibinfo{pages}{1--72}.
\newblock


\bibitem[\protect\citeauthoryear{Baltra and Heidemann}{Baltra and
  Heidemann}{2020}]%
        {baltra2020improving}
\bibfield{author}{\bibinfo{person}{Guillermo Baltra} {and}
  \bibinfo{person}{John Heidemann}.} \bibinfo{year}{2020}\natexlab{}.
\newblock \showarticletitle{Improving Coverage of Internet Outage Detection in
  Sparse Blocks}. In \bibinfo{booktitle}{\emph{Proceedings of the Passive and
  Active Measurement Conference}}. \bibinfo{publisher}{Springer},
  \bibinfo{address}{Eugene, Oregon, USA}.
\newblock


\bibitem[\protect\citeauthoryear{{BBC News Editors}}{{BBC News
  Editors}}{2020a}]%
        {BBC20a}
\bibfield{author}{\bibinfo{person}{{BBC News Editors}}.}
  \bibinfo{year}{2020}\natexlab{a}.
\newblock \showarticletitle{Student Flash Mob: Sparks in a pan or a raging
  fire}.
\newblock \bibinfo{journal}{\emph{BBC News}} (\bibinfo{date}{Feb.~28}
  \bibinfo{year}{2020}).
\newblock
\urldef\tempurl%
\url{https://www.bbc.com/thai/thailand-51640629}
\showURL{%
\tempurl}
\newblock
\shownote{(translated from Thai).}


\bibitem[\protect\citeauthoryear{{BBC News Editors}}{{BBC News
  Editors}}{2020b}]%
        {BBC20b}
\bibfield{author}{\bibinfo{person}{{BBC News Editors}}.}
  \bibinfo{year}{2020}\natexlab{b}.
\newblock \showarticletitle{Thailand protests: State of emergency lifted after
  days of rallies}.
\newblock \bibinfo{journal}{\emph{BBC News}} (\bibinfo{date}{Oct.~22}
  \bibinfo{year}{2020}).
\newblock
\urldef\tempurl%
\url{https://www.bbc.com/news/world-asia-54641775}
\showURL{%
\tempurl}


\bibitem[\protect\citeauthoryear{Beverly, Durairajan, Plonka, and
  Rohrer}{Beverly et~al\mbox{.}}{2018}]%
        {Beverly18a}
\bibfield{author}{\bibinfo{person}{Robert Beverly},
  \bibinfo{person}{Ramakrishnan Durairajan}, \bibinfo{person}{David Plonka},
  {and} \bibinfo{person}{Justin~P. Rohrer}.} \bibinfo{year}{2018}\natexlab{}.
\newblock \showarticletitle{In the {IP} of the Beholder: Strategies for Active
  {IPv6} Topology Discovery}. In \bibinfo{booktitle}{\emph{Proceedings of the
  ACM Internet Measurement Conference}} (johnh: pafile).
  \bibinfo{publisher}{{ACM}}, \bibinfo{address}{Boston, Massachusetts, USA},
  \bibinfo{pages}{308--321}.
\newblock
\urldef\tempurl%
\url{https://doi.org/10.1145/3278532.3278559}
\showDOI{\tempurl}


\bibitem[\protect\citeauthoryear{B{\"o}ttger, Ibrahim, and Vallis}{B{\"o}ttger
  et~al\mbox{.}}{2020}]%
        {Boettger20a}
\bibfield{author}{\bibinfo{person}{Timm B{\"o}ttger}, \bibinfo{person}{Ghida
  Ibrahim}, {and} \bibinfo{person}{Ben Vallis}.}
  \bibinfo{year}{2020}\natexlab{}.
\newblock \showarticletitle{How the Internet reacted to Covid-19---A
  perspective from Facebook's Edge Network}. In
  \bibinfo{booktitle}{\emph{Proceedings of the ACM Internet Measurement
  Conference}}. \bibinfo{publisher}{{ACM}}, \bibinfo{address}{Pittsburgh, PA,
  USA}, \bibinfo{pages}{34--41}.
\newblock
\urldef\tempurl%
\url{https://doi.org/10.1145/3419394.3423621}
\showDOI{\tempurl}


\bibitem[\protect\citeauthoryear{Bottger, Ibrahim, and Vallis}{Bottger
  et~al\mbox{.}}{2020}]%
        {bottger2020internet}
\bibfield{author}{\bibinfo{person}{Timm Bottger}, \bibinfo{person}{Ghida
  Ibrahim}, {and} \bibinfo{person}{Ben Vallis}.}
  \bibinfo{year}{2020}\natexlab{}.
\newblock \showarticletitle{How the Internet reacted to Covid-19: A perspective
  from Facebook's Edge Network}. In \bibinfo{booktitle}{\emph{Proceedings of
  the ACM Internet Measurement Conference}}. \bibinfo{pages}{34--41}.
\newblock


\bibitem[\protect\citeauthoryear{Brodkin}{Brodkin}{2020}]%
        {Brodkin20a}
\bibfield{author}{\bibinfo{person}{Jon Brodkin}.}
  \bibinfo{year}{2020}\natexlab{}.
\newblock \bibinfo{title}{{Netflix}, {YouTube} cut video quality in {Europe}
  after pressure from {EU} official}.
\newblock \bibinfo{howpublished}{Ars Technica
  \url{https://arstechnica.com/tech-policy/2020/03/netflix-and-youtube-cut-streaming-quality-in-europe-to-handle-pandemic/}}.
\newblock
\urldef\tempurl%
\url{https://arstechnica.com/tech-policy/2020/03/netflix-and-youtube-cut-streaming-quality-in-europe-to-handle-pandemic/}
\showURL{%
\tempurl}


\bibitem[\protect\citeauthoryear{Cai and Heidemann}{Cai and Heidemann}{2010}]%
        {Cai10a}
\bibfield{author}{\bibinfo{person}{Xue Cai} {and} \bibinfo{person}{John
  Heidemann}.} \bibinfo{year}{2010}\natexlab{}.
\newblock \showarticletitle{Understanding Block-level Address Usage in the
  Visible {Internet}}. In \bibinfo{booktitle}{\emph{Proceedings of the {ACM}
  SIGCOMM Conference}}. \bibinfo{publisher}{{ACM}}, \bibinfo{address}{New
  Delhi, India}, \bibinfo{pages}{99--110}.
\newblock
\urldef\tempurl%
\url{https://doi.org/10.1145/1851182.1851196}
\showDOI{\tempurl}


\bibitem[\protect\citeauthoryear{{CAIDA}}{{CAIDA}}{2007}]%
        {caidaArk}
\bibfield{author}{\bibinfo{person}{{CAIDA}}.} \bibinfo{year}{2007}\natexlab{}.
\newblock \bibinfo{title}{Archipelago (Ark) Measurement Infrastructure}.
\newblock \bibinfo{howpublished}{website
  \url{https://www.caida.org/projects/ark/}}.
\newblock
\urldef\tempurl%
\url{https://www.caida.org/projects/ark/}
\showURL{%
\tempurl}


\bibitem[\protect\citeauthoryear{{CAIDA}}{{CAIDA}}{2012}]%
        {caidaNetworkTelescope}
\bibfield{author}{\bibinfo{person}{{CAIDA}}.} \bibinfo{year}{2012}\natexlab{}.
\newblock \bibinfo{title}{Network Telescope}.
\newblock
\newblock
\urldef\tempurl%
\url{https://www.caida.org/projects/network_telescope/}
\showURL{%
\tempurl}


\bibitem[\protect\citeauthoryear{Candela, Luconi, and Vecchio}{Candela
  et~al\mbox{.}}{2020}]%
        {candela2020impact}
\bibfield{author}{\bibinfo{person}{Massimo Candela}, \bibinfo{person}{Valerio
  Luconi}, {and} \bibinfo{person}{Alessio Vecchio}.}
  \bibinfo{year}{2020}\natexlab{}.
\newblock \showarticletitle{Impact of the COVID-19 pandemic on the Internet
  latency: A large-scale study}.
\newblock \bibinfo{journal}{\emph{Computer Networks}}  \bibinfo{volume}{182}
  (\bibinfo{year}{2020}), \bibinfo{pages}{107495}.
\newblock


\bibitem[\protect\citeauthoryear{Cheung, Wong, and Ho-him}{Cheung
  et~al\mbox{.}}{2020}]%
        {hongkong1}
\bibfield{author}{\bibinfo{person}{Tony Cheung}, \bibinfo{person}{Natalie
  Wong}, {and} \bibinfo{person}{Chan Ho-him}.} \bibinfo{year}{2020}\natexlab{}.
\newblock \bibinfo{title}{Coronavirus: 'little, if any, possibility' Hong Kong
  schools resume fully on April 20, Lam says}.
\newblock
  \bibinfo{howpublished}{\url{https://www.scmp.com/news/hong-kong/education/article/3075521/coronavirus-little-if-any-possibility-hong-kong-schools}}.
\newblock


\bibitem[\protect\citeauthoryear{{China News Service WeChat Official
  Account}}{{China News Service WeChat Official Account}}{2020}]%
        {Chinanews20a}
\bibfield{author}{\bibinfo{person}{{China News Service WeChat Official
  Account}}.} \bibinfo{year}{2020}\natexlab{}.
\newblock
  \bibinfo{title}{多省份启动重大突发公共卫生事件一级响应
  意味着什么 (What does it mean for multiple provinces to initiate a
  first-level response to major public health emergencies?)}.
\newblock
  \bibinfo{howpublished}{\url{https://m.chinanews.com/wap/detail/zw/gn/2020/01-24/9068940.shtml}}.
\newblock
\urldef\tempurl%
\url{https://m.chinanews.com/wap/detail/zw/gn/2020/01-24/9068940.shtml}
\showURL{%
\tempurl}


\bibitem[\protect\citeauthoryear{Cleveland, Cleveland, McRae, and
  Terpenning}{Cleveland et~al\mbox{.}}{1990}]%
        {cleveland1990stl}
\bibfield{author}{\bibinfo{person}{Robert~B Cleveland},
  \bibinfo{person}{William~S Cleveland}, \bibinfo{person}{Jean~E McRae}, {and}
  \bibinfo{person}{Irma Terpenning}.} \bibinfo{year}{1990}\natexlab{}.
\newblock \showarticletitle{STL: A seasonal-trend decomposition}.
\newblock \bibinfo{journal}{\emph{Journal of Official Statistics}}
  \bibinfo{volume}{6}, \bibinfo{number}{1} (\bibinfo{year}{1990}),
  \bibinfo{pages}{3--73}.
\newblock


\bibitem[\protect\citeauthoryear{Crisis24}{Crisis24}{2020a}]%
        {iran1}
\bibfield{author}{\bibinfo{person}{Crisis24}.}
  \bibinfo{year}{2020}\natexlab{a}.
\newblock \bibinfo{title}{Iran: Natiowide lockdown implemented as over 11,300
  COVID-19 cases confirmed March 13 /update 12}.
\newblock
  \bibinfo{howpublished}{\url{https://www.garda.com/crisis24/news-alerts/322811/iran-natiowide-lockdown-implemented-as-over-11300-covid-19-cases-confirmed-march-13-update-12}}.
\newblock


\bibitem[\protect\citeauthoryear{Crisis24}{Crisis24}{2020b}]%
        {uae1}
\bibfield{author}{\bibinfo{person}{Crisis24}.}
  \bibinfo{year}{2020}\natexlab{b}.
\newblock \bibinfo{title}{UAE: Three-day lockdown scheduled March 26-29 /update
  16}.
\newblock
  \bibinfo{howpublished}{\url{https://www.garda.com/crisis24/news-alerts/326651/uae-three-day-lockdown-scheduled-march-26-29-update-16}}.
\newblock


\bibitem[\protect\citeauthoryear{Cuthbertson}{Cuthbertson}{2020}]%
        {france1}
\bibfield{author}{\bibinfo{person}{Anthony Cuthbertson}.}
  \bibinfo{year}{2020}\natexlab{}.
\newblock \bibinfo{title}{Coronavirus: France imposes 15-day lockdown and
  mobilises 100,000 police to enforce restrictions}.
\newblock
  \bibinfo{howpublished}{\url{https://www.independent.co.uk/news/world/europe/coronavirus-france-lockdown-cases-update-covid-19-macron-a9405136.html}}.
\newblock


\bibitem[\protect\citeauthoryear{Dainotti, Squarcella, Aben, Claffy, Chiesa,
  Russo, and Pescap{\'e}}{Dainotti et~al\mbox{.}}{2011}]%
        {dainotti2011analysis}
\bibfield{author}{\bibinfo{person}{Alberto Dainotti}, \bibinfo{person}{Claudio
  Squarcella}, \bibinfo{person}{Emile Aben}, \bibinfo{person}{Kimberly~C
  Claffy}, \bibinfo{person}{Marco Chiesa}, \bibinfo{person}{Michele Russo},
  {and} \bibinfo{person}{Antonio Pescap{\'e}}.}
  \bibinfo{year}{2011}\natexlab{}.
\newblock \showarticletitle{Analysis of country-wide internet outages caused by
  censorship}. In \bibinfo{booktitle}{\emph{Proceedings of the 2011 ACM SIGCOMM
  conference on Internet measurement conference}}. \bibinfo{pages}{1--18}.
\newblock


\bibitem[\protect\citeauthoryear{De~Livera, Hyndman, and Snyder}{De~Livera
  et~al\mbox{.}}{2011}]%
        {de2011forecasting}
\bibfield{author}{\bibinfo{person}{Alysha~M De~Livera}, \bibinfo{person}{Rob~J
  Hyndman}, {and} \bibinfo{person}{Ralph~D Snyder}.}
  \bibinfo{year}{2011}\natexlab{}.
\newblock \showarticletitle{Forecasting time series with complex seasonal
  patterns using exponential smoothing}.
\newblock \bibinfo{journal}{\emph{Journal of the American statistical
  association}} \bibinfo{volume}{106}, \bibinfo{number}{496}
  (\bibinfo{year}{2011}), \bibinfo{pages}{1513--1527}.
\newblock


\bibitem[\protect\citeauthoryear{Dhamdhere, Clark, Gamero-Garrido, Luckie, Mok,
  Akiwate, Gogia, Bajpai, Snoeren, and kc~claffy}{Dhamdhere
  et~al\mbox{.}}{2018}]%
        {Dhamdhere18a}
\bibfield{author}{\bibinfo{person}{Amogh Dhamdhere}, \bibinfo{person}{David~D.
  Clark}, \bibinfo{person}{Alexander Gamero-Garrido}, \bibinfo{person}{Matthew
  Luckie}, \bibinfo{person}{Ricky K.~P. Mok}, \bibinfo{person}{Gautam Akiwate},
  \bibinfo{person}{Kabir Gogia}, \bibinfo{person}{Vaibhav Bajpai},
  \bibinfo{person}{Alex~C. Snoeren}, {and} \bibinfo{person}{kc claffy}.}
  \bibinfo{year}{2018}\natexlab{}.
\newblock \showarticletitle{Inferring Persistent Interdomain Congestion}. In
  \bibinfo{booktitle}{\emph{Proceedings of the {ACM} SIGCOMM Conference}}
  (johnh: pafile). \bibinfo{publisher}{{ACM}}, \bibinfo{address}{Budapest,
  Hungary}, \bibinfo{pages}{1--15}.
\newblock
\urldef\tempurl%
\url{https://doi.org/10.1145/3230543.3230549}
\showDOI{\tempurl}


\bibitem[\protect\citeauthoryear{Duarte}{Duarte}{2020}]%
        {Duarte2020}
\bibfield{author}{\bibinfo{person}{Marcos Duarte}.}
  \bibinfo{year}{2020}\natexlab{}.
\newblock \bibinfo{title}{detecta: A {Python} module to detect events in data}.
\newblock \bibinfo{howpublished}{\url{https://github.com/demotu/detecta}}.
\newblock
\urldef\tempurl%
\url{https://doi.org/10.5281/zenodo.4598962}
\showDOI{\tempurl}
\newblock
\shownote{GitHub repository.}


\bibitem[\protect\citeauthoryear{Durumeric, Adrian, Mirian, Bailey, and
  Halderman}{Durumeric et~al\mbox{.}}{2015}]%
        {censys15}
\bibfield{author}{\bibinfo{person}{Zakir Durumeric}, \bibinfo{person}{David
  Adrian}, \bibinfo{person}{Ariana Mirian}, \bibinfo{person}{Michael Bailey},
  {and} \bibinfo{person}{J.~Alex Halderman}.} \bibinfo{year}{2015}\natexlab{}.
\newblock \showarticletitle{A Search Engine Backed by Internet-Wide Scanning}.
  In \bibinfo{booktitle}{\emph{Proceedings of the ACM Conference on Computer
  and Communications Security}}. \bibinfo{publisher}{{ACM}},
  \bibinfo{address}{Denver, CO, USA}, \bibinfo{pages}{542--553}.
\newblock
\urldef\tempurl%
\url{https://doi.org/10.1145/2810103.2813703}
\showDOI{\tempurl}


\bibitem[\protect\citeauthoryear{Durumeric, Wustrow, and Halderman}{Durumeric
  et~al\mbox{.}}{2013a}]%
        {zmap182948}
\bibfield{author}{\bibinfo{person}{Zakir Durumeric}, \bibinfo{person}{Eric
  Wustrow}, {and} \bibinfo{person}{J.~Alex Halderman}.}
  \bibinfo{year}{2013}\natexlab{a}.
\newblock \showarticletitle{ZMap: Fast Internet-wide Scanning and Its Security
  Applications}. In \bibinfo{booktitle}{\emph{22nd {USENIX} Security Symposium
  ({USENIX} Security 13)}}. \bibinfo{publisher}{{USENIX} Association},
  \bibinfo{address}{Washington, D.C.}, \bibinfo{pages}{605--620}.
\newblock
\showISBNx{978-1-931971-03-4}
\urldef\tempurl%
\url{https://www.usenix.org/conference/usenixsecurity13/technical-sessions/paper/durumeric}
\showURL{%
\tempurl}


\bibitem[\protect\citeauthoryear{Durumeric, Wustrow, and Halderman}{Durumeric
  et~al\mbox{.}}{2013b}]%
        {Durumeric13a}
\bibfield{author}{\bibinfo{person}{Zakir Durumeric}, \bibinfo{person}{Eric
  Wustrow}, {and} \bibinfo{person}{J.~Alex Halderman}.}
  \bibinfo{year}{2013}\natexlab{b}.
\newblock \showarticletitle{{ZMap}: Fast {Internet}-wide Scanning and Its
  Security Applications}. In \bibinfo{booktitle}{\emph{Proceedings of the
  {USENIX} Security Symposium}} (johnh: pafile). \bibinfo{publisher}{{USENIX}},
  \bibinfo{address}{Washington, DC, USA}, \bibinfo{pages}{605--620}.
\newblock
\urldef\tempurl%
\url{https://www.usenix.org/system/files/conference/usenixsecurity13/sec13-paper_durumeric.pdf}
\showURL{%
\tempurl}


\bibitem[\protect\citeauthoryear{Effenberger, Kronbichler, Shin, Mayer, Tilg,
  and Perco}{Effenberger et~al\mbox{.}}{2020}]%
        {effenberger2020association}
\bibfield{author}{\bibinfo{person}{Maria Effenberger}, \bibinfo{person}{Andreas
  Kronbichler}, \bibinfo{person}{Jae~Il Shin}, \bibinfo{person}{Gert Mayer},
  \bibinfo{person}{Herbert Tilg}, {and} \bibinfo{person}{Paul Perco}.}
  \bibinfo{year}{2020}\natexlab{}.
\newblock \showarticletitle{Association of the {COVID-19} pandemic with
  {Internet} search volumes: a {Google} {Trends} analysis}.
\newblock \bibinfo{journal}{\emph{International Journal of Infectious
  Diseases}}  \bibinfo{volume}{95} (\bibinfo{year}{2020}),
  \bibinfo{pages}{192--197}.
\newblock


\bibitem[\protect\citeauthoryear{Epstein and Nolan}{Epstein and Nolan}{2020}]%
        {Epstein20a}
\bibfield{author}{\bibinfo{person}{Reid~J. Epstein} {and} \bibinfo{person}{Kay
  Nolan}.} \bibinfo{year}{2020}\natexlab{}.
\newblock \showarticletitle{A Few Thousand Protest Stay-at-Home Order at
  {Wisconsin} State Capitol}.
\newblock \bibinfo{journal}{\emph{New York Times}} (\bibinfo{date}{Apr.~24}
  \bibinfo{year}{2020}).
\newblock
\urldef\tempurl%
\url{https://www.nytimes.com/2020/04/24/us/politics/coronavirus-protests-madison-wisconsin.html}
\showURL{%
\tempurl}


\bibitem[\protect\citeauthoryear{Feldmann, Gasser, Lichtblau, Pujol, Poese,
  Dietzel, and Wagner}{Feldmann et~al\mbox{.}}{2020}]%
        {Feldmann20a}
\bibfield{author}{\bibinfo{person}{Anja Feldmann}, \bibinfo{person}{Oliver
  Gasser}, \bibinfo{person}{Franziska Lichtblau}, \bibinfo{person}{Enric
  Pujol}, \bibinfo{person}{Ingmar Poese}, \bibinfo{person}{Christoph Dietzel},
  {and} \bibinfo{person}{Daniel Wagner}.} \bibinfo{year}{2020}\natexlab{}.
\newblock \showarticletitle{The Lockdown Effect: Implications of the {COVID-19}
  Pandemic on Internet Traffic}. In \bibinfo{booktitle}{\emph{Proceedings of
  the ACM Internet Measurement Conference}}. \bibinfo{publisher}{{ACM}},
  \bibinfo{address}{Pittsburgh, PA, USA}, \bibinfo{pages}{1--18}.
\newblock
\urldef\tempurl%
\url{https://doi.org/10.1145/3419394.3423658}
\showDOI{\tempurl}


\bibitem[\protect\citeauthoryear{Feldmann, Gasser, Lichtblau, Pujol, Poese,
  Dietzel, Wagner, Wichtlhuber, Tapiador, Vallina-Rodriguez,
  et~al\mbox{.}}{Feldmann et~al\mbox{.}}{2021}]%
        {feldmann2021year}
\bibfield{author}{\bibinfo{person}{Anja Feldmann}, \bibinfo{person}{Oliver
  Gasser}, \bibinfo{person}{Franziska Lichtblau}, \bibinfo{person}{Enric
  Pujol}, \bibinfo{person}{Ingmar Poese}, \bibinfo{person}{Christoph Dietzel},
  \bibinfo{person}{Daniel Wagner}, \bibinfo{person}{Matthias Wichtlhuber},
  \bibinfo{person}{Juan Tapiador}, \bibinfo{person}{Narseo Vallina-Rodriguez},
  {et~al\mbox{.}}} \bibinfo{year}{2021}\natexlab{}.
\newblock \showarticletitle{A year in lockdown: how the waves of COVID-19
  impact internet traffic}.
\newblock \bibinfo{journal}{\emph{Commun. ACM}} \bibinfo{volume}{64},
  \bibinfo{number}{7} (\bibinfo{year}{2021}), \bibinfo{pages}{101--108}.
\newblock


\bibitem[\protect\citeauthoryear{Foremski, Plonka, and Berger}{Foremski
  et~al\mbox{.}}{2016}]%
        {Foremski16a}
\bibfield{author}{\bibinfo{person}{Pawel Foremski}, \bibinfo{person}{David
  Plonka}, {and} \bibinfo{person}{Arthur Berger}.}
  \bibinfo{year}{2016}\natexlab{}.
\newblock \showarticletitle{Entropy/{IP}: Uncovering Structure in {IPv6}
  Addresses}. In \bibinfo{booktitle}{\emph{Proceedings of the ACM Internet
  Measurement Conference}} (johnh: pafile). \bibinfo{publisher}{{ACM}},
  \bibinfo{address}{Santa Monica, CA, USA}, \bibinfo{pages}{167--181}.
\newblock
\urldef\tempurl%
\url{https://doi.org/10.1145/2987443.2987445}
\showDOI{\tempurl}


\bibitem[\protect\citeauthoryear{{France 24}}{{France 24}}{2020}]%
        {russia1}
\bibfield{author}{\bibinfo{person}{{France 24}}.}
  \bibinfo{year}{2020}\natexlab{}.
\newblock \bibinfo{title}{{Moscow goes into lockdown, urges other regions to
  take steps to slow coronavirus}}.
\newblock
  \bibinfo{howpublished}{\url{https://www.france24.com/en/20200330-moscow-goes-into-lockdown-urges-other-russian-regions-to-take-measures-to-curb-coronavirus-spread}}.
\newblock
\newblock
\shownote{Online; accessed 29 January 2014.}


\bibitem[\protect\citeauthoryear{Gasser, Scheitle, Gebhard, and Carle}{Gasser
  et~al\mbox{.}}{2016}]%
        {Gasser16a}
\bibfield{author}{\bibinfo{person}{Oliver Gasser}, \bibinfo{person}{Quirin
  Scheitle}, \bibinfo{person}{Sebastian Gebhard}, {and} \bibinfo{person}{Georg
  Carle}.} \bibinfo{year}{2016}\natexlab{}.
\newblock \showarticletitle{Scanning the {IPv6} Internet: Towards a
  Comprehensive Hitlist}. In \bibinfo{booktitle}{\emph{Proceedings of the IFIP
  International Workshop on Traffic Monitoring and Analysis}} (johnh: pafile).
  \bibinfo{publisher}{IFIP}, \bibinfo{address}{Louvain La Neuve, Belgium}.
\newblock
\urldef\tempurl%
\url{http://tma.ifip.org/2016/papers/tma2016-final51.pdf}
\showURL{%
\tempurl}


\bibitem[\protect\citeauthoryear{Gont}{Gont}{2014}]%
        {Gont14a}
\bibfield{author}{\bibinfo{person}{F. Gont}.} \bibinfo{year}{2014}\natexlab{}.
\newblock \bibinfo{booktitle}{\emph{A Method for Generating Semantically Opaque
  Interface Identifiers with {IPv6} Stateless Address Autoconfiguration
  ({SLAAC})}}.
\newblock \bibinfo{type}{RFC} 7217. \bibinfo{institution}{Internet Request For
  Comments}.
\newblock
\urldef\tempurl%
\url{https://doi.org/10.17487/RFC7217}
\showDOI{\tempurl}


\bibitem[\protect\citeauthoryear{Gont and Chown}{Gont and Chown}{2016}]%
        {Gont16a}
\bibfield{author}{\bibinfo{person}{F. Gont} {and} \bibinfo{person}{T. Chown}.}
  \bibinfo{year}{2016}\natexlab{}.
\newblock \bibinfo{booktitle}{\emph{Network Reconnaissance in {IPv6}
  Networks}}.
\newblock \bibinfo{type}{RFC} 7707. \bibinfo{institution}{Internet Request For
  Comments}.
\newblock
\urldef\tempurl%
\url{https://doi.org/10.17487/RFC7707}
\showDOI{\tempurl}


\bibitem[\protect\citeauthoryear{Google}{Google}{2021}]%
        {Google21a}
\bibfield{author}{\bibinfo{person}{Google}.} \bibinfo{year}{2021}\natexlab{}.
\newblock \bibinfo{title}{Google {IPv6}}.
\newblock \bibinfo{howpublished}{Web page \url{https://www.google.com/ipv6/}}.
\newblock
\urldef\tempurl%
\url{https://www.google.com/intl/en/ipv6/statistics.html}
\showURL{%
\tempurl}


\bibitem[\protect\citeauthoryear{Graham, McMillan, and Tentler}{Graham
  et~al\mbox{.}}{2014}]%
        {Graham14a}
\bibfield{author}{\bibinfo{person}{Robert Graham}, \bibinfo{person}{Paul
  McMillan}, {and} \bibinfo{person}{Dan Tentler}.}
  \bibinfo{year}{2014}\natexlab{}.
\newblock \bibinfo{title}{Mass Scanning the Internet}.
\newblock \bibinfo{howpublished}{Presentation at Defcon 22}.
\newblock
\urldef\tempurl%
\url{https://defcon.org/images/defcon-22/dc-22-presentations/Graham-McMillan-Tentler/DEFCON-22-Graham-McMillan-Tentler-Masscaning-the-Internet.pdf}
\showURL{%
\tempurl}


\bibitem[\protect\citeauthoryear{Grover, Park, Srikanth~Sundaresan, Kim, and
  Feamster}{Grover et~al\mbox{.}}{2013}]%
        {Grover13a}
\bibfield{author}{\bibinfo{person}{Sarthak Grover}, \bibinfo{person}{Mi~Seon
  Park}, \bibinfo{person}{Sam~Burnett Srikanth~Sundaresan},
  \bibinfo{person}{Hyojoon Kim}, {and} \bibinfo{person}{Nick Feamster}.}
  \bibinfo{year}{2013}\natexlab{}.
\newblock \showarticletitle{Peeking Behind the {NAT}: An Empirical Study of
  Home Networks}. In \bibinfo{booktitle}{\emph{Proceedings of the ACM Internet
  Measurement Conference}} (johnh: pafile). \bibinfo{publisher}{{ACM}},
  \bibinfo{address}{Barcelona, Spain}.
\newblock
\urldef\tempurl%
\url{http://conferences.sigcomm.org/imc/2013/papers/imc061-groverA.pdf}
\showURL{%
\tempurl}


\bibitem[\protect\citeauthoryear{Guardian}{Guardian}{2020}]%
        {belgium1}
\bibfield{author}{\bibinfo{person}{The Guardian}.}
  \bibinfo{year}{2020}\natexlab{}.
\newblock \bibinfo{title}{Belgium enters lockdown over coronavirus crisis –
  in pictures}.
\newblock
  \bibinfo{howpublished}{\url{https://www.theguardian.com/world/gallery/2020/mar/18/belgium-enters-lockdown-over-coronavirus-crisis-in-pictures}}.
\newblock


\bibitem[\protect\citeauthoryear{Guillot, Fontugne, Winter, Merindol, King,
  Dainotti, and Pelsser}{Guillot et~al\mbox{.}}{2019}]%
        {guillot2019chocolatine}
\bibfield{author}{\bibinfo{person}{Andreas Guillot}, \bibinfo{person}{Romain
  Fontugne}, \bibinfo{person}{Philipp Winter}, \bibinfo{person}{Pascal
  Merindol}, \bibinfo{person}{Alistair King}, \bibinfo{person}{Alberto
  Dainotti}, {and} \bibinfo{person}{Cristel Pelsser}.}
  \bibinfo{year}{2019}\natexlab{}.
\newblock \showarticletitle{Chocolatine: Outage detection for internet
  background radiation}. In \bibinfo{booktitle}{\emph{2019 Network Traffic
  Measurement and Analysis Conference (TMA)}}. IEEE, \bibinfo{address}{Paris,
  France}, \bibinfo{numpages}{8}~pages.
\newblock


\bibitem[\protect\citeauthoryear{Guo and Heidemann}{Guo and Heidemann}{2018}]%
        {Guo18a}
\bibfield{author}{\bibinfo{person}{Hang Guo} {and} \bibinfo{person}{John
  Heidemann}.} \bibinfo{year}{2018}\natexlab{}.
\newblock \showarticletitle{Detecting {ICMP} Rate Limiting in the {Internet}}.
  In \bibinfo{booktitle}{\emph{Proceedings of the Passive and Active
  Measurement Conference}} (johnh: pafile). \bibinfo{publisher}{Springer},
  \bibinfo{address}{Berlin, Germany}, \bibinfo{pages}{to appear}.
\newblock
\urldef\tempurl%
\url{https://www.isi.edu/%7ejohnh/PAPERS/Guo18a.html}
\showURL{%
\tempurl}


\bibitem[\protect\citeauthoryear{Gustafsson}{Gustafsson}{2000}]%
        {gustafsson2000adaptive}
\bibfield{author}{\bibinfo{person}{Fredrik Gustafsson}.}
  \bibinfo{year}{2000}\natexlab{}.
\newblock \bibinfo{booktitle}{\emph{Adaptive Filtering and Change Detection}}.
  Vol.~\bibinfo{volume}{1}.
\newblock \bibinfo{publisher}{John Wiley {\&} Sons, Inc.}
\newblock
\showISBNx{9780471492870}
\urldef\tempurl%
\url{https://doi.org/10.1002/0470841613}
\showDOI{\tempurl}


\bibitem[\protect\citeauthoryear{Heidemann, Pradkin, Govindan, Papadopoulos,
  Bartlett, and Bannister}{Heidemann et~al\mbox{.}}{2008}]%
        {Heidemann08c}
\bibfield{author}{\bibinfo{person}{John Heidemann}, \bibinfo{person}{Yuri
  Pradkin}, \bibinfo{person}{Ramesh Govindan}, \bibinfo{person}{Christos
  Papadopoulos}, \bibinfo{person}{Genevieve Bartlett}, {and}
  \bibinfo{person}{Joseph Bannister}.} \bibinfo{year}{2008}\natexlab{}.
\newblock \showarticletitle{Census and Survey of the Visible Internet}. In
  \bibinfo{booktitle}{\emph{Proceedings of the ACM Internet Measurement
  Conference}}. \bibinfo{publisher}{ACM}, \bibinfo{address}{Vouliagmeni,
  Greece}, \bibinfo{pages}{169--182}.
\newblock
\urldef\tempurl%
\url{https://doi.org/10.1145/1452520.1452542}
\showDOI{\tempurl}


\bibitem[\protect\citeauthoryear{Hekking}{Hekking}{2020a}]%
        {Hekking20a}
\bibfield{author}{\bibinfo{person}{Morgan Hekking}.}
  \bibinfo{year}{2020}\natexlab{a}.
\newblock \showarticletitle{{COVID-19}: {Morocco} Declares State of Emergency}.
\newblock \bibinfo{journal}{\emph{Morocco World News}} (\bibinfo{date}{Mar.~19}
  \bibinfo{year}{2020}).
\newblock
\urldef\tempurl%
\url{https://www.moroccoworldnews.com/2020/03/296213/covid-19-morocco-declares-public-health-emergency}
\showURL{%
\tempurl}
\newblock
\shownote{\url{https://www.moroccoworldnews.com/2020/03/296213/covid-19-morocco-declares-public-health-emergency}.}


\bibitem[\protect\citeauthoryear{Hekking}{Hekking}{2020b}]%
        {Morocco1}
\bibfield{author}{\bibinfo{person}{Morgan Hekking}.}
  \bibinfo{year}{2020}\natexlab{b}.
\newblock \bibinfo{title}{COVID-19: Morocco Declares State of Emergency}.
\newblock
\newblock
\urldef\tempurl%
\url{https://www.moroccoworldnews.com/2020/03/296213/covid-19-morocco-declares-public-health-emergency}
\showURL{%
\tempurl}


\bibitem[\protect\citeauthoryear{Hyndman and Athanasopoulos}{Hyndman and
  Athanasopoulos}{2018}]%
        {hyndman2018forecasting}
\bibfield{author}{\bibinfo{person}{Rob~J Hyndman} {and} \bibinfo{person}{George
  Athanasopoulos}.} \bibinfo{year}{2018}\natexlab{}.
\newblock \bibinfo{booktitle}{\emph{Forecasting: principles and practice}}.
\newblock \bibinfo{publisher}{OTexts}.
\newblock


\bibitem[\protect\citeauthoryear{IANA}{IANA}{2021}]%
        {ianaipv4}
\bibfield{author}{\bibinfo{person}{IANA}.} \bibinfo{year}{2021}\natexlab{}.
\newblock \bibinfo{title}{IPv4 Address Space Registry}.
\newblock
\newblock
\urldef\tempurl%
\url{https://www.iana.org/assignments/ipv4-address-space/ipv4-address-space.xhtml}
\showURL{%
\tempurl}


\bibitem[\protect\citeauthoryear{Imana, Korolova, and Heidemann}{Imana
  et~al\mbox{.}}{2021}]%
        {Imana21c}
\bibfield{author}{\bibinfo{person}{Basileal Imana}, \bibinfo{person}{Aleksandra
  Korolova}, {and} \bibinfo{person}{John Heidemann}.}
  \bibinfo{year}{2021}\natexlab{}.
\newblock \showarticletitle{Institutional Privacy Risks in Sharing {DNS} Data}.
  In \bibinfo{booktitle}{\emph{Proceedings of the Applied Networking Research
  Workshop}} (johnh: pafile). \bibinfo{publisher}{{ACM}},
  \bibinfo{address}{Virtual}.
\newblock
\urldef\tempurl%
\url{https://www.isi.edu/%7ejohnh/PAPERS/Imana21c.html}
\showURL{%
\tempurl}


\bibitem[\protect\citeauthoryear{India}{India}{2020}]%
        {India1}
\bibfield{author}{\bibinfo{person}{UN India}.} \bibinfo{year}{2020}\natexlab{}.
\newblock \bibinfo{title}{COVID-19: Lockdown across India, in line with WHO
  guidance.}
\newblock
  \bibinfo{howpublished}{\url{https://news.un.org/en/story/2020/03/1060132}}.
\newblock


\bibitem[\protect\citeauthoryear{Kang, Alba, and Satariano}{Kang
  et~al\mbox{.}}{2020}]%
        {Kang20a}
\bibfield{author}{\bibinfo{person}{Cecilia Kang}, \bibinfo{person}{Davey Alba},
  {and} \bibinfo{person}{Adam Satariano}.} \bibinfo{year}{2020}\natexlab{}.
\newblock \showarticletitle{Surging Traffic Is Slowing Down Our {Internet}}.
\newblock \bibinfo{journal}{\emph{New York Times}} (\bibinfo{date}{Mar.~26}
  \bibinfo{year}{2020}).
\newblock
\urldef\tempurl%
\url{https://www.nytimes.com/2020/03/26/business/coronavirus-internet-traffic-speed.html}
\showURL{%
\tempurl}


\bibitem[\protect\citeauthoryear{Katz-Bassett, Madhyastha, John, Krishnamurthy,
  Wetherall, and Anderson}{Katz-Bassett et~al\mbox{.}}{2008}]%
        {katz2008studying}
\bibfield{author}{\bibinfo{person}{Ethan Katz-Bassett},
  \bibinfo{person}{Harsha~V Madhyastha}, \bibinfo{person}{John~P John},
  \bibinfo{person}{Arvind Krishnamurthy}, \bibinfo{person}{David Wetherall},
  {and} \bibinfo{person}{Thomas~E Anderson}.} \bibinfo{year}{2008}\natexlab{}.
\newblock \showarticletitle{Studying Black Holes in the Internet with Hubble}.
  In \bibinfo{booktitle}{\emph{NSDI}}, Vol.~\bibinfo{volume}{8}.
  \bibinfo{publisher}{{USENIX}}, \bibinfo{pages}{247--262}.
\newblock


\bibitem[\protect\citeauthoryear{Kreibich, Weaver, Nechaev, and
  Paxson}{Kreibich et~al\mbox{.}}{2010}]%
        {Kreibich10a}
\bibfield{author}{\bibinfo{person}{Christian Kreibich},
  \bibinfo{person}{Nicholas Weaver}, \bibinfo{person}{Boris Nechaev}, {and}
  \bibinfo{person}{Vern Paxson}.} \bibinfo{year}{2010}\natexlab{}.
\newblock \showarticletitle{{Netalyzr}: Illuminating The Edge Network}. In
  \bibinfo{booktitle}{\emph{Proceedings of the ACM Internet Measurement
  Conference}} (johnh: pafile). \bibinfo{publisher}{{ACM}},
  \bibinfo{address}{Melbourne, Victoria, Australia}, \bibinfo{pages}{246--259}.
\newblock
\urldef\tempurl%
\url{http://conferences.sigcomm.org/imc/2010/papers/p1.pdf}
\showURL{%
\tempurl}


\bibitem[\protect\citeauthoryear{Lutu, Perino, Bagnulo, Frias-Martinez, and
  Khangosstar}{Lutu et~al\mbox{.}}{2020}]%
        {Lutu20a}
\bibfield{author}{\bibinfo{person}{Andra Lutu}, \bibinfo{person}{Diego Perino},
  \bibinfo{person}{Marcelo Bagnulo}, \bibinfo{person}{Enrique Frias-Martinez},
  {and} \bibinfo{person}{Javad Khangosstar}.} \bibinfo{year}{2020}\natexlab{}.
\newblock \showarticletitle{A Characterization of the {COVID-19} Pandemic
  Impact on a Mobile Network Operator Traffic}. In
  \bibinfo{booktitle}{\emph{Proceedings of the ACM Internet Measurement
  Conference}}. \bibinfo{publisher}{{ACM}}, \bibinfo{address}{Pittsburgh, PA,
  USA}, \bibinfo{pages}{19--33}.
\newblock
\urldef\tempurl%
\url{https://doi.org/10.1145/3419394.3423655}
\showDOI{\tempurl}


\bibitem[\protect\citeauthoryear{Martins}{Martins}{2020}]%
        {Brazil1}
\bibfield{author}{\bibinfo{person}{Por~Valéria Martins}.}
  \bibinfo{year}{2020}\natexlab{}.
\newblock \bibinfo{title}{Casos de pacientes com coronavírus sobe para 197 em
  SC e governo prorroga quarentena}.
\newblock
  \bibinfo{howpublished}{\url{https://g1.globo.com/sc/santa-catarina/noticia/2020/03/29/casos-de-pacientes-com-coronavirus-sobe-para-197-em-sc-e-governo-prorroga-quarentena.ghtml}}.
\newblock


\bibitem[\protect\citeauthoryear{Maxmind}{Maxmind}{2020}]%
        {Maxmind20a}
\bibfield{author}{\bibinfo{person}{Maxmind}.} \bibinfo{year}{2020}\natexlab{}.
\newblock \bibinfo{title}{GeoLite City}.
\newblock \bibinfo{howpublished}{Web page
  \url{http://dev.maxmind.com/geoip/geolite}}.
\newblock
\urldef\tempurl%
\url{http://dev.maxmind.com/geoip/geolite}
\showURL{%
\tempurl}


\bibitem[\protect\citeauthoryear{Moura, an~Qasim~Lone, Poursaied, Asghari, and
  van Eeten}{Moura et~al\mbox{.}}{2015}]%
        {Moura15a}
\bibfield{author}{\bibinfo{person}{Giovane C.~M. Moura},
  \bibinfo{person}{Carlos~Ga{\~n}{\'a}n an Qasim~Lone}, \bibinfo{person}{Payam
  Poursaied}, \bibinfo{person}{Hadi Asghari}, {and} \bibinfo{person}{Michel van
  Eeten}.} \bibinfo{year}{2015}\natexlab{}.
\newblock \showarticletitle{How Dynamic is the {ISPs} Address Space? Towards
  {Internet}-Wide {DHCP} Churn Estimation}. In
  \bibinfo{booktitle}{\emph{Proceedings of the IFIP Networking}}.
  \bibinfo{publisher}{IFIP}, \bibinfo{pages}{1--9}.
\newblock
\urldef\tempurl%
\url{https://doi.org/10.1109/IFIPNetworking.2015.7145335}
\showDOI{\tempurl}


\bibitem[\protect\citeauthoryear{Mundo}{Mundo}{2020}]%
        {spain1}
\bibfield{author}{\bibinfo{person}{El Mundo}.} \bibinfo{year}{2020}\natexlab{}.
\newblock \bibinfo{title}{Pedro Sánchez anuncia el estado de alarma para
  frenar el coronavirus 24 horas antes de aprobarlo}.
\newblock
  \bibinfo{howpublished}{\url{https://www.elmundo.es/espana/2020/03/13/5e6b844e21efa0dd258b45a5.html}}.
\newblock


\bibitem[\protect\citeauthoryear{Murdock, Li, Bramsen, Durumeric, and
  Paxson}{Murdock et~al\mbox{.}}{2017}]%
        {Murdock17a}
\bibfield{author}{\bibinfo{person}{Austin Murdock}, \bibinfo{person}{Frank Li},
  \bibinfo{person}{Paul Bramsen}, \bibinfo{person}{Zakir Durumeric}, {and}
  \bibinfo{person}{Vern Paxson}.} \bibinfo{year}{2017}\natexlab{}.
\newblock \showarticletitle{Target Generation for Internet-wide {IPv6}
  Scanning}. In \bibinfo{booktitle}{\emph{Proceedings of the ACM Internet
  Measurement Conference}} (johnh: pafile). \bibinfo{publisher}{{ACM}},
  \bibinfo{address}{San Diego, CA, USA}, \bibinfo{pages}{242--253}.
\newblock
\urldef\tempurl%
\url{https://doi.org/10.1145/3131365.3131405}
\showDOI{\tempurl}


\bibitem[\protect\citeauthoryear{Padmanabhan, Dhamdhere, Aben, kc~claffy, and
  Spring}{Padmanabhan et~al\mbox{.}}{2016}]%
        {Padmanabhan16a}
\bibfield{author}{\bibinfo{person}{Ramakrishna Padmanabhan},
  \bibinfo{person}{Amogh Dhamdhere}, \bibinfo{person}{Emile Aben},
  \bibinfo{person}{kc claffy}, {and} \bibinfo{person}{Neil Spring}.}
  \bibinfo{year}{2016}\natexlab{}.
\newblock \showarticletitle{Reasons Dynamic Addresses Change}. In
  \bibinfo{booktitle}{\emph{Proceedings of the ACM Internet Measurement
  Conference}}. \bibinfo{publisher}{{ACM}}, \bibinfo{address}{Santa Monica, CA,
  USA}, \bibinfo{pages}{183--198}.
\newblock
\urldef\tempurl%
\url{https://doi.org/10.1145/2987443.2987461}
\showDOI{\tempurl}


\bibitem[\protect\citeauthoryear{Padmanabhan, Rula, Richter, Strowes, and
  Dainotti}{Padmanabhan et~al\mbox{.}}{2020}]%
        {Padmanabhan20a}
\bibfield{author}{\bibinfo{person}{Ramakrishna Padmanabhan},
  \bibinfo{person}{John~P. Rula}, \bibinfo{person}{Philipp Richter},
  \bibinfo{person}{Stephen~D. Strowes}, {and} \bibinfo{person}{Alberto
  Dainotti}.} \bibinfo{year}{2020}\natexlab{}.
\newblock \showarticletitle{{DynamIPs}: Analyzing address assignment practices
  in {IPv4} and {IPv6}}. In \bibinfo{booktitle}{\emph{Proceedings of the ACM
  Conference on Emerging Networking Experiments and Technologies}} (johnh:
  pafile). \bibinfo{publisher}{{ACM}}, \bibinfo{address}{xxx},
  \bibinfo{pages}{16}.
\newblock
\urldef\tempurl%
\url{https://doi.org/10.1145/3386367.3431314}
\showURL{%
\tempurl}


\bibitem[\protect\citeauthoryear{Press}{Press}{2020}]%
        {Wuhan1}
\bibfield{author}{\bibinfo{person}{The~Associated Press}.}
  \bibinfo{year}{2020}\natexlab{}.
\newblock \bibinfo{title}{Timeline: China's COVID-19 outbreak and lockdown of
  Wuhan}.
\newblock
  \bibinfo{howpublished}{\url{https://abcnews.go.com/Health/wireStory/timeline-chinas-covid-19-outbreak-lockdown-wuhan-75421357}}.
\newblock


\bibitem[\protect\citeauthoryear{Quan, Heidemann, and Pradkin}{Quan
  et~al\mbox{.}}{2013}]%
        {quan2013trinocular}
\bibfield{author}{\bibinfo{person}{Lin Quan}, \bibinfo{person}{John Heidemann},
  {and} \bibinfo{person}{Yuri Pradkin}.} \bibinfo{year}{2013}\natexlab{}.
\newblock \showarticletitle{Trinocular: Understanding {Internet} Reliability
  Through Adaptive Probing}. In \bibinfo{booktitle}{\emph{Proceedings of the
  {ACM} SIGCOMM Conference}}. \bibinfo{publisher}{{ACM}},
  \bibinfo{address}{Hong Kong, China}, \bibinfo{pages}{255--266}.
\newblock
\urldef\tempurl%
\url{https://doi.org/10.1145/2486001.2486017}
\showDOI{\tempurl}


\bibitem[\protect\citeauthoryear{Quan, Heidemann, and Pradkin}{Quan
  et~al\mbox{.}}{2014}]%
        {Quan14c}
\bibfield{author}{\bibinfo{person}{Lin Quan}, \bibinfo{person}{John Heidemann},
  {and} \bibinfo{person}{Yuri Pradkin}.} \bibinfo{year}{2014}\natexlab{}.
\newblock \showarticletitle{When the {Internet} Sleeps: Correlating Diurnal
  Networks With External Factors}. In \bibinfo{booktitle}{\emph{Proceedings of
  the ACM Internet Measurement Conference}}. \bibinfo{publisher}{{ACM}},
  \bibinfo{address}{Vancouver, BC, Canada}, \bibinfo{pages}{87--100}.
\newblock
\urldef\tempurl%
\url{https://doi.org/10.1145/2663716.2663721}
\showDOI{\tempurl}


\bibitem[\protect\citeauthoryear{Rekhter, Moskowitz, Karrenberg, de~Groot, and
  Lear}{Rekhter et~al\mbox{.}}{1996}]%
        {Rekhter96a}
\bibfield{author}{\bibinfo{person}{Y. Rekhter}, \bibinfo{person}{B. Moskowitz},
  \bibinfo{person}{D. Karrenberg}, \bibinfo{person}{G.~J. de Groot}, {and}
  \bibinfo{person}{E. Lear}.} \bibinfo{year}{1996}\natexlab{}.
\newblock \bibinfo{booktitle}{\emph{Address Allocation for Private Internets}}.
\newblock \bibinfo{type}{RFC} 1918. \bibinfo{institution}{Internet Request For
  Comments}.
\newblock
\urldef\tempurl%
\url{ftp://ftp.rfc-editor.org/in-notes/rfc1918.txt}
\showURL{%
\tempurl}


\bibitem[\protect\citeauthoryear{Reuters}{Reuters}{2020}]%
        {venezuela1}
\bibfield{author}{\bibinfo{person}{Reuters}.} \bibinfo{year}{2020}\natexlab{}.
\newblock \bibinfo{title}{Venezuela's to implement nationwide quarantine as
  coronavirus cases rise to 33}.
\newblock
  \bibinfo{howpublished}{\url{https://www.reuters.com/article/us-health-coronavirus-venezuela-maduro/venezuelas-to-implement-nationwide-quarantine-as-coronavirus-cases-rise-to-33-idUSKBN214015}}.
\newblock


\bibitem[\protect\citeauthoryear{Richter, Padmanabhan, Spring, Berger, and
  Clark}{Richter et~al\mbox{.}}{2018}]%
        {richter2018advancing}
\bibfield{author}{\bibinfo{person}{Philipp Richter},
  \bibinfo{person}{Ramakrishna Padmanabhan}, \bibinfo{person}{Neil Spring},
  \bibinfo{person}{Arthur Berger}, {and} \bibinfo{person}{David Clark}.}
  \bibinfo{year}{2018}\natexlab{}.
\newblock \showarticletitle{Advancing the art of internet edge outage
  detection}. In \bibinfo{booktitle}{\emph{Proceedings of the Internet
  Measurement Conference 2018}}. \bibinfo{pages}{350--363}.
\newblock


\bibitem[\protect\citeauthoryear{Richter, Smaragdakis, Plonka, and
  Berger}{Richter et~al\mbox{.}}{2016a}]%
        {Richter16b}
\bibfield{author}{\bibinfo{person}{Philipp Richter}, \bibinfo{person}{Georgios
  Smaragdakis}, \bibinfo{person}{David Plonka}, {and} \bibinfo{person}{Arthur
  Berger}.} \bibinfo{year}{2016}\natexlab{a}.
\newblock \showarticletitle{Beyond Counting: New Perspectives on the Active
  {IPv4} Address Space}. In \bibinfo{booktitle}{\emph{Proceedings of the ACM
  Internet Measurement Conference}} (johnh: pafile).
  \bibinfo{publisher}{{ACM}}, \bibinfo{address}{Santa Monica, CA, USA},
  \bibinfo{pages}{135--149}.
\newblock
\urldef\tempurl%
\url{https://doi.org/10.1145/2987443.2987473}
\showDOI{\tempurl}


\bibitem[\protect\citeauthoryear{Richter, Wohlfart, Vallina-Rodriguez, Allman,
  Bush, Feldmann, Kreibich, Weaver, and Paxson}{Richter et~al\mbox{.}}{2016b}]%
        {Richter16c}
\bibfield{author}{\bibinfo{person}{Philipp Richter}, \bibinfo{person}{Florian
  Wohlfart}, \bibinfo{person}{Narseo Vallina-Rodriguez}, \bibinfo{person}{Mark
  Allman}, \bibinfo{person}{Randy Bush}, \bibinfo{person}{Anja Feldmann},
  \bibinfo{person}{Christian Kreibich}, \bibinfo{person}{Nicholas Weaver},
  {and} \bibinfo{person}{Vern Paxson}.} \bibinfo{year}{2016}\natexlab{b}.
\newblock \showarticletitle{A Multi-perspective Analysis of Carrier-Grade {NAT}
  Deployment}. In \bibinfo{booktitle}{\emph{Proceedings of the ACM Internet
  Measurement Conference}} (johnh: pafile). \bibinfo{publisher}{{ACM}},
  \bibinfo{address}{Santa Monica, CA, USA}.
\newblock
\urldef\tempurl%
\url{https://doi.org/10.1145/2987443.2987474}
\showDOI{\tempurl}


\bibitem[\protect\citeauthoryear{Robel}{Robel}{2020}]%
        {Robel20a}
\bibfield{author}{\bibinfo{person}{Lauren Robel}.}
  \bibinfo{year}{2020}\natexlab{}.
\newblock \bibinfo{title}{A Letter to {Hoosiers}}.
\newblock \bibinfo{howpublished}{Student communications
  \url{https://provost.indiana.edu/statements/covid/for-students/march-19.html}}.
\newblock
\urldef\tempurl%
\url{https://provost.indiana.edu/statements/covid/for-students/march-19.html}
\showURL{%
\tempurl}


\bibitem[\protect\citeauthoryear{Schafer}{Schafer}{1998}]%
        {Schafer98a}
\bibfield{author}{\bibinfo{person}{Sarah Schafer}.}
  \bibinfo{year}{1998}\natexlab{}.
\newblock \showarticletitle{With Capital in Panic, Pizza Deliveries Soar}.
\newblock \bibinfo{journal}{\emph{The Washington Post}}
  (\bibinfo{date}{Dec.~19} \bibinfo{year}{1998}), \bibinfo{pages}{D1}.
\newblock
\urldef\tempurl%
\url{https://www.washingtonpost.com/wp-srv/politics/special/clinton/stories/pizza121998.htm}
\showURL{%
\tempurl}


\bibitem[\protect\citeauthoryear{Schulman and Spring}{Schulman and
  Spring}{2011}]%
        {Schulman11a}
\bibfield{author}{\bibinfo{person}{Aaron Schulman} {and} \bibinfo{person}{Neil
  Spring}.} \bibinfo{year}{2011}\natexlab{}.
\newblock \showarticletitle{Pingin' in the Rain}. In
  \bibinfo{booktitle}{\emph{Proceedings of the ACM Internet Measurement
  Conference}} (johnh: pafile). \bibinfo{publisher}{{ACM}},
  \bibinfo{address}{Berlin, Germany}, \bibinfo{pages}{19--25}.
\newblock
\urldef\tempurl%
\url{https://doi.org/10.1145/2068816.2068819}
\showDOI{\tempurl}


\bibitem[\protect\citeauthoryear{Seabold and Perktold}{Seabold and
  Perktold}{2010}]%
        {seabold2010statsmodels}
\bibfield{author}{\bibinfo{person}{Skipper Seabold} {and}
  \bibinfo{person}{Josef Perktold}.} \bibinfo{year}{2010}\natexlab{}.
\newblock \showarticletitle{Statsmodels: Econometric and statistical modeling
  with {Python}}. In \bibinfo{booktitle}{\emph{Proceedings of the 9th Python in
  Science Conference}}, Vol.~\bibinfo{volume}{57}. Austin, TX,
  \bibinfo{publisher}{\url{scipy.org}}, \bibinfo{pages}{61}.
\newblock


\bibitem[\protect\citeauthoryear{Setboonsarng}{Setboonsarng}{2020}]%
        {Setboonsarng20a}
\bibfield{author}{\bibinfo{person}{Chayut Setboonsarng}.}
  \bibinfo{year}{2020}\natexlab{}.
\newblock \bibinfo{title}{Hundreds join protest against ban of opposition party
  in Thailand}.
\newblock
  \bibinfo{howpublished}{\url{https://www.reuters.com/article/us-thailand-politics/hundreds-join-protest-against-ban-of-opposition-party-in-thailand-idUSKCN20G0EW}}.
\newblock


\bibitem[\protect\citeauthoryear{Shafiq, Ji, Liu, and Wang}{Shafiq
  et~al\mbox{.}}{2011}]%
        {Shafiq11a}
\bibfield{author}{\bibinfo{person}{M.~Zubair Shafiq}, \bibinfo{person}{Lusheng
  Ji}, \bibinfo{person}{Alex~X. Liu}, {and} \bibinfo{person}{Jia Wang}.}
  \bibinfo{year}{2011}\natexlab{}.
\newblock \showarticletitle{Characterizing and Modeling {Internet} Traffic
  Dynamics of Cellular Devices}. In \bibinfo{booktitle}{\emph{Proceedings of
  the {ACM} {SIGMETRICS}}} (johnh: pafile). \bibinfo{publisher}{{ACM}},
  \bibinfo{address}{San Jose, CA, USA}, \bibinfo{pages}{305--316}.
\newblock
\urldef\tempurl%
\url{https://doi.org/10.1145/1993744.1993776}
\showDOI{\tempurl}


\bibitem[\protect\citeauthoryear{Shah, Fontugne, Aben, Pelsser, and Bush}{Shah
  et~al\mbox{.}}{2017}]%
        {shah2017disco}
\bibfield{author}{\bibinfo{person}{Anant Shah}, \bibinfo{person}{Romain
  Fontugne}, \bibinfo{person}{Emile Aben}, \bibinfo{person}{Cristel Pelsser},
  {and} \bibinfo{person}{Randy Bush}.} \bibinfo{year}{2017}\natexlab{}.
\newblock \showarticletitle{Disco: Fast, good, and cheap outage detection}. In
  \bibinfo{booktitle}{\emph{2017 Network Traffic Measurement and Analysis
  Conference (TMA)}}. IEEE, \bibinfo{pages}{1--9}.
\newblock


\bibitem[\protect\citeauthoryear{Slovenija}{Slovenija}{2020}]%
        {svn1}
\bibfield{author}{\bibinfo{person}{Radiotelevizija Slovenija}.}
  \bibinfo{year}{2020}\natexlab{}.
\newblock \bibinfo{title}{Staršem 50 odstotkov plače, varstvo za nujno
  potrebne poklice}.
\newblock
  \bibinfo{howpublished}{\url{https://www.rtvslo.si/zdravje/novi-koronavirus/starsem-50-odstotkov-place-varstvo-za-nujno-potrebne-poklice/516908}}.
\newblock


\bibitem[\protect\citeauthoryear{Stewart, Mason, and Dodd}{Stewart
  et~al\mbox{.}}{2020}]%
        {uk1}
\bibfield{author}{\bibinfo{person}{Heather Stewart}, \bibinfo{person}{Rowena
  Mason}, {and} \bibinfo{person}{Vikram Dodd}.}
  \bibinfo{year}{2020}\natexlab{}.
\newblock \bibinfo{title}{Boris Johnson orders UK lockdown to be enforced by
  police}.
\newblock
  \bibinfo{howpublished}{\url{https://www.theguardian.com/world/2020/mar/23/boris-johnson-orders-uk-lockdown-to-be-enforced-by-police}}.
\newblock


\bibitem[\protect\citeauthoryear{Stutz, Pradkin, Song, and Heidemann}{Stutz
  et~al\mbox{.}}{2021a}]%
        {Stutz21b}
\bibfield{author}{\bibinfo{person}{Erica Stutz}, \bibinfo{person}{Yuri
  Pradkin}, \bibinfo{person}{Xiao Song}, {and} \bibinfo{person}{John
  Heidemann}.} \bibinfo{year}{2021}\natexlab{a}.
\newblock \bibinfo{title}{{ANT} Evaluation of {COVID-19} Internet Downtrends}.
\newblock \bibinfo{howpublished}{\url{https://covid.ant.isi.edu/}}.
\newblock
\urldef\tempurl%
\url{https://covid.ant.isi.edu/}
\showURL{%
\tempurl}


\bibitem[\protect\citeauthoryear{Stutz, Pradkin, Song, and Heidemann}{Stutz
  et~al\mbox{.}}{2021b}]%
        {Stutz21a}
\bibfield{author}{\bibinfo{person}{Erica Stutz}, \bibinfo{person}{Yuri
  Pradkin}, \bibinfo{person}{Xiao Song}, {and} \bibinfo{person}{John
  Heidemann}.} \bibinfo{year}{2021}\natexlab{b}.
\newblock \showarticletitle{Visualizing {Internet} Measurements of {Covid-19}
  Work-from-Home}. In \bibinfo{booktitle}{\emph{Proceedings of the National
  Symposium for NSF REU Research in Data Science, Systems, and Security (REU
  2021 Symposium)}} (johnh: pafile). \bibinfo{address}{Virtual Workshop},
  \bibinfo{pages}{5633--5638}.
\newblock
\urldef\tempurl%
\url{https://doi.org/tbd}
\showDOI{\tempurl}


\bibitem[\protect\citeauthoryear{Sulaymaniyah}{Sulaymaniyah}{2020}]%
        {iraq1}
\bibfield{author}{\bibinfo{person}{Erbil Sulaymaniyah}.}
  \bibinfo{year}{2020}\natexlab{}.
\newblock \bibinfo{title}{raq's KRG eases coronavirus lockdown after two
  months}.
\newblock
  \bibinfo{howpublished}{\url{https://www.aa.com.tr/en/middle-east/iraqs-krg-eases-coronavirus-lockdown-after-two-months/1838231}}.
\newblock


\bibitem[\protect\citeauthoryear{Sylvers and Legorano}{Sylvers and
  Legorano}{2020}]%
        {italy1}
\bibfield{author}{\bibinfo{person}{Eric Sylvers} {and}
  \bibinfo{person}{Giovanni Legorano}.} \bibinfo{year}{2020}\natexlab{}.
\newblock \bibinfo{title}{As Virus Spreads, Italy Locks Down Country}.
\newblock
  \bibinfo{howpublished}{\url{https://www.wsj.com/articles/italy-bolsters-quarantine-checks-after-initial-lockdown-confusion-11583756737}}.
\newblock


\bibitem[\protect\citeauthoryear{Söder}{Söder}{2020}]%
        {germany1}
\bibfield{author}{\bibinfo{person}{Markus Söder}.}
  \bibinfo{year}{2020}\natexlab{}.
\newblock \bibinfo{title}{Lockdown was started in Freiburg, Baden-Württemberg
  and Bavaria on 20 March 2020. Two days later, it was expanded to the whole of
  Germany.}
\newblock
  \bibinfo{howpublished}{\url{https://pbs.twimg.com/media/ETjc738WoAI5zAt?format=jpg&name=large}}.
\newblock


\bibitem[\protect\citeauthoryear{Talabong}{Talabong}{2020}]%
        {Metro1}
\bibfield{author}{\bibinfo{person}{Rambo Talabong}.}
  \bibinfo{year}{2020}\natexlab{}.
\newblock \bibinfo{title}{Metro Manila to be placed on lockdown due to
  coronavirus outbreak}.
\newblock
  \bibinfo{howpublished}{\url{https://www.rappler.com/nation/metro-manila-placed-on-lockdown-coronavirus-outbreak}}.
\newblock


\bibitem[\protect\citeauthoryear{Tanwar}{Tanwar}{2020}]%
        {Tanwar20a}
\bibfield{author}{\bibinfo{person}{Sangeeta Tanwar}.}
  \bibinfo{year}{2020}\natexlab{}.
\newblock \showarticletitle{Delhi chief minister seeks {Indian} Army's help
  after riots claim 20 lives over two days}.
\newblock \bibinfo{journal}{\emph{Quartz India}} (\bibinfo{date}{Feb.~25}
  \bibinfo{year}{2020}).
\newblock
\urldef\tempurl%
\url{https://qz.com/india/1808434/arvind-kejriwal-wants-curfew-army-to-quell-delhi-riots/}
\showURL{%
\tempurl}


\bibitem[\protect\citeauthoryear{Times}{Times}{2020}]%
        {malaysia1}
\bibfield{author}{\bibinfo{person}{New~Straits Times}.}
  \bibinfo{year}{2020}\natexlab{}.
\newblock \bibinfo{title}{Covid-19: Movement Control Order imposed with only
  essential sectors operating}.
\newblock
  \bibinfo{howpublished}{\url{https://www.nst.com.my/news/nation/2020/03/575177/covid-19-movement-control-order-imposed-only-essential-sectors-operating}}.
\newblock


\bibitem[\protect\citeauthoryear{Ukani, Mirian, and Snoeren}{Ukani
  et~al\mbox{.}}{2021}]%
        {UkaniIMC21Covid}
\bibfield{author}{\bibinfo{person}{Alisha Ukani}, \bibinfo{person}{Ariana
  Mirian}, {and} \bibinfo{person}{Alex~C. Snoeren}.}
  \bibinfo{year}{2021}\natexlab{}.
\newblock \showarticletitle{Locked-in during Lock-down: Undergraduate Life on
  the Internet in a Pandemic}. In \bibinfo{booktitle}{\emph{Proceedings of the
  21st ACM Internet Measurement Conference}} (Virtual Event)
  \emph{(\bibinfo{series}{IMC '21})}. \bibinfo{publisher}{Association for
  Computing Machinery}, \bibinfo{address}{New York, NY, USA},
  \bibinfo{pages}{480–486}.
\newblock
\showISBNx{9781450391290}
\urldef\tempurl%
\url{https://doi.org/10.1145/3487552.3487828}
\showDOI{\tempurl}


\bibitem[\protect\citeauthoryear{van~der Toorn, Sommese, Sperotto, van
  Rijswijk{-}Deij, and Jonker}{van~der Toorn et~al\mbox{.}}{2022}]%
        {Toorn22Covid}
\bibfield{author}{\bibinfo{person}{Olivier van~der Toorn},
  \bibinfo{person}{Raffaele Sommese}, \bibinfo{person}{Anna Sperotto},
  \bibinfo{person}{Roland van Rijswijk{-}Deij}, {and} \bibinfo{person}{Mattijs
  Jonker}.} \bibinfo{year}{2022}\natexlab{}.
\newblock \showarticletitle{Saving Brian's Privacy: the Perils of Privacy
  Exposure through Reverse {DNS}}.
\newblock \bibinfo{journal}{\emph{CoRR}}  \bibinfo{volume}{abs/2202.01160}
  (\bibinfo{year}{2022}).
\newblock
\showeprint[arXiv]{2202.01160}
\urldef\tempurl%
\url{https://arxiv.org/abs/2202.01160}
\showURL{%
\tempurl}


\bibitem[\protect\citeauthoryear{Wan, Izhikevich, Adrian, Yoshioka, Holz,
  Rossow, and Durumeric}{Wan et~al\mbox{.}}{2020}]%
        {Wan20a}
\bibfield{author}{\bibinfo{person}{Gerry Wan}, \bibinfo{person}{Liz
  Izhikevich}, \bibinfo{person}{David Adrian}, \bibinfo{person}{Katsunari
  Yoshioka}, \bibinfo{person}{Ralph Holz}, \bibinfo{person}{Christian Rossow},
  {and} \bibinfo{person}{Zakir Durumeric}.} \bibinfo{year}{2020}\natexlab{}.
\newblock \showarticletitle{On the Origin of Scanning: The Impact of Location
  on Internet-Wide Scans}. In \bibinfo{booktitle}{\emph{Proceedings of the ACM
  Internet Measurement Conference}}. \bibinfo{publisher}{{ACM}},
  \bibinfo{address}{Pittsburgh, PA, USA}, \bibinfo{pages}{662--679}.
\newblock
\urldef\tempurl%
\url{https://doi.org/10.1145/3419394.3424214}
\showDOI{\tempurl}


\bibitem[\protect\citeauthoryear{Welle}{Welle}{2020}]%
        {ukr1}
\bibfield{author}{\bibinfo{person}{Deutsche Welle}.}
  \bibinfo{year}{2020}\natexlab{}.
\newblock \bibinfo{title}{Coronavirus: What are the lockdown measures across
  Europe?}
\newblock
  \bibinfo{howpublished}{\url{https://www.dw.com/en/coronavirus-what-are-the-lockdown-measures-across-europe/a-52905137}}.
\newblock


\bibitem[\protect\citeauthoryear{Wikipedia}{Wikipedia}{2022}]%
        {Wikipedia22a}
\bibfield{author}{\bibinfo{person}{Wikipedia}.}
  \bibinfo{year}{2022}\natexlab{}.
\newblock \bibinfo{title}{Timeline of the {COVID-19} pandemic in {Thailand}}.
\newblock
  \bibinfo{howpublished}{\url{https://en.wikipedia.org/wiki/Timeline_of_the_COVID-19_pandemic_in_Thailand}}.
\newblock
\urldef\tempurl%
\url{https://en.wikipedia.org/wiki/Timeline_of_the_COVID-19_pandemic_in_Thailand}
\showURL{%
\tempurl}
\newblock
\shownote{accessed 2022-05-16.}


\bibitem[\protect\citeauthoryear{Xie, Yu, Achan, Gillum, Goldszmidt, and
  Wobber}{Xie et~al\mbox{.}}{2007}]%
        {Xie07a}
\bibfield{author}{\bibinfo{person}{Yinglian Xie}, \bibinfo{person}{Fang Yu},
  \bibinfo{person}{Kannan Achan}, \bibinfo{person}{Eliot Gillum},
  \bibinfo{person}{Moises Goldszmidt}, {and} \bibinfo{person}{Ted Wobber}.}
  \bibinfo{year}{2007}\natexlab{}.
\newblock \showarticletitle{How Dynamic are {IP} Addresses?}. In
  \bibinfo{booktitle}{\emph{Proceedings of the {ACM} SIGCOMM Conference}}.
  \bibinfo{publisher}{{ACM}}, \bibinfo{address}{Kyoto, Japan},
  \bibinfo{pages}{301--312}.
\newblock
\urldef\tempurl%
\url{https://doi.org/10.1145/1282380.1282415}
\showDOI{\tempurl}


\end{thebibliography}
\end{CJK*}

\appendix

\section{Research Ethics}
	\label{sec:research_ethics}

\reviewfix{S22X2}

In developing a new measurement technique,
  we must consider potential new risks it creates for individuals and
  for organizations.
Our overall goal is to identify human behavior.
Our general way to minimize risk is to focus on aggregate behavior
  and avoid identification of individuals
  through a combination of technical and policy methods.

\emph{Risks:}
The primary risk to individuals is that data may
  expose the behavior of specific users---work schedules of individuals
  may be sensitive.
We avoid this risk by separating what we learn about IP addresses
  from knowledge of specific individuals.
We do not have any knowledge of which individuals use which IP addresses,
  so our data does not, by itself, pose any risk to individuals.

Of course, \emph{other} datasets associate IP addresses with individuals.
ISPs and organizations may track user-to-IP mapping for accounting
  and accountability.
Combining such mapping data with our data might pose a risk.
We minimize this risk by
  aggregating data by /24 prefix early in our processing pipeline,
  and handling pre-aggregated data with largely automated procedures.
Our analysis require specific IP addresses
  only through the reconstruction phase (\autoref{sec:accumulating_addreses}).

Beyond individuals,
  WFH behavior may be sensitive to organizations~\cite{Imana21c}.
For example, increased deliveries correlates with longer work
  and may indicate unusual events~\cite{Schafer98a}.
In some countries and regions, Covid response has become politicized,
  and there WFH or its absence may be supporting or resisting government rules.
While protecting privacy of individuals is important,
  the reputation of organizations or regions must be balanced
  with the public's need to understand the choices people make.

\emph{Benefits:}
The benefits of our work are to provide a new method of identifying WFH
  on a global scale by reanalyzing existing data.
The Covid pandemic is an ongoing global health crisis
  resulting in millions of deaths
  with broad social and economic impacts over the last two years.
We believe
  that our analysis can provide a new perspective on actual human behavior,
  thereby contributing to discussions about public health response.
Although measurements of the Internet have their own limitations,
  we see it as providing a valuable complement to traditional
  tools to understand public health, such a as surveys,
  institutional reporting, and wastewater monitoring.

In our view these benefits outweigh the risks.

\emph{Data collection:}
Most of our analysis is new evaluation
  of the existing dataset listed in \autoref{sec:datasets_more_data}.
We are now  exploring new data collection
  (\autoref{sec:alg_reconstruction}).
Although some risk is created by
  new information available through our analysis,
  the underlying data and therefore the potential of such risk is not new.

\emph{Data distribution:}
We are committed to providing the results of our data to the research community
  and no charge.
However, to ensure that researchers do not create additional privacy risks,
  we distribute data under terms-of-use that forbids de-anonymization
  and redistribution.

\emph{IRB review:}
Our work was reviewed by our university's
  Institutional Review Board
  and because it does not identify individuals,
  it was classified as non-human-subjects research
  (USC \#UP-20-00909).

\section{Datasets Used in This Paper}
	\label{sec:datasets_more_data}

\begin{table*}
	\begin{small}
		\begin{tabular}{lllr}
			\textbf{abbr.} & \textbf{dataset name} & \textbf{start} & \textbf{duration} \\
			2019q4-w & internet\_outage\_adaptive\_a38w-20191001 & 2019-10-01 & 12 weeks \\
			2020q1-w & internet\_outage\_adaptive\_a39w-20200101 & 2020-01-01 & 12 weeks \\
			2020q2-w &internet\_outage\_adaptive\_a40w-20200401 & 2020-04-01 & 12 weeks \\
			\hline
			2020q1-j &internet\_outage\_adaptive\_a39j-20200101 & 2020-01-01 & 12 weeks \\
			2020q2-j &internet\_outage\_adaptive\_a40j-20200401 & 2020-04-01 & 12 weeks \\
			\hline
			2020q1-n &internet\_outage\_adaptive\_a39n-20200101 & 2020-01-01 & 12 weeks\\
			2020q2-n &internet\_outage\_adaptive\_a40n-20200401 & 2020-04-01 & 12 weeks \\
			\hline
			2020q1-e &internet\_outage\_adaptive\_a39e-20200101 & 2020-01-01 & 12 weeks \\
			2020q2-e &internet\_outage\_adaptive\_a40e-20200401 & 2020-04-01 & 12 weeks \\
			\hline
			2020it89-w &internet\_address\_survey\_reprobing\_it89w-20200219 & 2020-02-19 & 2 weeks
		\end{tabular}
	\end{small}
	\caption{We use Internet outage datasets collected by Trinocular from \SitesWord vantage points.
           Letter indicates VP's location:
           e: ISI-East, near Washington, DC;
           j: Japan, Keio University, near Tokyo;
           n: Netherlands data from SurfNet;
           w: ISI-West, Los Angeles.
		\reviewfix{I21A10}
	}
	\label{tab:datasets}
\end{table*}

\autoref{tab:datasets} provides a full list of datasets used this this paper.

\begin{figure*}
	\centering
        \subfloat[Anon.~university~A with WFH.]{
		\includegraphics[width=0.3\textwidth]{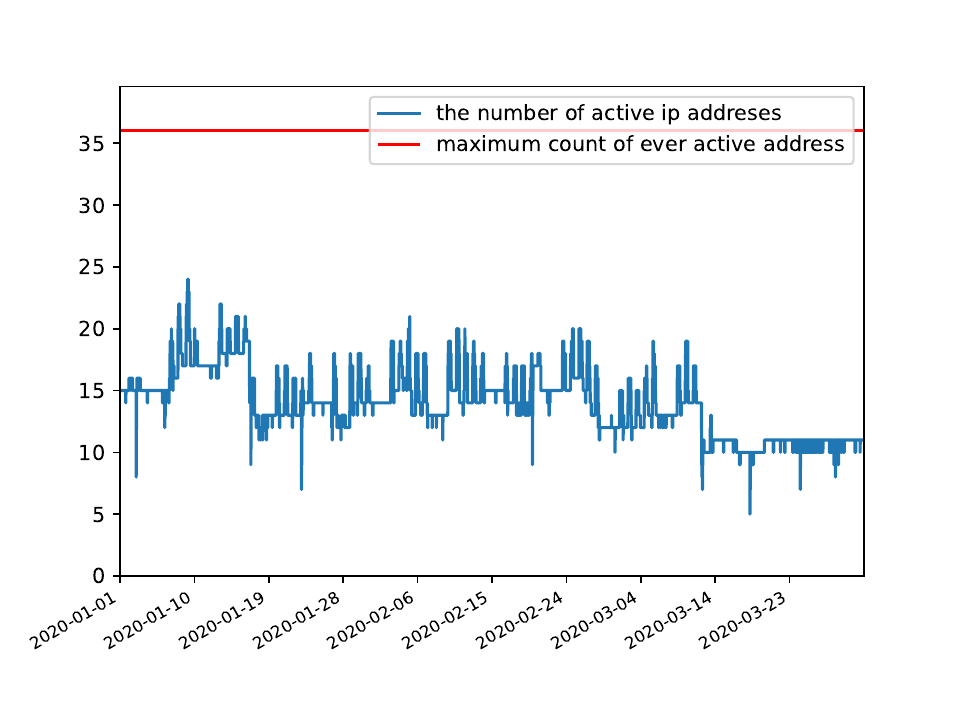}
            }
            \quad 
        \subfloat[Anon.~university~B with WFH.]{
		\includegraphics[width=0.3\textwidth]{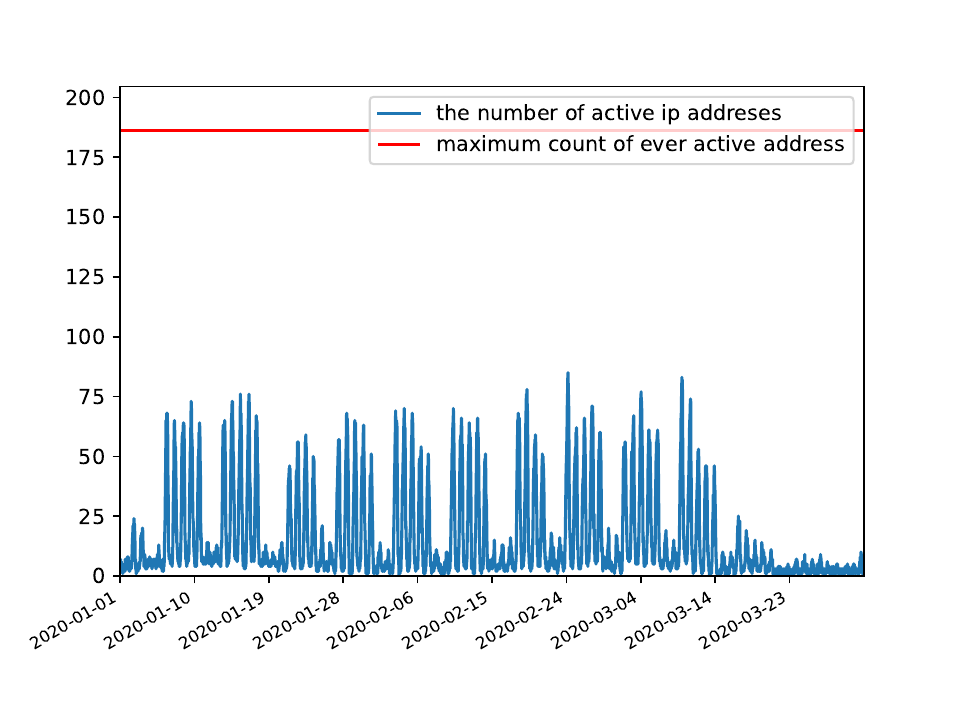}
         }
         \quad 
        \subfloat[Anon.~university~C (no change).]{
		\label{fig:block_958e1a00}
		\includegraphics[width=0.3\textwidth]{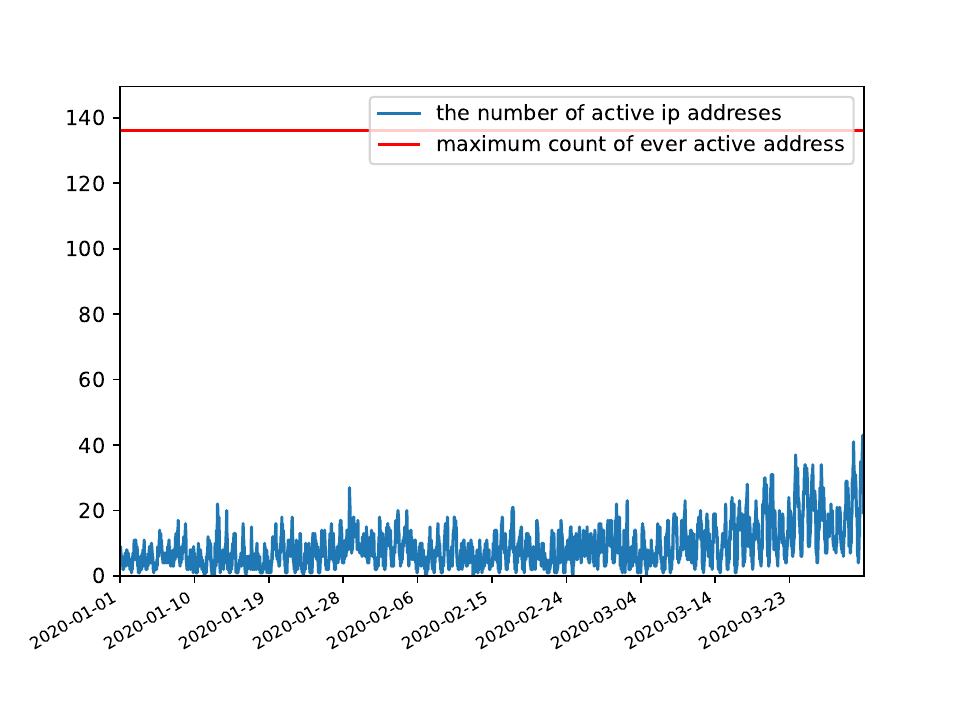}
                }
	\caption{Sample blocks from three anonymized U.S.~universities.}
                \label{fig:additional_universities}
\end{figure*}

\section{Case Study of Specific Blocks}
  \label{sec:case_study_specific_blocks}

We use \autoref{fig:isi_lab_example} as a running example
  in \autoref{sec:methodology} to show our methodology.
It was one of many that we examined when developing our approach.
We next review additional blocks
  to provide more representatives of changes that we observed.

\subsection{Detections in Two Additional Blocks}

\begin{figure}
	\subfloat[The anon.~block with only dirunal behavior. ]{
		\label{fig:other.blocks.zero}
		\mbox{\includegraphics[width=0.9\columnwidth,clip]{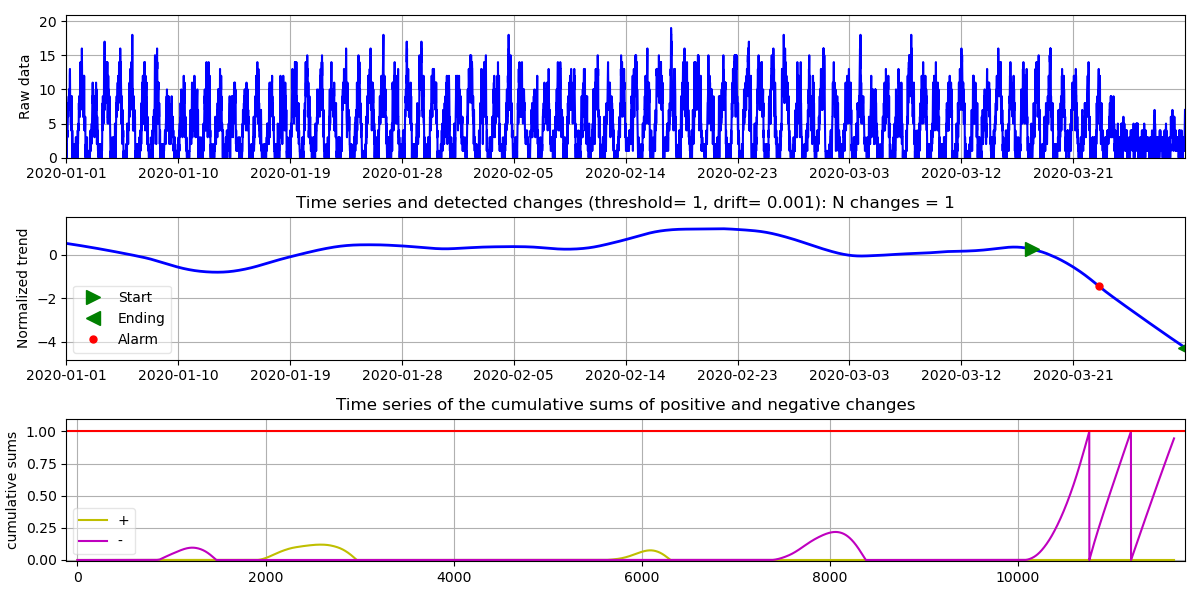}}
	} \\
	\subfloat[The anon.~block with a non Covid-related change. ]{
	\label{fig:other.blocks.outages}
	\mbox{\includegraphics[width=0.9\columnwidth,clip]{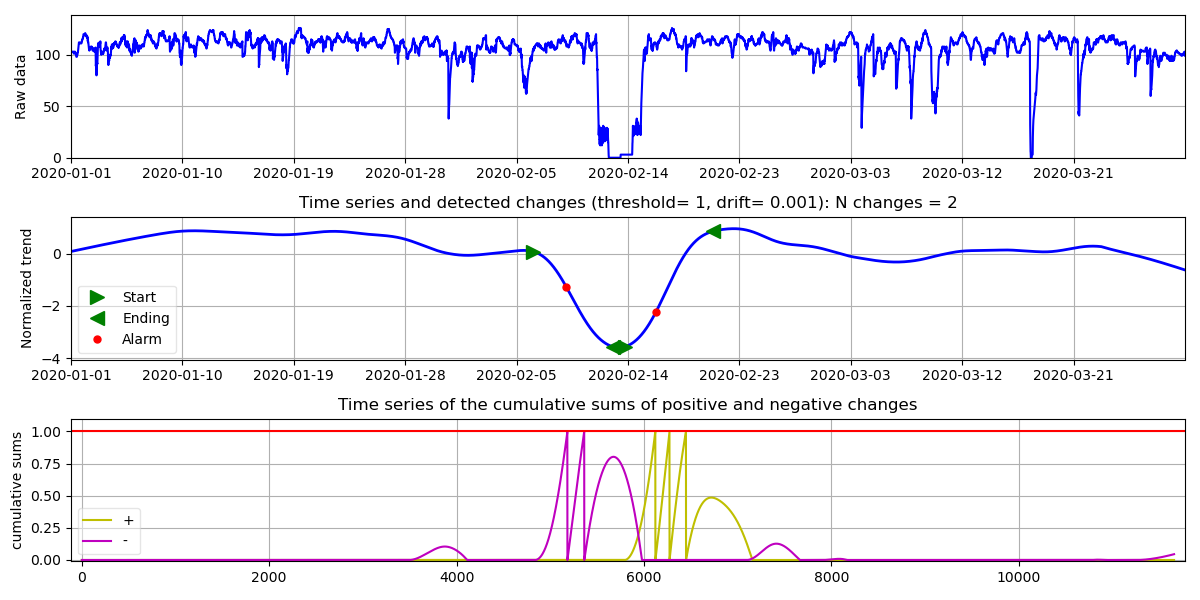}}}
	\caption{Two representative change-sensitive blocks.}
	\label{fig:other.cs.blocks}
\end{figure}

We first show a diurnal block that appears to have a Covid-related change.
In \autoref{fig:other.blocks.zero} we see
  active addresses change from 0 to 20 over each 24 hours,
  a trend that occurs all days of the week, including weekends.
The diurnal behavior disappears on 2020-03-20, suggesting a Covid-related lockdown.
This block is in the U.A.E.,
  so this event roughly matches news reports (see \autoref{sec:validating_discoverability}).

A different block is shown in \autoref{fig:other.blocks.outages}.
This block is change-sensitive,
  with a small but detectable diurnal swing.
This block shows a small decrease in trend
  in the last few days of March,
  but not enough to trigger detection.
It also shows a large change in mid-February,
  with active address dropping from around 100 to 0.
We believe this event corresponds to a network outage or ISP-based reassignment
  of users to another address block.
The pair of a downward trend followed up an upward detection
  shown in the bottom bar is typical of this kind of event.
This example is consistent with the pair of downward and upward trends
  that occur over many blocks 
  in Beijing on 2002-04-15 and -18, shown in \autoref{fig:summary_line_chart_Wuhan}.

\subsection{A Pre-Covid VPN}

\begin{figure}
	\subfloat[Active addresses over 3 months (input data).]{
		\label{fig:diurnal_block_807d3400}
		\mbox{\includegraphics[width=0.75\columnwidth]{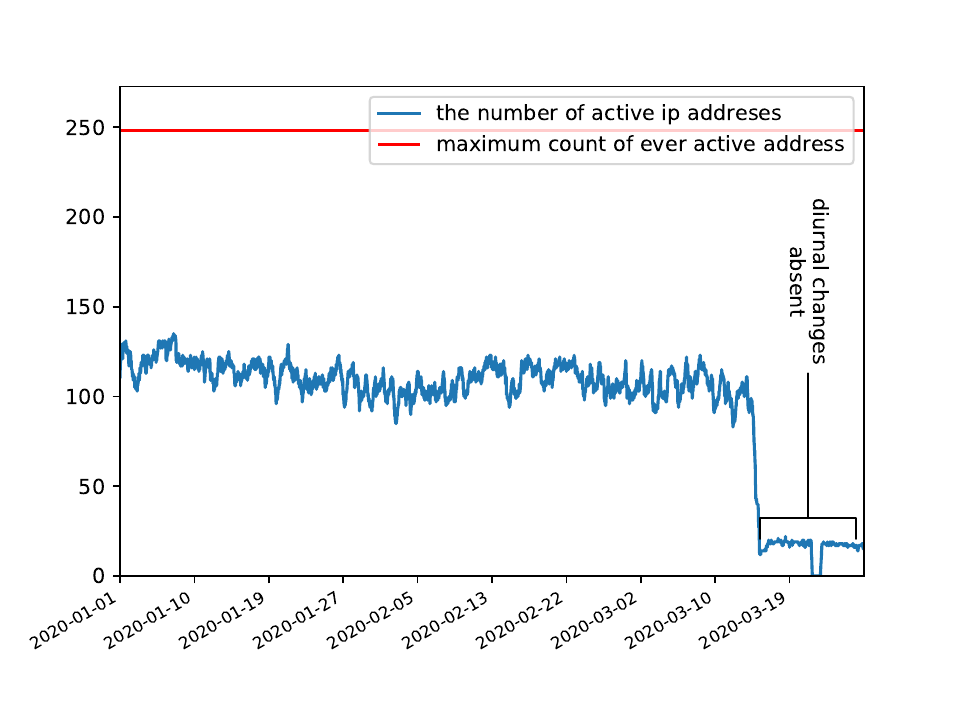}}
	} \\
	\subfloat[STL trend with annotations (top) and CUSUM sums (bottom).]{
		\label{fig:CUSUM_detrend_block_807d3400_a39w}
		\mbox{\includegraphics[width=0.75\columnwidth,trim=0 0 0 235,clip]{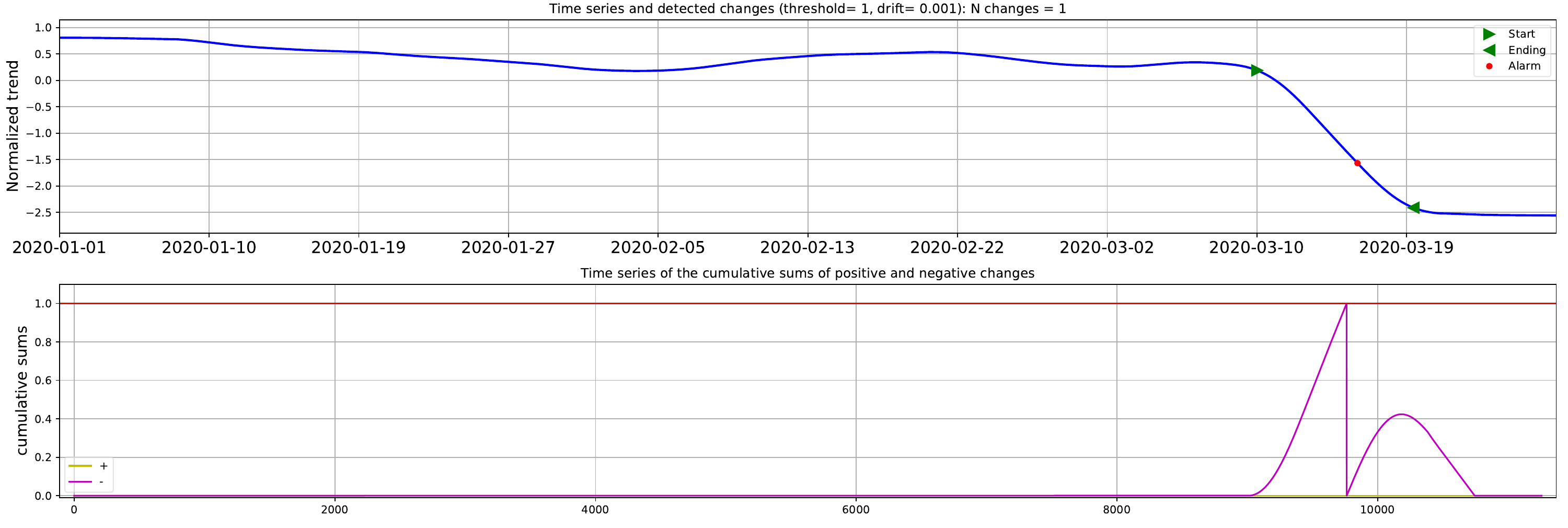}}
	}
	\caption{A VPN block (\AnonOrClear{192.0.3.0/24}{128.125.52.0/24}) and detection.
	}
	\label{fig:vpn_example}
\end{figure}

We next consider a /24 block
\AnonOrClear{(prefix anonymized)}{(128.125.52.0/24)}
that is part of \OurU's VPN
(\autoref{fig:vpn_example}).
We initially determined it was VPN based on reverse DNS addresses,
and then later confirmed this use with \OurU network operators.

\autoref{fig:diurnal_block_807d3400} shows the number of active addresses over time.
We see that after 10 weeks of steady use, 
  the address use drops off significantly,
  just as WFH begins.
This outcome seems backwards from what one would expect
  ---VPN use should go up with WFH.
\OurU network operators explained that
  they shifted the campus VPN to a newer, larger address space
  because of an anticipated increase in use.
Thus use of address space went down because use shifted to another block.

\autoref{fig:CUSUM_detrend_block_807d3400_a39w} 
  confirms that the usage change 
  is found with change-point detection.
This block is classified as change-sensitive block.
We observe the number of active IP addresses has a significant drop around 
  2020-03-15 based on \autoref{fig:diurnal_block_807d3400} 
  and the upper bar chart of \autoref{fig:CUSUM_detrend_block_807d3400_a39w}.
The change point detected around 2020-03-15 reflects 
  ground truth that WFH begins at \OurU.

Our detection algorithms finds this block,
  although it cannot (by itself) 
  show that these users have moved to a different address block.

\subsection{Other University Blocks}
\label{sec:some_university_blocks}

\autoref{fig:additional_universities} shows three additional blocks from U.S.~universities,
  the first two of which are change sensitive and show show WFH in March 2020.
\autoref{fig:block_958e1a00} is change sensitive but does not show WFH\@.

Many of the change-sensitive blocks in the U.S.~are at universities.
We expand on an additional case 
  in \autoref{sec:case_study_indiana}.

\section{A Reconstruction Challenge}
\label{sec:counterexample}

\begin{figure}
	\mbox{\includegraphics[width=0.8\columnwidth]{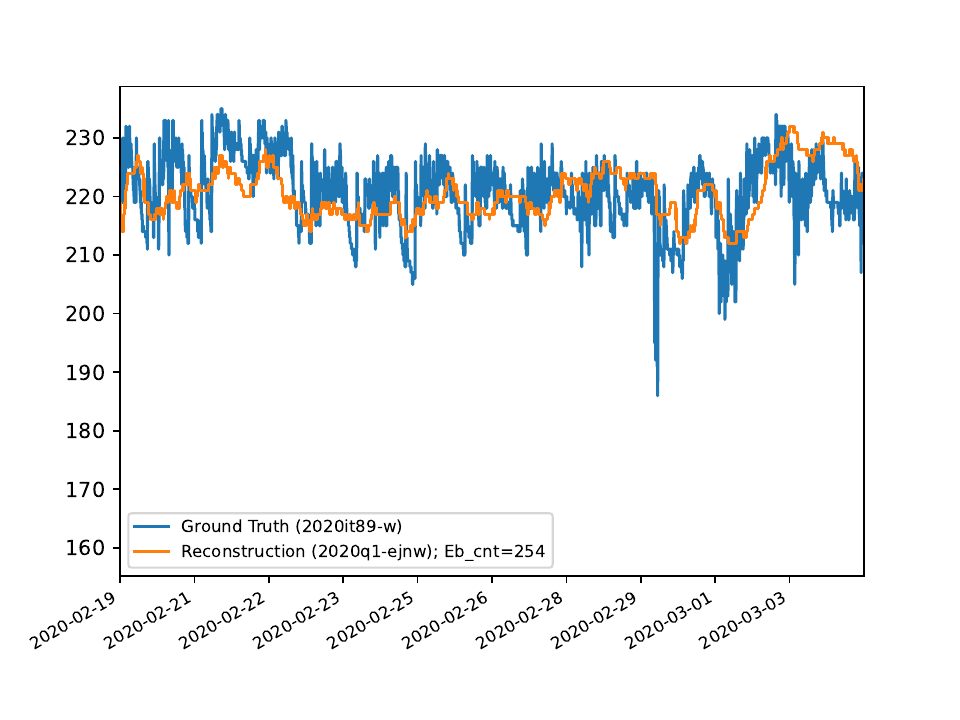}}
	\caption{A /24 block that reconstruction tags as not
		change-sensitive whereas ground truth does.}
	\label{fig:counterexample}
\end{figure}

We explore address reconstruction in \autoref{sec:accumulating_addreses},
  extending it with multiple observers (\autoref{sec:multiple_observers})
  and additional observations (\autoref{sec:alg_reconstruction}).

We showed two examples of 4-way reconstruction compared to ground truth
  in \autoref{fig:trin_survey_block_cbc3c100_c136bb00}.
We next present a /24 block where ground truth
  is change sensitive, but the reconstruction fails to preserve this status.
In \autoref{fig:counterexample} the 4-way reconstruction (the orange line)
  loses daily changes and is not identified as diurnal,
  but the ground truth (blue line) is.
This block has many active addresses ($E(b)$ is 254)
  and most are in use (the mean is around 220 active addresses),
  placing this bock at the upper-right corner of \autoref{fig:counterexample}.
Although Trinocular's stop-on-success rule leaves this block undersampled,
  additional probing will help.

\section{Validating Discoverability: Additional Details}
	\label{sec:discoverability_more_data}

Here we provide graphs documenting our
  random samples used in validating discoverability in  \autoref{sec:validating_discoverability}.

\begin{figure}
	\subfloat[The fraction of blocks that decrease (light red) or increase (dark blue) usage at
	24N,54E in The United Arab Emirates for 2020q1 and q2.]{
		\includegraphics[width=0.7\columnwidth]
		{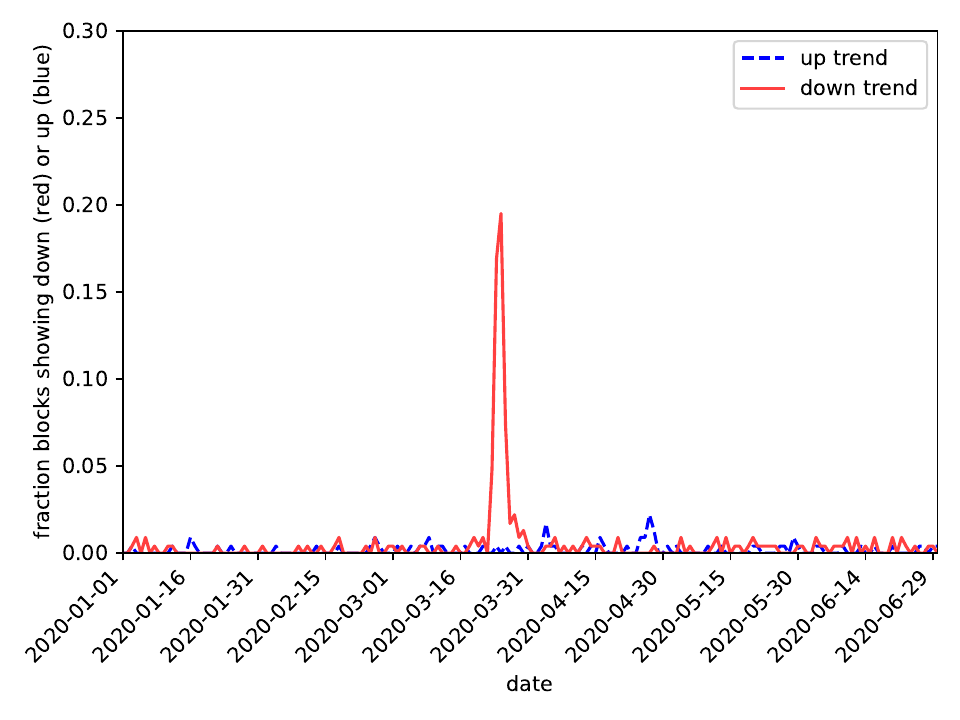}
		\label{fig:summary_line_chart_24N_54E.ARE}
	} \\
	\subfloat[The fraction of blocks that decrease (light red) or increase (dark blue) usage at
	46N,14E in Slovenia for 2020q1 and q2.]{
		\includegraphics[width=0.7\columnwidth]
		{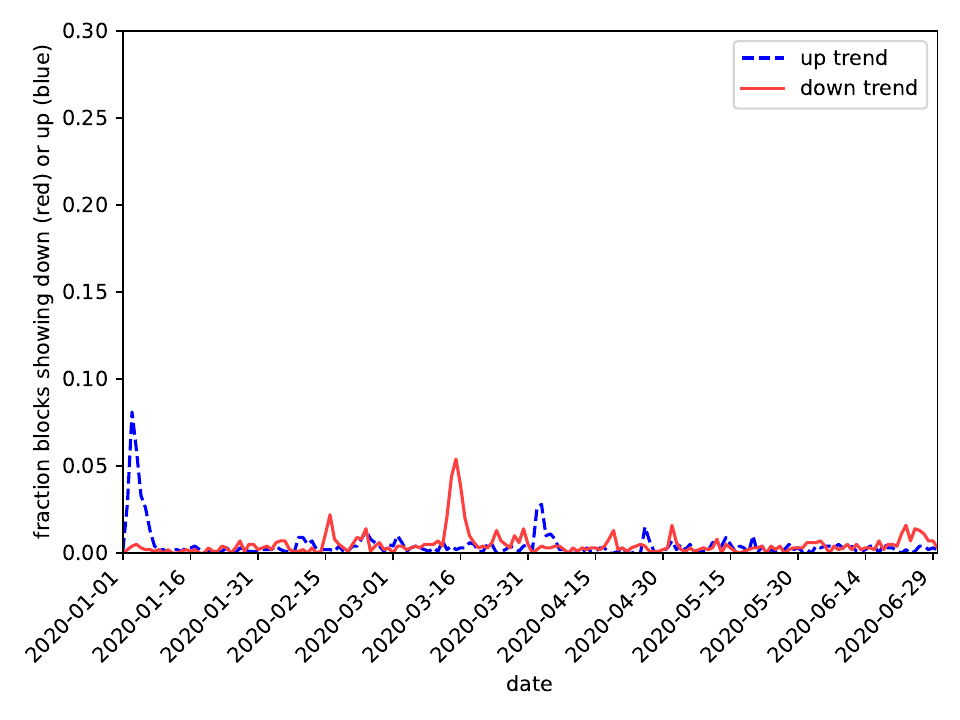}
		\label{fig:summary_line_chart_46N_14E.SVN}
	}
	\caption{The line-chart of block counts, as a function of date ($x$-axis) and the fraction of blocks showing changes ($y$-axis).}
	\label{fi1g:reconstruction.compare}
\end{figure}

\autoref{fig:summary_line_chart_24N_54E.ARE} shows
  the large change near the curfew date, with 
  16.5\% or 38 of change-sensitive blocks showing a decrease in use on 2020-03-24.
This example from UAE confirms our ability to discover Covid-related changes.

Again, the Covid-related changes stand out from noise in
  \autoref{fig:summary_line_chart_46N_14E.SVN}, 
  with 6.1\% or 57 of blocks showing a decrease in use on 2020-03-15 and 
  3.8\% of blocks on 2020-03-16.
This example from Slovenia confirms our approach as well.  

\section{Case Study: More Discovered Events}
  \label{sec:more_wfh_detections}

We examined three real-world events in \autoref{sec:results},
  but we looked at several more.
We look at two events in Thailand and Morocco.

\subsection{Thailand on 2020-02-22}
	\label{sec:CaseStudy:Thailand}

\begin{figure}
		\subfloat[2x2$^{\circ}$ grid dotmap for 2020-02-22.]{
			\label{fig:global_2020-02-22}
			\mbox{\begin{annotationimage}{height=4cm,trim=630 252.7 225 210,clip}
			{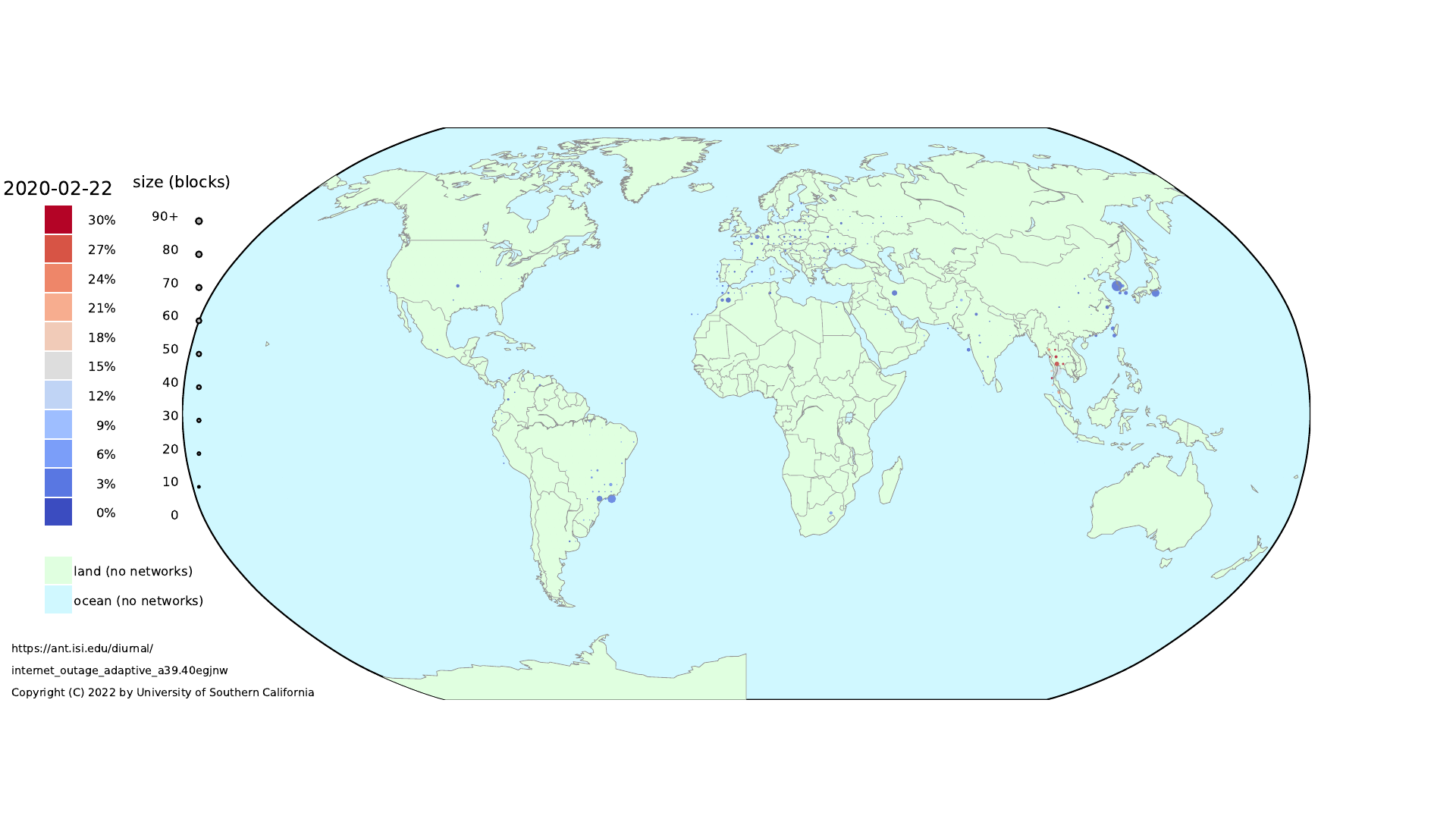}
			\path (0.25,0.2)node(x){Bangkok}(0.6,0.55)node(y){};
			\draw[->,black,font=1pt] (x) -- (y);
			\end{annotationimage}}
		} \\
		\subfloat[Changes for (12N, 100E: Bangkok), 2020h1.]{
			\label{fig:summary_line_chart_12N_100E.Thailand}
			\includegraphics[width=0.7\columnwidth]{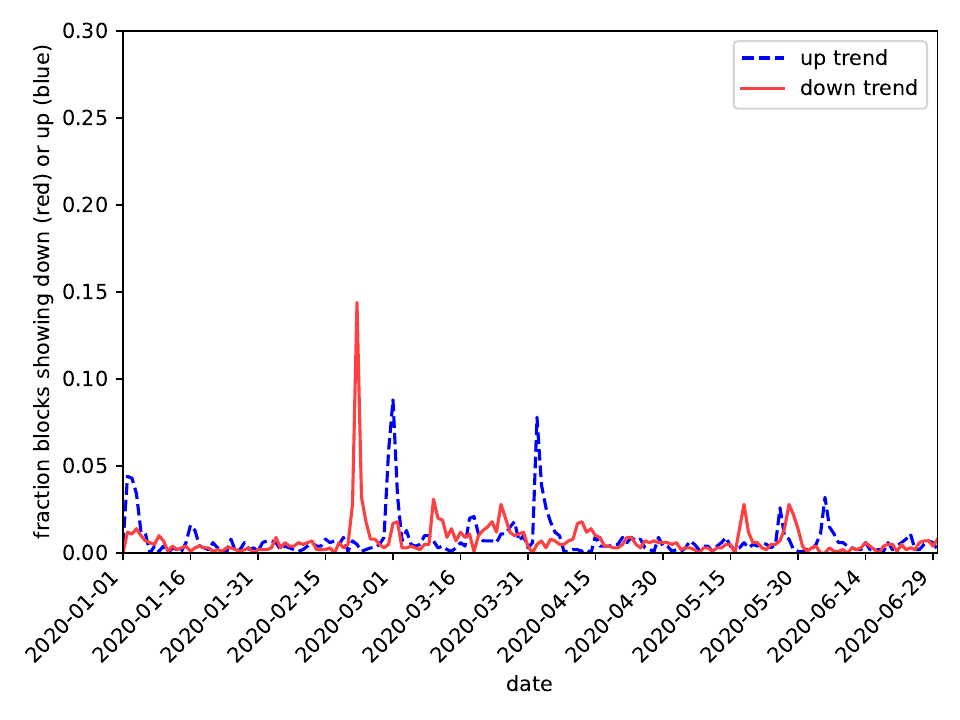}
		}
		\caption{Thailand}
		\label{fig:Thailand_changes}
\end{figure}

We next examine an event in Thailand.
Like India (\autoref{sec:CaseStudy:NewDelhi})
  we discovered this event by browsing our website,
  and when we looked for root causes we found civil unrest.
\autoref{fig:global_2020-02-22} shows the grid 
  with the large number of networks (12N, 100E),
  many of which change on 2020-02-22
  ( 14.3\% or 144 of 1,003 blocks).
Several other nearby grid cells have many fewer blocks,
  but have 50\% or more of them showing changes (the smaller but dark red circles).

\reviewfix{S22A91}
This date corresponds with student protests
  about the dissolution 
  of an opposition party~\cite{Setboonsarng20a,BBC20a}.
Covid-19-related lockdowns did not begin until March and April, 2020~\cite{Wikipedia22a}.
News reports confirm government-imposed curfews due to protests in October 2020~\cite{BBC20b}.
This combination suggests protest-related disruptions keeping
  students at home,
  an event that is not Covid-related,
  but something we count as a true positive

This case study confirms that we detect changes in network usage
  that happen as a result of lockdowns, but that there are
  causes of lockdown other than Covid-19.
In fact,
  \autoref{fig:summary_line_chart_12N_100E.Thailand}
  shows this event was the largest downward change
  for this grid cell in 2020h1.

\subsection{Morocco on 2020-03-24}
\label{sec:CaseStudy:Morocco}

\begin{figure}
	\subfloat[A  $2\times 2^{\circ}$ grid dotmap on 2020-03-23.]{
		\label{fig:global_2020-03-24_morocco}
		\mbox{\includegraphics[height=4cm,trim=40 125 825 130,clip]
		{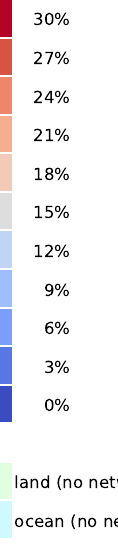}}
		\mbox{\begin{annotationimage}{height=4cm,trim=390 250 380 150,clip}
		{FIG/SWITCH/2020-03-24.fsdb.4sites.pdf}
		\path (0.25,0.2)node(x){Morocco}(0.48,0.67)node(y){};
		\draw[->,black,font=1pt] (x) -- (y);
		\end{annotationimage}}
	} \\
	\subfloat[Block changes over time for the Morocco grid cell.]{
		\label{fig:summary_line_chart_32N_6W_morocco}
		\includegraphics[width=0.7\columnwidth]{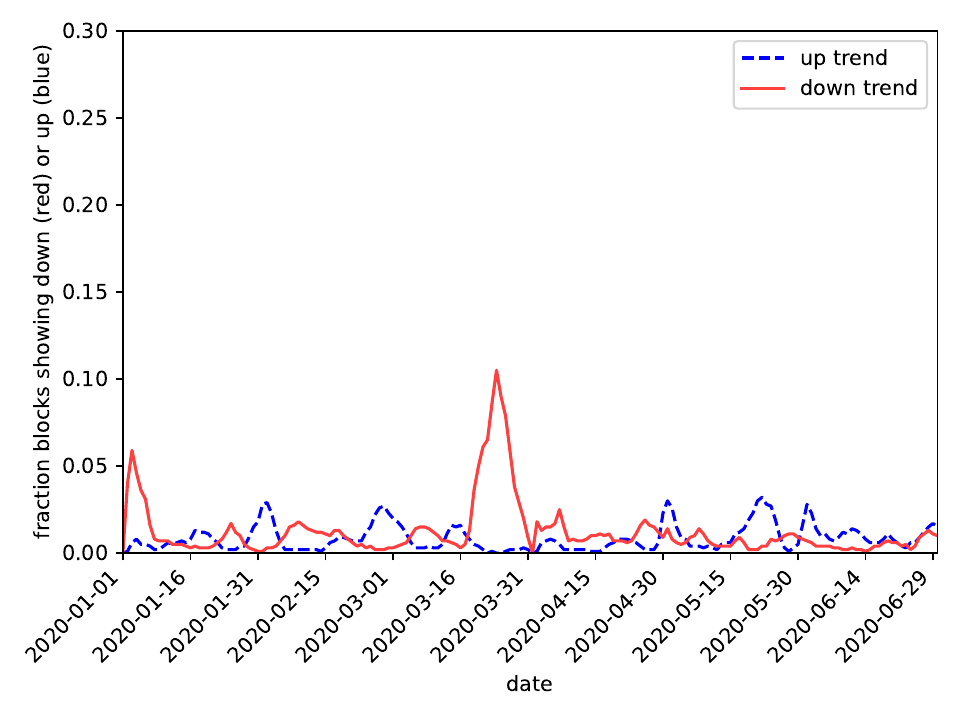}
	}
	\caption{Fraction of block changes in Morocco with Covid-related events on 2020-03-24.}
	\label{fig:Morocco_changes}
\end{figure}

Finally, we examine Morocco. 
We see a large peak of changes in March corresponding with 
  announcement of a Covid-19 emergency in the country.

\autoref{fig:global_2020-03-24_morocco} shows our 2x2\,degree grid of 
  changes on 2020-03-24.
We observe that a grid cell consisting of 13,989 change-sensitive blocks 
  in Morocco (at 32N,6W) is triggered by a large change --- about
  10.5\% or 1,472
  change-sensitive blocks showing downward trends,
  as confirmed in \autoref{fig:summary_line_chart_32N_6W_morocco} showing 
  a series of downward trends detected starting from 2020-03-19 to 2020-03-27.

These changes are related to activities in Morocco.:
  A state of medical emergency responded to Covid-19 was declared on 2020-03-19 to 
  take effect on 2020-03-20 at 6\,pm local time~\cite{Morocco1}.

\subsection{Indiana on 2020-03-15}
	\label{sec:case_study_indiana}

\begin{figure}
  \includegraphics[width=3.3in]{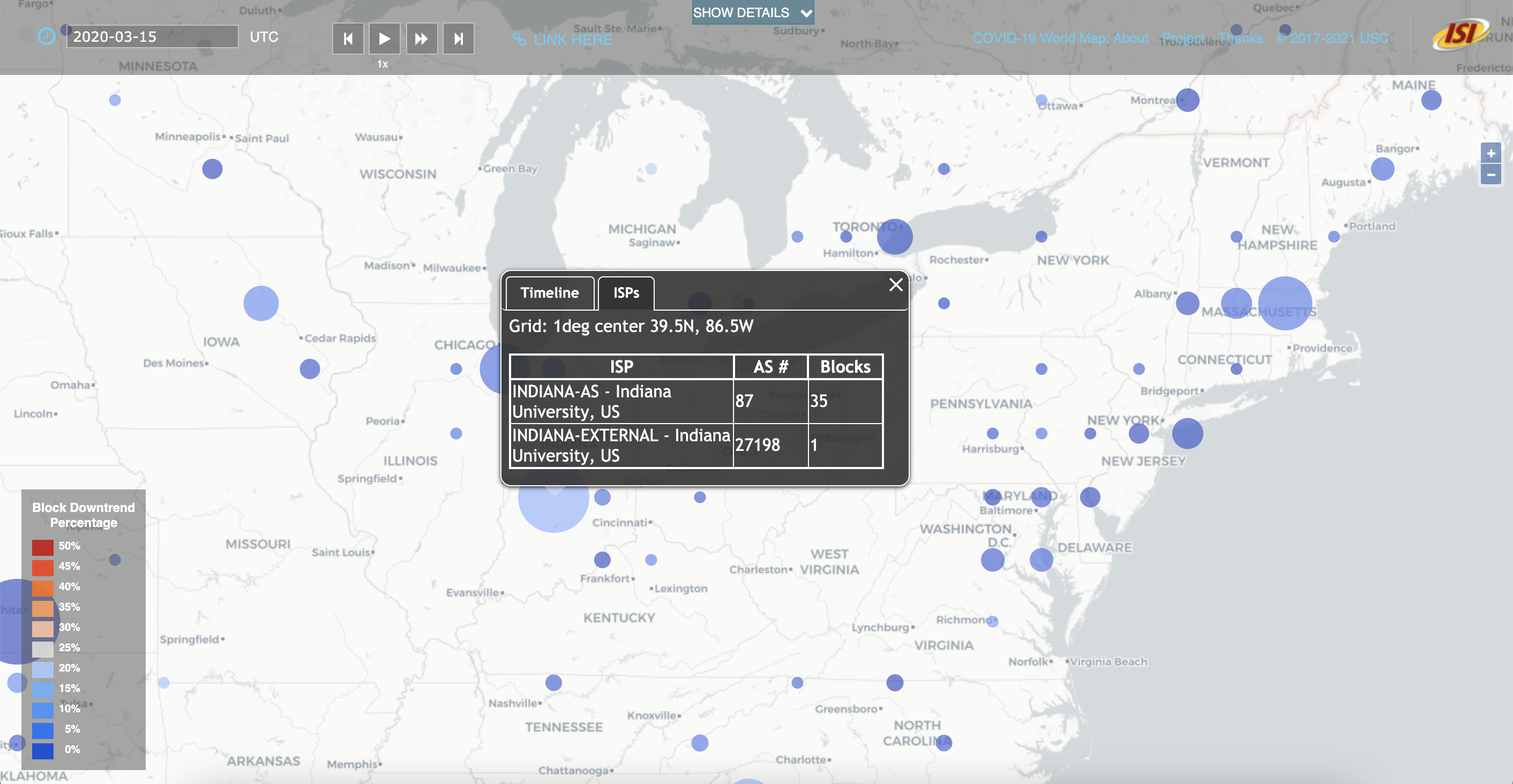}
\caption{An event discovered through our interactive website. Source:~\cite{Stutz21a}, Figure 8.}
\label{fig:Indiana-2020-03-15}
\end{figure}

To understand applicability of our approach in North America
  we explored WFH events in the United States
  through our website~\cite{Stutz21b}.
We observed a moderately large event in Indiana,
  appearing as a large, light-blue circle in \autoref{fig:Indiana-2020-03-15}.

Our website supports examination of the underlying blocks.
we can see that 36 blocks in Indiana University (AS87 and AS27198)
  were detected as WFH on 2020-03-15.
This data corresponds with the beginning of spring break (Friday, 2020-03-13)
  followed by remote learning beginning on 2020-03-19~\cite{Robel20a}.

This example shows the use of our algorithms and website to discover
  an event unknown to us.
It also shows the role of universities for having change-sensitive networks.
Universities often have large IPv4 allocations
  (as Autonomous System 87, IU was an early adopter)
  and so are able to use public IP addresses for dynamic use.
  
\label{page:last_page}

\clearpage

\end{document}